\documentclass[]{elsart}
\usepackage{natbib}
\usepackage{amssymb}
\usepackage{amsmath}
\usepackage{amsopn}
\journal{Physics Reports}
\usepackage{epsfig}
\newcommand{\lanf}{\textquotedblleft}
\newcommand{\ranf}{\textquotedblright}
\newcommand{\remark}[1]{}

\newcommand{\ket}[1]{\ensuremath{|#1\rangle}}

\newcommand{\braket}[2]{\ensuremath{\langle #1|#2\rangle}}

\newcommand{\new}[1]{#1}
\newcommand{\newnew}[1]{}
\newcommand{\old}[1]{}

\begin{document}

\begin{frontmatter}

% Title, authors and addresses

% use the thanksref command within \title, \author or \address for footnotes;
% use the corauthref command within \author for corresponding author footnotes;
% use the ead command for the email address,
% and the form \ead[url] for the home page:
% \title{Title\thanksref{label1}}
% \thanks[label1]{}
% \author{Name\corauthref{cor1}\thanksref{label2}}
% \ead{email address}
% \ead[url]{home page}
% \thanks[label2]{}
% \corauth[cor1]{}
% \address{Address\thanksref{label3}}
% \thanks[label3]{}

\title{Quantum computing with trapped ions}

% use optional labels to link authors explicitly to addresses:
% \author[label1,label2]{}
% \address[label1]{}
% \address[label2]{}

%\author{H.~H{\"a}ffner$^{1,2}$, C.F.~Roos$^{1,2}$, W.~H{\"a}nsel$^{1}$, R.~Blatt$^{1,2}$}

%\address{$^1$ Institut f{\"u}r Experimentalphysik, Universit{\"a}t
%Innsbruck, Technikerstra{\ss}e 25, A-6020 Innsbruck, Austria}
%\address{$^2$ Institut f\"ur Quantenoptik und Quanteninformation,
%\"Osterreichische Akademie der Wissenschaften, Technikerstra{\ss}e 21a, A-6020 Innsbruck, Austria}

\author[OEAW,EXP,BERK,LBNL]{H.~H{\"a}ffner}
\corauth[cor]{Corresponding author.}
\ead{hartmut.haeffner@uibk.ac.at}
\author[OEAW,EXP]{C.~F.~Roos}
\author[OEAW,EXP]{R.~Blatt}

\address[OEAW]{Institut f\"ur Quantenoptik und Quanteninformation,
\"Osterreichische Akademie der Wissenschaften, Technikerstra{\ss}e 21a, A-6020 Innsbruck, Austria}

\address[EXP]{Institut f{\"u}r Experimentalphysik, Universit{\"a}t
Innsbruck, Technikerstra{\ss}e 25, A-6020 Innsbruck, Austria}

\address[BERK]{\new{Dept. of Physics, University of California, Berkeley, CA 94720, USA}}
\address[LBNL]{\new{Materials Sciences Division, Lawrence Berkeley National Laboratory, Berkeley, CA 94720, USA}}

\date{\today}

\begin{abstract}
Quantum computers hold the promise to solve certain computational task much more efficiently
than classical computers.
We review the recent experimental advancements towards a quantum computer with trapped ions. In particular, various implementations of qubits, quantum gates and some key experiments are discussed. Furthermore, we review some implementations of quantum algorithms such as a deterministic teleportation of quantum information and an error correction scheme.
\end{abstract}

\begin{keyword}
% keywords here, in the form: keyword \sep keyword

Quantum computing and information, entanglement, ion traps

%\pacs{}

%03.67.Lx Quantum computation
%03.65.Ud Entanglement and quantum nonlocality (e.g. EPR paradox, Bell's
%inequalities, GHZ states, etc.)
%32.80.Qk   Coherent control of atomic interactions with photons

% PACS codes here, in the form: \PACS code \sep code

\end{keyword}

\end{frontmatter}

\tableofcontents

\section{Introduction}
\remark{Preface}
The aim of this article is to review the recent development of ion trap quantum computing. The field evolved rapidly in the recent decade. Thus the many facets of experimental ion trap quantum computing and its techniques cannot be covered all. Instead, we want to present here a coherent picture of the most important experimental issues  and refer the reader to the original publications for the details. We also describe some of the milestones achieved so far in ion trap quantum computing, like teleportation of quantum states and quantum error correction. Still much of the work especially towards shuttling ions with segmented traps is only touched upon.

% such important advancements as for instance  the implementation of the \lanf Semiclassical Fourier transformation\ranf \citep{Chiaverini:2005} and \lanf Purification of two-atom entanglement\ranf \citep{Reichle2006a} are not covered.

\remark{Motivation}
A quantum computer uses the principles of quantum mechanics to solve certain mathematical problems faster than normal computers.  Such a quantum computer processes quantum information whose most basic unit is called a quantum bit (qubit).  Already a small quantum computer, consisting of forty qubits\footnote{To describe the (arbitrary) state of a forty qubit system, $2^{40}$ complex numbers are necessary. Already this requirement exceeds the capacities of current super computers.}, could solve quantum mechanical problems that are intractable with current computers. In particular, the study of quantum mechanical many body systems would benefit considerably from such a device \citep{Feynman1982,Lloyd1996}.
\remark{Development of quantum computation}
\new{In 1989, David Deutsch discovered a mathematical problem which can be solved faster with quantum mechanical means than with classical ones \citep{Deutsch1989}. But it was a few years later when the rapid development of quantum computation set in, marked by Peter Shor's discovery of a quantum algorithm with which large numbers can be factored much faster than with today's classical algorithms \citep{Shor1994}. }

Shortly afterwards, Ignacio Cirac and Peter Zoller found a physical system on which such quantum algorithms could be implemented \citep{Cirac1995}: single trapped ions were supposed to carry the quantum information, which is manipulated and read out with focused laser beams. Already within a year's time, David Wineland's group at National Institute of Standards and Technology demonstrated the heart of such an ion-trap quantum computer \citep{Monroe1995}, a controlled bit flip on a single ion. Without exaggeration one can say that those two publications mark the birth of experimental quantum computation. Rapidly also other implementations were considered. In particular, liquid-state nuclear magnetic resonance was used to demonstrate a quantum algorithm \citep{Gershenfeld1997,Chuang1998a,Jones1998} and later even Shor's algorithm was implemented on a seven qubit register \citep{Vandersypen2001}. On the theoretical side, many new interesting implications were found. For quantum computation, the most relevant implication was the discovery of quantum error correction protocols by Peter Shor \citep{Shor1995} and Andrew Steane \citep{Steane1996}. These protocols allow for the implementation of arbitrary long quantum algorithms without perfect control.

\remark{Basics of quantum information}
The basic unit of quantum information (a qubit) can be implemented with a two-level system such as the electron's spin in a magnetic field or using two levels of an atom. A simple quantum computation initializes the qubits, manipulates them and finally reads out the final states. Any physical implementation of quantum computation must be able to perform these tasks. Thus the physical system must satisfy the requirements laid out in Sec.~\ref{sec:DiVincenzo-criteria} to qualify as a universal quantum computer \citep{DiVincenzo2001}.
\newnew{*********************************************** removed ******************}
\begin{enumerate}
  \item It must be a scalable physical system with well characterized qubits.
  \item It must be possible to initialize the qubits.
  \item The qubits must have a coherence time much longer than the operation time.
  \item There has to be a universal set of quantum gates. In the most simple case one considers single-qubit and entangling two-qubit gates.
  \item A qubit-specific measurement must be attainable.
\end{enumerate}
Furthermore, to build up a quantum network, one requires:
\begin{enumerate}
\addtocounter{enumi}{5}
\item The ability to interconvert stationary and flying qubits.
\item The ability to faithfully transmit flying qubits between specified locations.
\end{enumerate}
\newnew{*********************************************** removed *********************}
\remark{Different experimental approaches to a quantum computer}
In principle, these requirements can be fulfilled with a number of physical approaches, like nuclear magnetic resonance \citep{Gershenfeld1997}, cavity quantum electrodynamics (cavity-QED) \citep{Raimond2001}, Josephson junctions \citep{Makhlin2001}, a combination of circuit-QED and Josephson junctions \citep{Wallraff2004,Majer2007} and quantum dots \citep{Loss1998,Petta2005}. Methods using linear optics have also been proposed \citep{Knill2001a} and actually were used for quantum information processing \citep{Walther2005b,Lanyon:2008}, however, here the emphasis is naturally more on the transmission of quantum information rather than processing it.

Using nuclear magnetic resonance, quite a number of impressive demonstration experiments have been performed. Unfortunately, the state of the quantum register (molecules) can only be poorly initialized (except in special cases \citep{Jones2000}), making the scaling properties of NMR quantum computation not very promising \citep{Warren1997,Jones2000,Linden2001}. At the moment Josephson junctions seem to be very appealing, especially when combined with superconducting strip line cavities \citep{Wallraff2004,Majer2007}. Recently, \citet{Steffen2006a} and \citet{Plantenberg2007} demonstrated Bell-states and a controlled-NOT, respectively, with pairs of Josephson junction qubits.

\section{Ion trap quantum computers}
\remark{Why ion-traps for quantum computing?}
Of the many approaches that have been proposed for constructing a quantum computer, trapped ions are currently one of the most advanced \citep{ARDA2005}.
This is also reflected in the fact that around the year 2000 only half a dozen groups pursued experimental ion-trap quantum computing while in 2008 there are  more than 25 groups working in the field.

Already long before the idea of quantum computation was picked up by experimentalists, four out of the five core criteria required by DiVincenzo for a quantum computer were demonstrated with trapped ions in the laboratory:
initialization \citep{Wineland1980} and read-out  of the internal electronic states of trapped ions \citep{Nagourney1986,Sauter1986,Bergquist1986}, extremely long coherence times \citep{Bollinger1991} and laser cooled ion crystals with many ions \citep{Diedrich1987,Wineland1987,Raizen1992a,Raizen1992b} serving as a qubit register. Then \citet{Cirac1995} realized that  a quantum computation can be carried out by coupling the ions via a collective motional degree of freedom.
% by addressing the individual ions in a string with cleverly chosen laser pulses
Thus, a route to implement the missing conditional evolution of physically separated qubits (a two-qubit interaction) was introduced. In addition they proved that the size of the resources necessary to control trapped ions does not increase exponentially with the number of qubits \citep{Cirac1995}.

\remark{Brief history of ion trap QC}
Soon after the proposal by Cirac and Zoller in 1995, the NIST ion storage group around David Wineland implemented the key idea of the proposal ---a conditioned phase shift--- with a single Be$^+$-ion  \citep{Monroe1995}. Furthermore they demonstrated a few other two-qubit gate candidates \citep{Sackett2000,DeMarco2002,Leibfried2003a}, entangled up to four ions \citep{Sackett2000}, demonstrated a so-called decoherence free subspace \citep{Kielpinski2001} and simulated a nonlinear beam-splitter \citep{Leibfried2002}.

In our group in Innsbruck, the Deutsch-Josza algorithm was demonstrated with a single Ca$^+$-ion \citep{Gulde2003}, followed by the first implementation of a set of universal gates on a two-ion string \citep{Schmidt-Kaler2003}. In addition, the creation of various entangled states and the partial read-out of an entangled
quantum register was demonstrated \citep{Roos2004a}.

Further milestones in ion-trap quantum computing were experiments on quantum teleportation by both groups \citep{Barrett2004,Riebe2004}, an error correction protocol by the NIST-group \citep{Chiaverini:2004a}, entanglement
of six and eight particles \citep{Leibfried2005,Haeffner2005a} by both groups, respectively and entanglement purification \citep{Reichle2006a}. Recently,ions in separate traps have been also entangled using ion-photon entanglement by the Ann-Arbor group \citep{Blinov2004,Maunz2007,Moehring2007,Matsukevich2008}.
\old{The Ann-Arbor group demonstrated an entangling two-qubit gate \citep{Haljan:2005a} with cadmium ions and implemented Grover's search
algorithm \citep{Brickman2005}.  Finally, the Oxford group implemented an entangling gate with Ca$^+$-ions \citep{Home:2006a}.
}

Currently, miniaturization and integration of segmented
ion traps is rapidly progressing. Already for some time, the NIST group has been successfully using  microfabricated segmented traps to relieve the difficulties of single ion addressing.
Inspired by this success, in the mean time virtually all ion trap groups started to develop segmented trap technologies. Furthermore, U.S. funding bodies initiated contacts between the various groups to further ion trap related technologies  and established
contacts to microfabrication laboratories, such as Lucent Technologies and Sandia National Laboratories. In Europe the 'Specific Targeted Research Project' MICROTRAPS with six participating groups has been formed to further microfabricated ion trap technologies.
% \begin{itemize}
% \item D. Wineland (Boulder)
% \item R. Blatt (Innsbruck)
% \item A. Steane (Oxford)
% \item C. Monroe (Ann Arbor)
% \item M. Drewsen (Aarhus)
% \item I.L. Chuang (MIT)
% \item F. Schmidt-Kaler (Ulm)
% \item J. Eschner (Barcelona)
% \item T. Sch\"atz (MPQ)
% \item S. Urabe (Osaka)
% \item G. Werth (Mainz)
% \item D. Kielpinsky (Griffith)
% \item Los Alamos
% \item P. Gill (NPL)
% \item R. Slusher (Lucent)
% \item C. Wunderlich (Siegen)
% \item W. Lange / W. Hensinger (Sussex)
% \item Sandia
% \item N. Fortson (Seattle)
% \item P.C. Haljan (Simon Fraser)
% \end{itemize}

\subsection{Principles of ion-trap quantum computers}\label{sec:DiVincenzo-criteria}
A excellent overview and detailed account of the fundamental issues of ion trap quantum computing is given by \citet{Wineland1998} and by \citet{Sasura2002}. Furthermore,
\citet{Leibfried2003b} review the progress towards the manipulation and control of single ions. Very recently, the generation and applications of entangled ions were reviewed by \citet{Blatt:2008}.

We start here by summarizing how ion-trap quantum computers fulfill the DiVincenzo  criteria mentioned above \citep{DiVincenzo2001}:
\begin{enumerate}

\item \textbf{A scalable physical system with well characterized qubits:} long-lived internal levels of the ions serve as the qubits (see Sec.~\ref{sec:ion-qubits}).
The qubit register is formed by strings of ions in a
(linear) trap. \new{While this approach is in principle scalable, it is desirable to distribute the ions among multiple traps \citep{Wineland1998,Kielpinski2002}). Thus complications, due to the increased mass and more the complicated mode structure of large ion strings can be circumvented. First steps in this direction will be briefly discussed in Sec.~\ref{sec:scaling}.}

\item \textbf{The ability to initialize the state of the qubits:} this \new{is most easily} achieved by optical pumping to a well-defined electronic state. Fidelities of 0.99 are typical, in Sec.~\ref{sec:optical-pumping} also methods to achieve much higher fidelities will be discussed.

\item \textbf{A coherence time much longer than the operational
time:} in current quantum computing experiments, typically
coherence times of \new{a few milliseconds} are achieved which are
about one to two orders of magnitude longer than the time scale
for quantum operations  (see Sec.~\ref{sec:decoherence} for
details). The coherence time is often limited by magnetic field
fluctuations. on the other hand coherence times of more than 10~s
have been demonstrated with Raman-transitions between
magnetic-field insensitive transitions \citep{Langer2005}. Similar
observations were made in $^{43}$Ca$^+$ by the Oxford and
Innsbruck groups \citep{Lucas:2007,Benhelm:2008c}. Furthermore,
\citet{Bollinger1991} demonstrated a coherence time of more than
10 minutes using a microwave drive instead of a Raman-laser setup.
However, so far only one experiment demonstrated the combination
of high-fidelity quantum gates and magnetic field insensitive
transitions \citep{Haljan:2005b}.

\item \textbf{A universal set of quantum gates:}\begin{enumerate}

\item Single qubit gates \new{are} implemented by driving Rabi oscillations between the two qubit levels with resonant laser
pulses (see Sec.~\ref{sec:single-qubit-gates}). The gates can be represented as rotations of the Bloch sphere where the axis of rotation can be selected by changing the phase of the exciting laser field (single photon transition)
or the phase difference of the two Raman beams (see Sec.~\ref{sec:single-qubit-gates}).
A single-qubit phase-gate can be produced directly by an off-resonant laser via an AC-Stark shift (see Sec.~\ref{sec:single-qubit-gates}). \new{}
\item For two-qubit gates usually the long range interaction due to the Coulomb force is employed. In the original proposal by
Cirac and Zoller, the quantum information of one ion is swapped to the common motional degree of freedom of the ion string \citep{Cirac1995}. Then
an operation conditioned on the motional state can be carried out on a second ion before the quantum information is swapped back from the motion to the first ion.
\new{Section~\ref{sec:two-qubit-gates} details this idea as well as other methods to implement multi-qubit gates. Implementations of some of the two-qubit gate recipes will be discussed in Secs.~ \ref{sec:cz-experiments} and \ref{sec:entangled-states}}.
\end{enumerate}

\item\textbf{ A qubit-specific measurement:} one of the qubit levels \old{can be}\new{is} excited on a strong transition to a higher lying auxiliary short-lived level while the other qubit level remains untouched. Thus,  fluorescence from the decay is detected only if the qubit is projected to the qubit level which is coupled to the auxiliary transition (see Sec.~\ref{sec:state-detection}).
\end{enumerate}
Additionally DiVincenzo requires:
\begin{enumerate}
\addtocounter{enumi}{5}

\item\textbf{The ability to interconvert stationary and flying qubits: }
ions can be stored in a high-finesse cavity. Thus, the ions' internal state can be mapped onto a photonic state \citep{Cirac1997}.

\item \textbf{ The ability to faithfully transmit flying qubits between specified
locations: } a photon can be transmitted through a fiber and at the target location coupled via another
high-finesse cavity to the target ion.
\end{enumerate}
There exist further possibilities to connect two distant ion trap quantum computers. One of them will be briefly discussed in Sec.~\ref{sec:entanglement-purification}.

For practical applications, the DiVincenzo criteria have not only to be fulfilled, but also the fidelity and speed of the implementations have to be considered (see Sec.~\ref{sec:threshold}). Furthermore, it is highly desirable to implement all operations as parallel as possible.

\subsection{The basic Hamiltonian}\label{sec:basic-hamiltonian}
\begin{figure}[tb!]
\begin{center}
\includegraphics[width=0.7\textwidth]{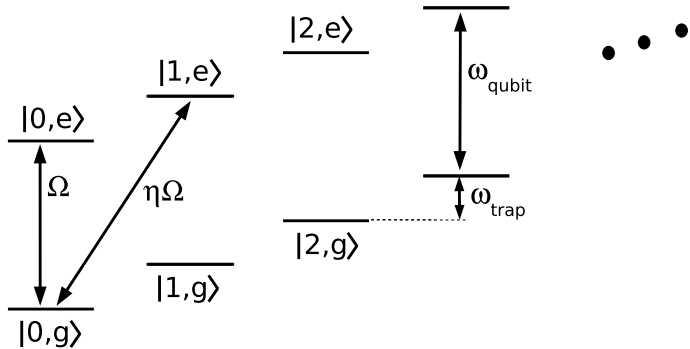}
\caption{\label{fig:ion-tensor-oscillator}
Energy level scheme of a single trapped ion with a ground ($\ket{g}$) and an excited ($\ket{e}$) level in a harmonic trap (oscillator states are labeled $\ket{0},\ket{1},\ket{2},\cdots$). $\Omega$ denotes the
carrier Rabi frequency. The Rabi frequency on the blue sideband transition $\ket{0,e}\leftrightarrow\ket{1,g}$ transition is reduced by the Lamb-Dicke factor $\eta$ as compared to the carrier transition (see
Eq.~\protect\ref{eq:blue-rabi-frequency}
). The symbols $\omega_{\rm qubit}$ and $\omega_{\rm t}$ denote the qubit and the trap frequency, respectively.}
\end{center}
\end{figure}

We will now briefly discuss the Hamiltonian of  two-level systems interacting with a quantized harmonic oscillator via laser light. For more detailed discussions, we refer to Refs.~\citep{Wineland1998} and \citep{Leibfried2003b}.
The basic level scheme of such a system is displayed in Fig.~\ref{fig:ion-tensor-oscillator}.
We start out by writing the Hamiltonian for a trapped single ion interacting with near resonant laser light, taking into account only two levels of the ion and one vibrational mode \new{(taken to be oriented along the $z$ direction)}:
\begin{equation}\label{eq:ion-trap-hamiltonian-no-Lamb-Dicke}
H=\hbar \Omega  \sigma_+ e^{-i (\Delta \, t - \varphi)}\exp\left( i \eta \left[a e^{-i \omega_{\rm t} t } + a^\dagger e^{i\omega_{\rm t} t} \right] \right)  + {\rm h.c.} \:.
\end{equation}
Here, $\sigma_\pm$ is either the atomic raising or the atomic lowering operator, while $a^\dagger$ and $a$ denote the creation and annihilation operator for a motional quantum, respectively. $\Omega$ characterizes the strength of the laser field in terms of the so-called Rabi frequency, $\varphi$ denotes the phase of the field with respect to the atomic polarization and
$\Delta$ is the laser-atom detuning. $\omega_{\rm t}$ denotes the trap frequency, $\eta=k_{z} z_0$ is the Lamb-Dicke parameter \new{with $k_{z}$ being the projection of the laser field's  wavevector along the $z$ direction} and $z_0=\sqrt{\hbar/(2m\omega_{\rm t})}$ is the spatial extension of the ion's ground state wave function in the harmonic oscillator ($m$ is here the ion's mass). We mention also that the rotating wave approximation has been applied which assumes that both the laser detuning and Rabi frequency are much smaller than optical frequencies. A similar treatment can be carried out for qubits based on Raman-transitions by eliminating the virtual level through which the two qubits are coupled.  Please note that in our definition the Rabi frequency measures the frequency with which the population is exchanged in contrast to the definition used by \citet{Wineland1998} and \citet{Leibfried2003b}.

Using the Lamb-Dicke approximation ($\eta\sqrt{\langle (a+a^\dagger)^2\rangle} \ll 1$) which is almost always valid for cold tightly bound ion strings, we can rewrite Eq.~\ref{eq:ion-trap-hamiltonian} \citep{Leibfried2003b,Jonathan2000}:
\begin{eqnarray}\label{eq:ion-trap-hamiltonian}
 H&=&\hbar \Omega\left\{\sigma_+e^{-i(\Delta\, t - \varphi)} + \sigma_-e^{i( \Delta \, t - \varphi)}\right. \\ \nonumber
 & & \left.+ i \eta (\sigma_+ e^{-i(\Delta\, t - \varphi)} - \sigma_-e^{i( \Delta \, t - \varphi)})\left(a e^{-i\omega_{\rm t} t} + a^\dagger e^{i\omega_{\rm t} t} \right)\right\}\:.
\end{eqnarray}
Three cases of the laser detuning $\Delta$ are of particular interest (see Fig.~\ref{fig:ion-tensor-oscillator}): $\Delta=0$ and $\Delta=\pm \omega_{\rm t}$.
This becomes apparent if a second rotating wave approximation is carried out where it is assumed that only one transition is relevant at a time. Discarding time dependent terms, we thus arrive at
\begin{enumerate}
 \item  the Hamiltonian describing the carrier transition ($\Delta=0$)
\begin{equation}\label{eq:hamiltonian-carrier}
 H^{\rm car}=\hbar \Omega (\sigma_+e^{i\varphi}+\sigma_-e^{-i\varphi})\:.
\end{equation}
 Here only the electronic states $\ket{g}$ and $\ket{e}$ of the ion are changed.
\item  the Hamiltonian describing the blue sideband transition ($\Delta=\omega_{\rm t}$)
\begin{equation}\label{eq:hamiltonian-blue}
H^+=i\hbar \Omega \eta(\sigma_+a^\dagger e^{i\varphi}-\sigma_- a e^{-i\varphi})\:.
\end{equation}
Simultaneously to exciting the electronic state of the ion, in this case a motional quantum (a phonon) is created. Within this two-level system, Rabi flopping with the Rabi frequency
\begin{equation}
 \label{eq:blue-rabi-frequency}
\Omega_{n,n+1}= \sqrt{n+1}\eta  \:  \Omega
\end{equation}
  occurs, where $n$ describes the number of motional quanta (phonons). For convenience, we define the blue sideband Rabi frequency $\Omega_+ = \Omega_{0,1}$ which describes the flopping frequency between the $\ket{g,0}$ and the $\ket{e,1}$ state.
\item  the Hamiltonian describing the red sideband transition ($\Delta=-\omega_{\rm t}$)
\begin{equation}\label{eq:hamiltonian-red}
H^-=i\hbar \Omega \eta(\sigma_- a^\dagger e^{-i\varphi}+\sigma_+ a e^{i\varphi})\:.
\end{equation}
Simultaneously to exciting the electronic state, here a phonon is destroyed and Rabi flopping with the Rabi frequency
\begin{equation}
 \label{eq:red-rabi-frequency}
\Omega_{n,n-1}= \sqrt{n} \eta  \: \Omega
\end{equation}
 takes place.
\end{enumerate}

Naturally, one is  interested in driving the sideband transitions as fast as possible to speed up the quantum operations based on the sideband transitions. However, especially for small Lamb-Dicke factors, the second rotating wave  approximation performed to obtain Eqs.~\ref{eq:hamiltonian-carrier}-\ref{eq:hamiltonian-red} is then not satisfied for Rabi frequencies $\Omega$ comparable to the trap frequency $\omega_{\rm t}$: as  can be seen from Eq.~\ref{eq:blue-rabi-frequency} for ions in the motional ground state, the carrier transition is stronger by a factor of $1/\eta$ as compared to the sideband transitions. Therefore, driving the weak sideband transitions with strong laser fields, Stark shifts and off-resonant excitations arise \citep{Steane2000}. For example, to achieve a side-band Rabi frequency $\Omega_+$ of a fraction $f$ of the trap frequency $\omega_{\rm t}$, we need a laser field of strength $\Omega=\Omega_+/\eta= f \omega_{\rm t}/\eta$.
The Stark shift $\Delta E$ of the qubit transition due to the presence of the carrier transition is given by
\begin{equation}\label{eq:carrier-Stark-shift}
\frac{\Delta E}{\hbar} =  \frac{\Omega^2}{2\Delta} =
\frac{1}{2\omega_{\rm t}}\frac{f\omega_{\rm t} \Omega_+}{\eta^2} =  \frac{f}{2\eta^2}\Omega_+   \;,
\end{equation}
The phase evolution  $\Delta E t/\hbar$ due to the AC-Stark shift becomes comparable to the desired Rabi flopping with frequency $\Omega_+$ already for \label{sec:AC-Stark-limit}
$f \sim \eta^2$ and thus typically for $f\sim0.01$ ($\eta\sim0.1$). Similarly, off-resonant excitations on the carrier transition might spoil the fidelity in this regime. In particular,
Rabi oscillations on the carrier transition occur with amplitude $A$ \citep{Steane2000}
\begin{equation}
\label{eq:off-resonant-excitations}
A=\frac{\Omega^2}{{\Omega^2+\omega_{\rm t}^2}}\approx \frac{\Omega^2}{\omega_{\rm t}^2} = \frac{f^2}{\eta^2} \;,
\end{equation}
where we assumed $\Omega \ll \omega_{\rm t} $, which is justified by the conclusions from Eq.~\ref{eq:carrier-Stark-shift}. Both the AC-Stark shift and the off-resonant excitations, can be at least  partially canceled with  methods described in Sec.~\ref{sec:imperfect-control}. However, it still remains very difficult to drive Rabi flops on sidebands much faster than $\eta \, \omega_{\rm t}$. Thus, it is hard to implement a two-qubit gate based on sideband transitions within one trap period.

\subsection{Choice of qubit ions}
\begin{figure}
\begin{center}
\includegraphics[width=0.6\textwidth]{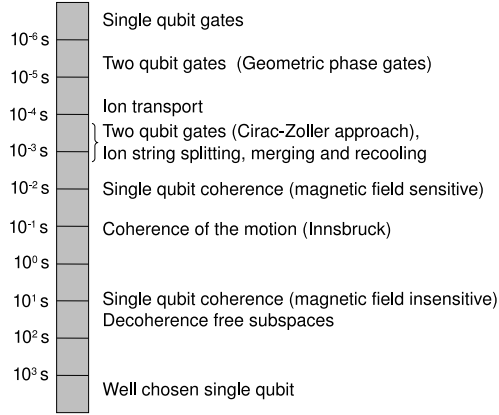}
\caption{\label{fig:time-scales} Currently achieved time scales in ion trap quantum computing. All operations can be implemented faster than the relevant decoherence mechanisms.}
\end{center}
\end{figure}

A good qubit candidate must meet certain criteria. In the case of trapped ions it is usually sufficient to concentrate on long coherence times
 on the qubit transition compared with the manipulation times and on the technical feasibility of the required lasers.
The scale to which the coherence time has to be compared with is given by the gate operation
times of typically $0.1 - 500\:\mu$s (c.f.~Fig~\ref{fig:time-scales}).

\begin{figure}
\begin{center}
\includegraphics[width=0.6\textwidth]{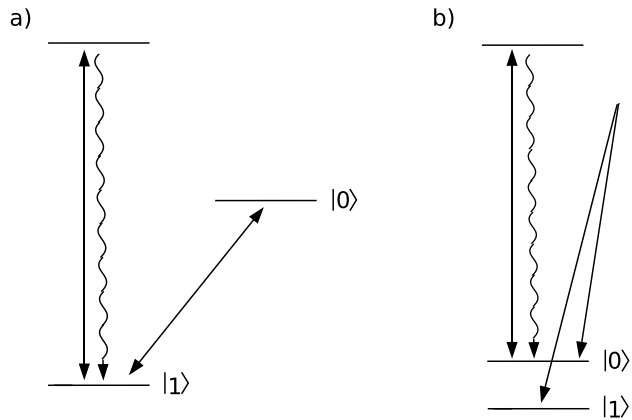}
\caption{\label{fig:qubits} Generic level schemes of atoms for optical qubits (left) and radio-frequency qubits (right). In addition to the two qubit levels $\{\ket{0},\ket{1}\}$ usually a third rapidly decaying level is used for laser cooling and state read-out. While the optical qubit is typically manipulated on a quadrupole transition, the radio-frequency qubit levels are connected with Raman-transitions.}
\end{center}
\end{figure}

\label{sec:ion-qubits}
Currently two different schemes are used to store quantum information in trapped ions: In the first scheme,
superpositions
between the electronic ground state and an excited metastable electronic state provide the two-level system
(optical qubit) (see Figs.~\ref{fig:qubits}a and \ref{fig:ca-40-level-scheme}).
The excited $D_{5/2}$-level in Ca$^+$ ---similarly to Sr$^+$ and Ba$^+$---  has a life time of more than one second
\citep{Barton2000,Kreuter2004,Kreuter2005a}.
In a second scheme, even larger coherence times can be achieved with superpositions encoded
in the electronic ground-state of the ions
(radio-frequency qubits) (see Fig.~\ref{fig:qubits}). Here, either the Zeeman or the hyperfine structure can be used.
The lifetimes of
these states are estimated to be much larger than any currently experimentally relevant
timescales.

%Thus here, the coherence times are purely limited by the phase coherence of the qubits rather than
%actual population decay.

There are different aspects to be considered for optical and and radio-frequency qubits: to fully exploit the potential of the optical qubit provided by  the $S_{1/2}$ and $D_{5/2}$-level in Ca$^+$, a laser line width of less than $200$~mHz is required. This
fractional stability of about
$10^{-15}$ is technically quite demanding, however, it has already been implemented for metrology purposes \citep{Young1999a,Young1999b,Hall2006}.
The transition frequencies for radio-frequency
qubits are usually below 10~GHz. Thus it is much easier to get the required stable phase
reference.

Radio-frequency qubits can be manipulated either directly by microwaves
\citep{Mintert2001} or ---as it is
usually the case--- by a Raman-process \citep{Monroe1995}. Here, a single atom absorbs a
photon from one laser field
and emits a second one into a second laser field. The frequency difference of the two laser
fields
provides the necessary energy for the population transfer and thus only this frequency
difference is important. The two laser fields are either derived from the same laser or a
phase locking between
two lasers can be implemented without resorting to optical frequency standards. Using optical fields instead of microwaves has the advantage that a much better spatial concentration  of the power can be achieved and much higher Rabi-frequencies can be attained as with microwaves.
Furthermore, by driving
radio-frequency qubits optically with two anti-parallel beams, two photons can transfer their recoil onto the atom. Thus, the coupling to the ion motion is increased as compared to single photon transitions (larger Lamb-Dicke factors $\eta$) and as a consequence the speed of two-qubit gates can be higher. Choosing  co-propagating beams, the coupling to the motion can be inhibited efficiently which has the benefit of suppressing the sensitivity to the ion motion.

For optical as well as for rf-qubits, the coherence times are often limited by fluctuations of the magnetic fields.
The reason is that usually the qubit basis states employed have different magnetic moments such that they experience an additional phase evolution due to the (fluctuating) magnetic field. Strategies to avoid these decoherence sources will be discussed in Sec.~\ref{sec:phase-coherence}.

We note here that a good qubit must not necessarily combine a large coherence time to manipulation time ratio with high fidelity initialization and read-out capabilities: initialization  and read-out can be implemented with an additional auxiliary ion of a different ion species. This idea has already been demonstrated by the NIST group \citep{Schmidt2005}: In this experiment aimed at implementing a frequency standard, the state of an individual Al$^+$ ion was detected via a Be$^+$ ion. However, for practical purposes it is desirable to initialize and read out the qubit ion directly.

Finally, a good qubit candidate must have all the relevant transitions in an accessible frequency regime. Generally, laser sources, fibers and detectors for short wavelengths are more expensive, less efficient and often more fault-prone than for longer wavelengths. However, ions tend to have short wavelength transitions. Thus there are only a very limited number of ions which have strong transitions in the visible frequency range. In the quantum computing context, calcium, strontium and ytterbium ions appear to be attractive due to their relative large wavelength transitions. However, beryllium and magnesium have a relatively small atomic mass which leads to large Lamb-Dicke factors ($\eta \approx 0.3$ in some of the NIST experiments) and eases coupling of the electronic and motional degrees of freedom  and thus makes them attractive in spite of difficulties with laser radiation generation, manipulation and fiber optics.

\subsection{Initialization and read-out}
\label{sec:optical-pumping}
Prior to the implementation of a quantum algorithm, the qubits must be initialized in a well-defined state. For atoms, in general, this can be most conveniently achieved by optical pumping. This idea was introduced by \citet{Kastler1950} and first implemented by \citet{Brossel1952} and by \citet{Hawkins1953}.
For reviews on optical pumping see Refs.~\citep{Happer1972}, \citep{Weber1977} and \citep{Wineland1980}.

The general idea of optical pumping is that an atom is driven until it decays into a state where the drive does not act any longer. Usually circularly polarized light is used to pump the atom into one of the extreme Zeeman-levels.
% For instance driving  the $S_{1/2}\rightarrow P_{1/2}$-transition in $^{40}$Ca$^+$ (for a level scheme of $^{40}$Ca$^+$ see Fig.~\ref{fig:ca-40-level-scheme}) with $\sigma^+$-polarized light, the atom is exited to the $P_{1/2} (m_j=+1/2$)-state by the light as long as it is in the $S_{1/2} (m_j=-1/2$)-state. From the $P_{1/2}$-state it has now the possibility to decay either to the $S_{1/2} (m_j=-1/2)$-state or the $S_{1/2} (m_j=+1/2)$-state. In the former case the atom is re-exited until it finally decays into the $S_{1/2} (m_j=+1/2)$-state in which it then remains. For $^{40}$Ca$^+$-ions this procedure is complicated due to the presence of the meta-stable $D_{3/2}$-level (see Fig.~\ref{fig:ca-40-level-scheme}).
% Pumping into this state can be prevented by shining in laser light on the $D_{3/2}\rightarrow P_{1/2}$-transition.
% The pumping speed can be estimated from the average number of cycles required to reach the target state and the time the atom spends in the excited state(s).
Typically, the target state is occupied in less than 1~$\mu$s with a probability larger than $0.99$. The initialization fidelity is usually limited by the quality of the polarization of the driving laser along the preferred axis of the qubit. In most cases, this axis is given by the direction of the magnetic field.

For fault-tolerant quantum computing, however, it appears that initialization fidelities exceeding 0.9999 are desirable. It is not clear whether such high preparation fidelities can be achieved with the methods described above.
%For instance birefringence in the vacuum windows due to stress induced by the air pressure can degrade the polarization.
Furthermore, there exist situations where it is not possible to achieve a pure circular polarization
along the quantization axis. The latter might happen for instance, because there is no optical access
along the direction of the magnetic field to send a laser in. In these cases, a frequency rather
than a polarization selection can be used for optical pumping if the ions offer a spectrally narrow transition \citep{Roos2006}.
\begin{figure}
\begin{center}
\includegraphics[width=0.75\textwidth]{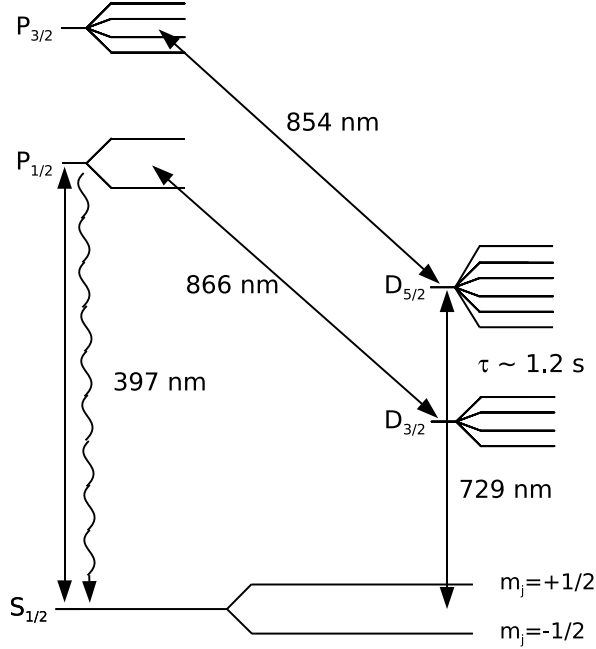}
\caption{\label{fig:ca-40-level-scheme} Level scheme of $^{40}$Ca$^+$, with Zeeman substructure and required laser wavelengths for manipulation of the calcium ions.}
\end{center}
\end{figure}

\begin{figure}
\begin{center}
\includegraphics[width=0.6\textwidth]{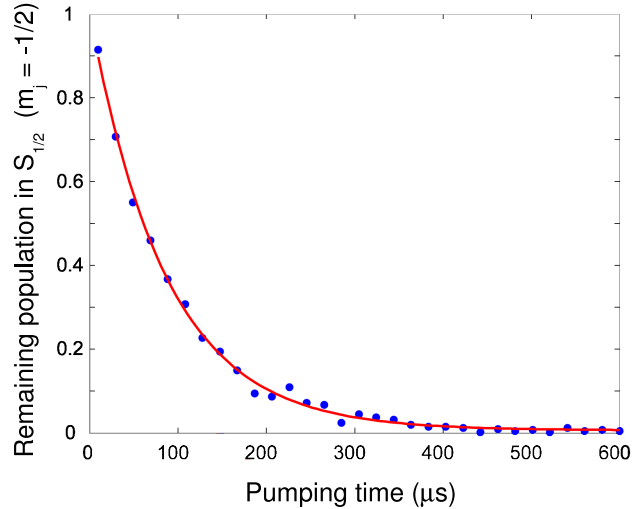}
\caption{\label{fig:729-optical-pumping} Optical pumping of the $^{40}$Ca$^+$  S$_{1/2}\: (m_j=+1/2)$ level using frequency selection rather than polarization selection.}
\end{center}
\end{figure}

We illustrate this procedure here for the $^{40}$Ca$^+$-ion: with a narrow band laser the ion is excited on the S$_{1/2}\:(m_j=+1/2) \longleftrightarrow \textrm{D}_{5/2} (m_j=-3/2)$ transition (see Fig.~\ref{fig:ca-40-level-scheme}) and simultaneously the D level is coupled to the P$_{3/2}$ level with a broad band laser at 854~nm. In this way the population of the S$_{1/2}\:(m_j=+1/2)$ state is effectively coupled to the rapidly decaying P$_{3/2}$~($m_j = \{-3/2,-1/2$\}) levels  while the S$_{1/2}\:(m_j=+1/2)$ level is not touched. Fig.~\ref{fig:729-optical-pumping} shows the depletion of the initially fully populated
 S$_{1/2}\: (m_j=+1/2)$ level with time.
Pumping time constants of about $10\;\mu$s have been demonstrated in Innsbruck.

In principle, the attainable pumping efficiency is only limited by the strongest Rabi frequency and pumping time. Assuming a typical transition splitting of the various Zeeman transitions of a few MHz, Rabi frequencies $\Omega_{\rm Rabi}$ lower than
$2\pi \times 10$~kHz are required to keep the off-resonant excitations $P_{\rm off} \approx \Omega^2/\Delta^2$ (see Eq.~\ref{eq:off-resonant-excitations}) of the unwanted transitions below $10^{-4}$. From that, we estimate that within 1~ms an initialization fidelity of 0.9999 can be reached. In experiments in Innsbruck, preparation fidelities exceeding 0.999 have already been observed \citep{Roos2006}.

\label{sec:state-detection}
At the end of a quantum algorithm the quantum register needs to measured. Radiation coupling to only one of the qubit levels can be used here. This idea was introduced  by \citet{Dehmelt1975} and is often termed electron shelving. Experimentally it was first implemented by \citet{Nagourney1986}, \citet{Sauter1986}, and by \citet{Bergquist1986}.

Referring to the level scheme of $^{40}$Ca$^+$ in
Fig.~\ref{fig:ca-40-level-scheme}, the ion does not fluoresce
under irradiation of light on the $S_{1/2}\rightarrow P_{1/2}$ and
$D_{3/2}\rightarrow P_{1/2}$ transitions if its valence electron
is in the $D_{5/2}$-state. If the electron, however, is in either
the $S_{1/2}$-, $P_{1/2}$- or $D_{3/2}$-state, the ion will
scatter approximately $10^7 - 10^8$ photons/s. A lens system
collects typically $10^{-3}$--$10^{-2}$ of this fluorescence light
such that with a photomultiplier tube (quantum efficiency $~\sim$
30\%) about 30 photons/ms can be detected. As typical background
count rates of photomultiplier tubes are usually well below 1
count/ms, one expects exceedingly high state detection fidelities
when collecting fluorescence for detection times of more than one
millisecond. However, this reasoning holds only if the ion  has a
negligible probability to decay either from the  $P_{1/2}$ to the
$D_{5/2}$ or from the  $D_{5/2}$ to the $S_{1/2}$ or the $D_{3/2}$
level   during the detection time. In practice, these two
constraints lead to an optimal detection time where the error due
to the Poissonian spread of the number of scattered photons is
balanced with the relaxation time scales
\citep{Roos2000phd,Acton2006,Myerson:2008a}.
Typical detection
times are about one millisecond with detecion fidelities exceeding 0.99.
\new{The fidelity can be further increased by anaysing the arrival times
of the photons with a maximum likelihood method and thus identifying some events when the
$D_{5/2}$ level decayed during the detection \citep{Myerson:2008a}.}

\new{Other
qubits can be detected very similarly. If the energy separation of the two qubits, however, is not large enough
to allow for selection via the laser frequency, the polarisation of the laser field can be used. For instance for $^9$Be$^+$, circular polarization ensures that predominantly only one of the two qubit states
scatters photons \citep{Sackett2000,Langer2006phd}.}

The detection efficiency can be further increased when the quantum state is mapped onto auxiliary qubits prior to its measurement:  for this, one prepares first an additional qubit (the ancilla qubit)  in state $\ket{0}$. A controlled-NOT operation maps then the qubit $\alpha \ket{0} + \beta \ket{1}$ onto the combined state $\alpha \ket{00} + \beta \ket{11}$  of the two qubits. Measuring both qubits yields an improved fidelity as compared to measuring a single qubit if the fidelity of the ancilla qubit state preparation and the controlled-NOT operation are high enough. \citet{Schaetz2005} demonstrated this procedure, albeit preparing the ancilla ion in $(\ket{0}+\ket{1})/\sqrt{2}$ and replacing the contolled-NOT with a controlled phase gate (see~Sec.~\ref{sec:geometric-phase-gate}) followed by a single-qubit rotation on the ancilla.

\citet{Hume2007} present another variant to improve the state detection fidelity by measuring the qubit repeatedly. In this experiment, the qubit is encoded in superpositions of the ground state and the excited qubit level of $^{27}$Al$^+$. This qubit state is first transferred to a $^9$Be$^+$ ion by a series of laser pulses \citep{Schmidt2005}: a first pulse on the red sideband on an auxiliary transition of the Al$^+$ ion  inserts one motional quantum only if the Al$^+$ ion is in the ground state. This motional excitation is transferred to the $^9$Be$^+$ ion with a red sideband pulse on the $^9$Be$^+$ ion which is then detected via the usual state dependent fluorescence method. Most importantly, the electronic auxiliary state of the Al$^+$ ion decays back to the ground state on a timescale of about 300~$\mu$s and not to the excited qubit state. Thus the information in which state the qubit is projected is still available in the  Al$^+$ ion and this procedure can be repeated. Based on the photon counts from  the $^9$Be$^+$ ion and previous detection results, another detection step is carried out. Repeating this procedure on average 6.5 times, \citet{Hume2007} achieve a detection efficiency of 0.9994. The accuracy of this method is only limited by processes which couple the excited qubit state of the Al$^+$ ion  ($ \sim 21$~s) to one of the electronic levels which does not decay to the ground state.

\new{Very recently, \citet{Myerson:2008a} took adavantage of the arrival times of the photons and combined it with the adaptive scheme just discussed. Thus they reached within an average detection time of 145~$\mu$s efficiencies of about 0.9999 of a qubit encoded in the $S-D$ manifold of a single $^{40}$Ca$^+$.}

\subsection{Single-qubit gates}
\label{sec:single-qubit-gates}
 It can be shown that all quantum algorithms can be broken down into a sequence of
single-qubit operations plus a specific two-qubit operation, e.g. a conditional phase
gate, a controlled-NOT gate or a $\sqrt{\rm SWAP}$ gate \citep{Deutsch1989,Nielsen2000}.
In ion traps, single-qubit operations are usually easy to implement and thus it is reasonable
to use this approach to attain universal quantum computing.

\begin{figure}[tb!]
\begin{center}
\includegraphics[width=0.7\textwidth]{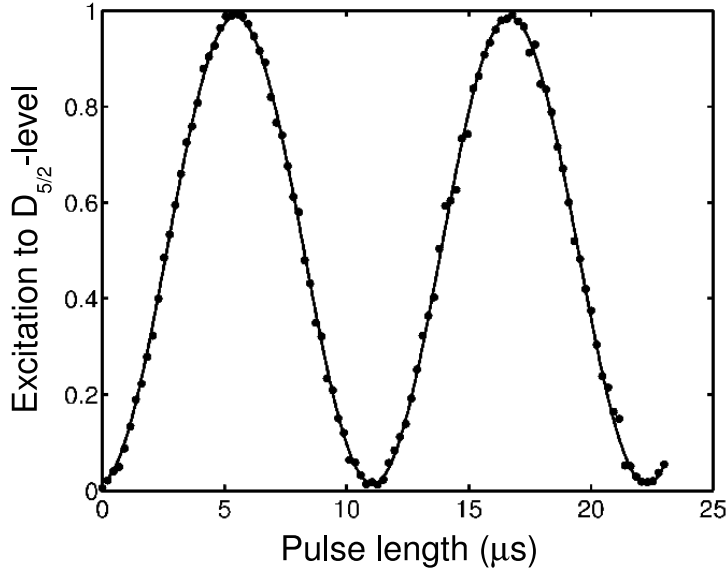}
\end{center}
\caption{Rabi oscillations of a single Ca$^+$ ion. Each dot represents 1000 experiments, each consisting of initialization, application of laser light on the qubit transition and state detection.}
\label{fig:Carrier-Rabi-oscillations}
\end{figure}

Rabi oscillations between the two qubit levels (see Fig.~\ref{fig:Carrier-Rabi-oscillations}) implement such single-qubit operations. Mathematically, we can describe the effect of resonant radiation inducing such a coupling by a rotation $R^C(\theta,\varphi)$  acting on the state vector $\alpha\ket{0} + \beta \ket{1}$ (c.f. \citet{Nielsen2000}):
\begin{eqnarray} \label{eq:single-qubit-operation}
 R^C(\theta,\phi) &=&\exp\left(i\theta/2\left( e^{i \varphi}\sigma_+ + e^{-i
\varphi}\sigma_- \right) \right) = I\:\cos \theta/2 + i (\sigma_x\cos \varphi -  \sigma_y\sin \varphi) \sin \theta/2  \nonumber \\
&=&  \left(\begin{array}{cc}
            \cos \theta/2 &  i e^{i \varphi} \sin \theta /2
            \\ i e^{-i \varphi} \sin \theta /2 & \cos \theta/2
     \end{array} \right),
 \end{eqnarray} where  %$\sigma_+=\ket{0}\bra{1}$ and $\sigma_-=\ket{1}\bra{0}$
$\sigma_+=\left( \begin{array}{cc}
0 & 1\\ 0 & 0\\
\end{array}\right)$ and $\sigma_-=\left( \begin{array}{cc}
0 & 0\\ 1 & 0\\
\end{array}\right)$
are the respective atomic raising and lowering operators. $\sigma_x=\left( \begin{array}{cc}
0 & 1\\ 1 & 0\\
\end{array}\right)$ and $\sigma_y=\left( \begin{array}{cc}
0 & -i\\ i & 0\\
\end{array}\right)$  are the Pauli-spin matrices. The angles $\theta$ and $\varphi$ specify the rotation.

\begin{figure}[tb!]
\begin{center}
\includegraphics[width=0.4\textwidth]{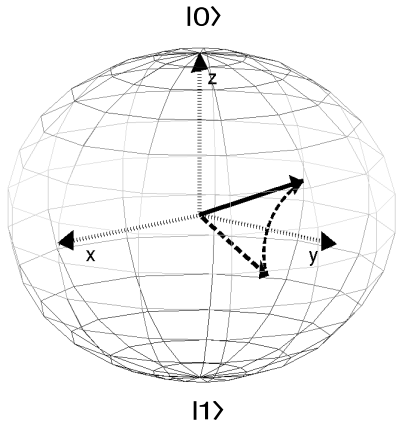}
\end{center}
\caption{Rotation around the $y$ axis visualized on the Bloch-sphere.}
\label{fig:XandYgate}
\end{figure}

Often single-qubit operations are visualized by use of the so-called Bloch sphere. We identify the north pole of the Bloch sphere with logical $\ket{0}$ (the energetically higher state) and the south pole with $\ket{1}$  (see Fig.~\ref{fig:XandYgate}). In the Bloch picture, the angle $\varphi$ specifies the axis of rotation in the equatorial plane and $\theta$ the rotation angle (pulse area), and thus any linear combination of $\sigma_x$ and $\sigma_y$ operations can be implemented with laser pulses.

Rotations around the
$z$ axis can be decomposed into rotations around the $x$ and the $y$ axis. Alternatively, all following rotations on that qubit can be shifted by $-\Delta \varphi$ to achieve effectively a rotation about the $z$ axis by $\Delta \varphi$. Finally, a far detuned laser beam can shift the relative energy $\Delta E$ of the eigenstates due to an AC-Stark effect by $\Delta E = \Omega^2/2\Delta$ (c.f.~Eq.\ref{eq:carrier-Stark-shift}). Thus, after a time $t=\hbar \Delta \varphi / \Delta E$  the desired phase shift is acquired.

The relevant control parameters in the ion trap experiments are the pulse area $\theta=\Omega \tau$ (Rabi frequency  $\Omega$, pulse length $\tau$) and the phase of the laser field $\varphi$. These control parameters can be conveniently controlled with an acousto-optical modulator in double-pass configuration (c.f. \citet{Donley2005}) by changing the amplitude and phase of the RF field driving the acousto-optical modulator.

We will give here a quick interpretation of the phase $\varphi$: After optical pumping the experiments start with the ion qubits in an energy eigenstate. Thus the electric field of the resonant excitation laser builds up a dipole (or quadrupole) moment oscillating in phase with the field at the laser frequency. In this way the first laser pulse (whose length is not a multiple of $\pi$) sets the phase reference for all subsequent operations on that ion. Thus, it becomes very intuitive to see e.g. that shifting the phase of a second excitation field by $\pi$ inverses the evolution.

In ion traps single-qubit manipulations are routinely carried out
with fidelities exceeding 0.99 \citep{Knill:2007}.
Single-qubit-gate fidelities are usually limited by laser
intensity fluctuations and in the case of single photon
transitions by the finite temperature of the ion crystal and in
the case of Raman transitions also by spontaneous emission from
the intermediate level. In the Lamb-Dicke limit (i.e. the
extension of the ground-state wave function is much smaller than
the projection of wavelength of the light onto the motion), the
effective Rabi frequency $\Omega_{\rm eff}$ of the transition is
given by  \citep{Wineland1998}
\begin{equation}\label{eq:carrier-rabi-incl-all-modes}
\Omega_{\rm eff}\approx\Omega\prod_{i}(1-n_i\eta_i^2)\:,
\end{equation}
where $i$ labels all motional modes, $n_i$ is their vibrational
quantum number and $\eta_i$ denotes the corresponding Lamb-Dicke factor. Here, in contrast to Eq.~\ref{eq:ion-trap-hamiltonian}, the second order in $\eta$ was taken into account. $\Omega$ is the Rabi frequency for an ion (string) completely cooled to the ground state of the trap. For Raman transitions, $\eta_i$ can be made quite small ($\sim 10^{-7}$) by using co-propagating beams such that the Rabi frequency is given just by $\Omega$ (the same reasoning holds for microwaves). Thus in practice, only single-qubit operations on optical qubits suffer from finite temperature effects.

The speed of single-qubit rotations is often limited by the acousto-optical modulator used to control the light field: during switching the light field, phase and amplitude chirps appear which can
spoil gate fidelities considerably. Further speed limitations are set by the amount of available laser power and more fundamentally by  transitions close by, e.g. due to other Zeeman levels. Excitations on unwanted transitions due to strong laser fields transfer the population out of the computational subspace and thus spoil the fidelity.

\subsubsection{Individual addressing of ion qubits}
\newnew{This chapter has been moved to here:???}:
\label{sec:ion-addressing}
Individual addressing of qubits is particularly difficult if
multi and single-qubit gates have to be carried out in the same trap: reasonable gate times are only possible for high trap frequencies resulting in small ion-ion spacings, which in turn require very well-shaped laser beams to address the ions.
%\citet{Wineland1998} discuss various possibilities to achieve effective single ion addressing.

The most straightforward way to achieve single ion addressing is to focus the qubit manipulation laser strongly. In the Innsbruck experiments a waist of 2~$\mu$m FWHM reliably distinguishes between ions spaced by 5~$\mu$m. Including deviations from a Gaussian beam shape, a 1000-fold reduction of the light intensity at the position of the adjacent ions as compared to the addressed ion can be achieved. This corresponds to an unwanted Rabi frequency of 0.03 on the adjacent ions with respect to the addressed ion (addressing error $\epsilon=\frac{\Omega_{\rm neighboring}}{\Omega_{\rm addressed}}=0.03$). \new{A complication arises from the fact that one would like to irrdiate the ion string with the laser beam perpendicular to the trap axis to resolve the ion positions. This in turn leads to a strongly reduced coupling of the laser field to the ion motion along the axial direction. In Innsbruck, an angle of about 68$^\circ$ between the laser beam and the trap symmetry axis was choosen as a compromise between the two diametrial requirements.}

To fulfill
the stringent requirements for quantum computation, one could use only every second ion thus increasing the
qubit--qubit distance. However, with an increasing number of ions, the Lamb-Dicke factor tends to become smaller which in turn results in slower two-qubit operations. Furthermore, the more complicated normal mode spectrum might reduce the obtainable fidelity.

\begin{figure}[tb!]
\begin{center}
\includegraphics[width=0.8\textwidth]{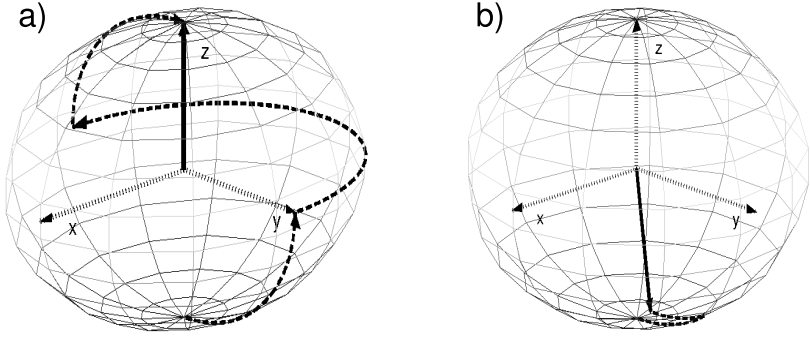}
\end{center}
\caption{\label{fig:addressing-error} Evolution of the Bloch vector on the Bloch sphere for a bit flip with the pulse sequence $R^{C}(\pi/2,\pi) R_z(\pi) R^{C}(\pi/2,0)$ (left-hand side, see text for an explanation). The right-hand side displays the state evolution for a Rabi frequency corresponding to 0.3 of the one on the left-hand side (pulse sequence $R^{C}(0.3\pi/2,\pi) R_z(0.3^2\pi) R^{C}(0.3\pi/2,0)$). }
\end{figure}

A more favorable method to reduce the
effect of addressing imperfections is to use composite pulses discussed also in Sec.~\ref{sec:composite-pulses}: here a single pulse is split into several smaller pulses. As
an example, we consider here the realization of an $R^{C}(\pi,0)$ operation with the pulse sequence $R^{C}(\pi/2,\pi) R_z(\pi) R^{C}(\pi/2,0)$ (see Fig.~\ref{fig:addressing-error}).
 The first pulse  ($R(\pi/2,0)$) moves the Bloch vector of the addressed qubit into the equatorial plane, while the Bloch vectors of
the unaddressed qubit $i$ is rotated by the addressing error angle $\epsilon_i \pi/2$. The AC-Stark pulse $R_z(\pi)$ (e.g. implemented with the laser tuned  off resonance) rotates the phase of the addressed qubit by $\pi$, while the phases of all other qubits are rotated only by $\epsilon_i^2 \pi$. The last $R(\pi/2,\pi)$ pulse finalizes then the rotation on the addressed qubit, while on the other qubits the effect of the first pulse is undone to a large extent for sufficiently small $\epsilon_i^2$. Fig.~\ref{fig:addressing-error}b illustrates this behavior on the Bloch sphere for a rather large addressing error of 0.3. We note that
if the addressing imperfection of the operations is well-known, much better results can be achieved. For instance \citet{Haljan:2005b} choose the length of a spatially inhomogeneous $R_z$ pulse such that  two qubits acquire a phase difference of $\pi$ \citep{Lee2006phd}. In this way one can afford addressing errors $\epsilon$ close to unity. Similar techniques were also used by \citet{Kielpinski2001}.

%Using an inhomogeneous laser beam detuned by about 200~GHz, a differential
% phase shift of $\pi$ can be imprinted onto the two ions within 10~$\mu$s. Thus one ion can acquire effectively a phase of $\pi$, while the other one has acquired a multiple of $2\pi$. Inserting this
% between two microwave pulses of length $\pi/4$ realizes thus an effective $\pi/2$-pulse on the latter
% ion, while the two microwave pulses cancel each other on the first ion. In this way arbitrary single qubit operations can be carried out on a small number of ions \citep{Lee2006phd}. A similar method \citep{Schaetz2004} to address single ions has been earlier applied in \citet{Rowe2001}.

This composite pulse technique uses the fact that single-qubit operations commute on different ions. This is not the case for sideband transitions as multiple ions interact simultaneously with the a vibrational mode. However, it
can be shown that by splitting sideband pulses into a sufficiently large number of
pulses ($\sim 5$) errors in the population of the vibrational mode are undone before they get large \citep{Haensel:2003}. Thus, addressing errors can also be suppressed in this case with composite pulses. A related technique is addressing in frequency space \citep{Staanum2002,Schrader2004,Haljan:2005b}: here an inhomogeneous laser or a magnetic field gradient induces different transition frequencies for each qubit. Thus the laser predominantly interacts with the qubit whose transition frequency matches the laser frequency. The disadvantage, however, is that in this case one must keep track of the phases of all qubits individually. In addition, it is necessary
to control the strength of the gradient field sufficiently well to avoid dephasing of the
qubits. This can be a serious problem when the qubit register is large and some of the qubits experience a very fast phase evolution due to the gradient field.

There are various other possibilities to achieve effective addressing (see e.g. \citet{Wineland1998}):
One particularly useful trick consists in changing the distance of the ions between the quantum operations by altering the trap stiffness (see e.g. Refs.~\citet{Rowe2001,Reichle2006a}). In this way the relative phase of the operations on the individual ions can be changed allowing one to address even individual groups of ions for non-local operations.
In addition, it was demonstrated that a two-ion string can be placed in such a way in the trap that the ions have a different micromotion amplitudes. This in turn leads to different
transition strengths and thus to single particle addressing capabilities \citep{Monroe1999,King1999phd}.
% %
% Many of these techniques work only for a limited number of ions. However, having in mind an ion trap architecture based on segmented traps (see \cite{Kielpinsky2002} and Sec.~\ref{sec:scaling}) only a moderate number of ions will have to be stored in the same trap at any given time. Thus basically all of the

\subsection{Two-qubit gates}
\label{sec:two-qubit-gates}
%In Sec.~\ref{sec:single-qubit-gates} we have seen how single qubits can be manipulated with fidelities exceeding 0.99.
One route to achieve a universal set of gates \citep{Deutsch1989}, is to complement single-qubit operations with two-qubit gate operations. These operations are one of the most important ingredients of a
quantum computer as they provide the
possibility to entangle two qubits. In combination with single-qubit operations they allow for implementation of any unitary operation \citep{Barenco1995}.

In many implementations of quantum algorithms with trapped ions, the fidelity of the whole algorithm was limited by the fidelity with which the two-qubit operations were implemented. Thus, currently the implementation of high-fidelity entangling gates is of crucial importance. In the following, we will discuss the Cirac-Zoller gate  (Sec.~\ref{sec:cz-gate}), the M{\o}lmer-S{\o}rensen gate (Sec.~\ref{sec:Moelmer-Soerensen-gate}) and the so-called geometric phase gate (Sec.~\ref{sec:geometric-phase-gate}), before we briefly summarize various additional proposals for two-qubit gates (Sec.~\ref{sec:other-gates}).

\subsubsection{Motion of ion strings}
\begin{figure}[tb!]
\begin{center}
\includegraphics[width=0.74\textwidth]{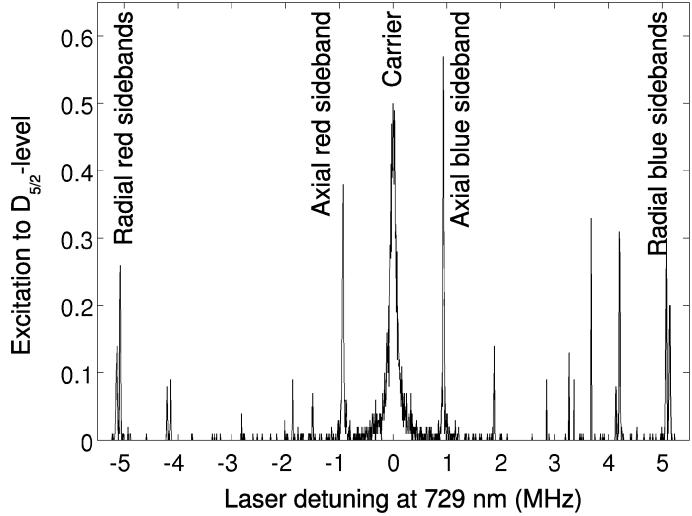}
\end{center}
\caption{Spectrum of a single trapped $^{40}$Ca$^+$ ion cooled close to the Doppler limit (see text). On resonance ($\Delta=0$) the strong carrier transition appears as well as for positive (negative) detunings $\Delta$ blue (red) sidebands are visible. The motional frequencies of the single ion can be deduced from the spectrum to amount to $\omega_z \approx 2\pi \times 1$~MHz (axial frequency) and $\omega_{{r}_{x,y}} \approx 2\pi \times 5$~MHz (radial frequencies).  In addition, higher order modes are visible at  $\Delta= m \omega_z \pm m \omega_r)$). }
\label{fig:1-ion-spectrum}
\end{figure}

The interaction between the ionic qubits can be mediated by motional degrees of freedom that serve as a quantum bus for distributing quantum information between the ions. Therefore, we will discuss first the
manipulation of the motion of single ions and ion strings \new{(see also Ref.~\citet{Leibfried2003b})}. The basic level scheme representing a single ion coupled to a single motional
mode is depicted in Fig.~\ref{fig:ion-tensor-oscillator}.
A laser field resonant with the carrier transition of frequency $\omega_{\rm qubit}$ drives transitions with Rabi frequency $\Omega_{\rm eff}$ where the motional state is
not changed (see Eq.~\ref{eq:carrier-rabi-incl-all-modes}). If the laser field, however, is detuned by the trap frequency towards higher energies, the
so-called blue sideband is excited (see Eq.~\ref{eq:blue-rabi-frequency}) and the operation
\new{
\begin{equation}
\label{eq:Rblue}
R^+(\theta,\varphi)=\exp\left[i \frac{\theta}{2} \left( e^{i \varphi} \sigma^+ a^\dagger + e^{-i \varphi} \sigma^- a \right) \right]\;.
\end{equation}
is carried out.
Here $\sigma^{\pm}$ are the atomic flip operators which act on the electronic quantum state of an ion by inducing transitions from the $|g\rangle$ to $|e\rangle$ state and vice versa (notation: $\sigma^+=|e\rangle \langle g|$). The operators $a$ and $a^\dagger$ denote the annihilation and creation of a phonon at the trap frequency $\omega$, i.e. they act on the motional quantum state. As in Eq.~\ref{eq:single-qubit-operation}, the parameter $\theta$ depends on the strength and the duration of the applied pulse, and $\varphi$ denotes the relative phase between the optical field and the atomic polarization.}
Importantly, the electronic and motional degree of freedom change simultaneously. Similarly for the opposite detuning the red sideband can be excited.
Fig.~\ref{fig:1-ion-spectrum} shows a spectrum of a single trapped $^{40}$Ca$^+$ ion near the qubit transition.
At a detuning of $\pm$~1~MHz, the red and the blue axial sidebands appear, respectively. In addition, radial sidebands (detuning $\sim \pm 5$~MHz) and higher order sidebands are visible. For the applied laser power and excitation time, the carrier transition is strongly saturated while the sidebands are only weakly saturated. This indicates that the sideband transitions are weaker than the carrier transition as it is expected from Eq.~\ref{eq:blue-rabi-frequency} and Eq.~\ref{eq:red-rabi-frequency}. For the experiment shown in Fig.~\ref{fig:1-ion-spectrum},
the single $^{40}$Ca$^+$-ion was cooled to a temperature of
$(\bar{n}_x,\bar{n}_y,\bar{n}_z) \approx (3,3,16)$ via Doppler cooling, while the Lamb-Dicke factors were $(\eta_x,\eta_y,\eta_z) \approx (0.01,0.01,0.08)$ (in this particular experiment the angle of the laser beam was given by $(49.2°,49.2°,22.5°)$).
% In this situation, the amplitude of the ion motion in the direction of the laser radiation is still much smaller than the wavelength of the radiation and thus the ion is in the Lamb-Dicke regime \citep{Wineland1998}. In this regime the strength of the red and blue sideband Rabi frequencies $\Omega_-$ and $\Omega_+$, respectively, are given by
% \begin{eqnarray}\label{eq:red-rabi-frequency}
% \Omega_-&=&\sqrt{n_i}\eta_i \Omega_0\:
% %\end{equation}
% \\
% %\begin{equation}
% \label{eq:blue-rabi-frequency}
% \Omega_+&=&\sqrt{n_i+1}\eta_i \Omega_0\:.
% \end{eqnarray}
%  Here $n_i$ is the phonon occupation number of the motion in question, $\eta_i=k_i x_0$ is the
% Lamb-Dicke factor ($k_i$ is  projection of the laser fields wavevector along the motion,
% $x_0=\sqrt{\hbar/(2 m \omega_i)}$ is the size of the ground state wave paket) and finally $\Omega_0$ characterizes the strength of the laser field
% in terms of the carrier Rabi frequency.

\begin{figure}[tb!]
\begin{center}
\includegraphics[width=0.7\textwidth]{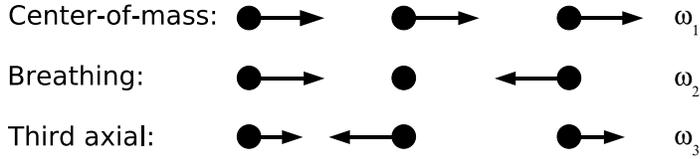}
\end{center}
\caption{Normal modes of a three-ion crystal \new{along the axial direction with motional frequencies  $\omega_i$}.}
\label{fig:normal-modes}
\end{figure}

\begin{figure}[tb!]
\begin{center}
\includegraphics[width=0.8\textwidth]{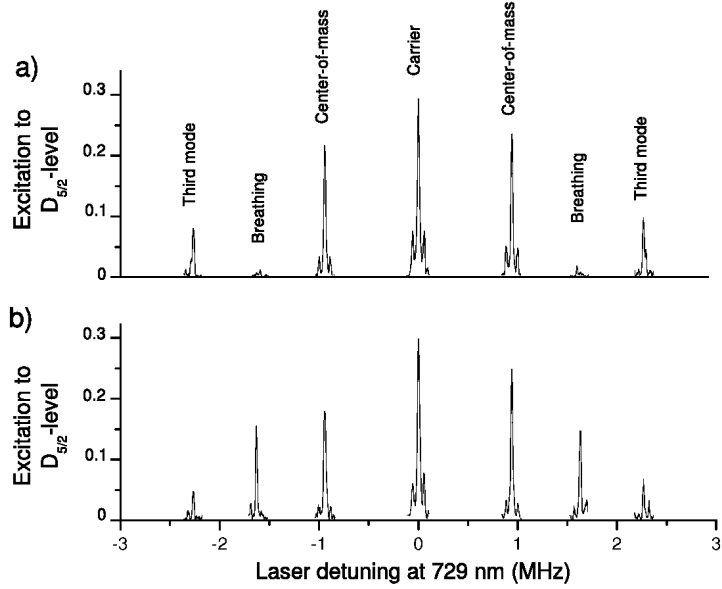}
\end{center}
\caption{Excitation spectra of a three-ion string. In {\protect Fig.~\ref{fig:3-ion-spectrum}a)}, the exciting laser is addressed to the center ion, while in   {\protect Fig.~\ref{fig:3-ion-spectrum}b)} the laser is addressed to the left ion. Visible are excitations of the motional modes as well as of the carrier transition. If the center ion is addressed {\protect Fig.~\ref{fig:3-ion-spectrum}a)}, the breathing mode can not be excited efficiently. The residual excitation of the breathing mode is most likely due to laser light interacting with the outer ions \new{(see {\protect Sec.~\ref{sec:ion-addressing}} on imperfect addressing).}}
\label{fig:3-ion-spectrum}
\end{figure}

For quantum logic experiments, however, the case of multiple ions is more interesting. Since the Coulomb interaction
couples the ions strongly together, it is useful to find the normal modes of the ion string \citep{Steane1997,James1998,Sasura2002}.  Taking a three ion string as an example and concentrating on the axial direction, we
find three normal modes: center-of-mass, breathing (or stretch) and an additional axial mode. Fig.~\ref{fig:normal-modes} illustrates the motion of each ion associated with each mode. Denoting the center-of-mass
modes frequency $\omega_{\rm 1}$, the breathing mode has a frequency of $\omega_{\rm 2} = \sqrt{3}\:\omega_{\rm 1}$ and the third axial mode has a frequency of $\omega_{\rm 3} = \sqrt{29/5}\:\omega_{\rm 1}$. Fig.~\ref{fig:normal-modes} indicates the relative motions of the ions.
The strength with which each mode couples to the motion is characterized by the eigenvectors $(1,1,1)/\sqrt{3}$ for the
center-of-mass, $(1,0,-1)/\sqrt{2}$ for the breathing and $(1,-2,1)/\sqrt{6}$ for the third axial mode \citep{James1998}. This means that the center
ion does not couple to the breathing mode and that the left and right ions couple with opposite phase factors to it.
Fig.~\ref{fig:3-ion-spectrum} illustrates that the breathing mode cannot be efficiently excited when the center ion is addressed (see Sec.~\ref{sec:ion-addressing} for addressing of individual ions).
For further details of the normal modes of ion strings (including larger ion strings), we refer to \citet{James1998}.

While measuring the spectra of ion strings needs only reasonably cold temperatures (e.g. ions cooled to the Lamb-Dicke limit), coherent operations on the sideband need usually ground state
cooling of this particular mode. The reason can be seen already in Eq.~\ref{eq:blue-rabi-frequency} which predicts that the sideband Rabi frequency strongly depends on the vibrational quantum number of the
corresponding motional mode and thus excitation on the sideband is incoherent for a finite temperature of this motional mode. On the other hand, this strong sensitivity of the sideband excitations to the motion will be the key to couple the ion qubits. To allow for coherent sideband operations, the motion of the mode in question is cooled to the ground state, although one could also imagine using any well-defined motional quantum number.
Finally, we extend Eq.~\ref{eq:blue-rabi-frequency} to take all motional modes into account (c.f.~Eq.~\ref{eq:carrier-rabi-incl-all-modes} and \citet{Wineland1998}):
\begin{equation}
\label{eq:blue-rabi-incl-all-modes}
\Omega^+_{\rm eff}=\Omega (\sqrt{n_{\rm bus}}\eta_{\rm bus}) \prod_m (1-n_m\eta_m^2)
\end{equation}
Here $\Omega^+_{\rm eff}$ is the effective Rabi frequency on the blue sideband, $\Omega$ the ideal carrier Rabi frequency and $n_{\rm bus}$ denotes the quantum number of the bus mode (i.e. the mode which will be used to couple the qubits), while $\eta_{\rm bus}$ labels its corresponding Lamb-Dicke factor. The index $m$ runs here over all modes except the bus mode. Since the influence of these modes on the sideband Rabi frequency $\Omega^+_{\rm eff}$ is strongly reduced, they are often termed spectator modes.

As discussed in Refs.~\citep{Neuhauser1978,Wineland1979,Marzoli1994}, ground state cooling in ion traps can be achieved by driving the red sideband (c.f. Fig.~\ref{fig:ion-tensor-oscillator}). Each absorption of a photon on the red sideband followed by a spontaneous emission on the carrier transition takes out one motional quantum of energy \citep{Stenholm1986,Eschner2003}.
% \remark{Das folgende ist zu vereinfachend:}In reality, however, a few things can go wrong: For instance, the re-emission can take place on a motional sideband. For ions outside the Lamb-Dicke regime this is actually the preferred way, thus sideband cooling works only efficiently if the ion string is well in the Lamb-Dicke regime.
It usually takes too long to wait for the spontaneous
decay of the upper level. Therefore in most experiments, the lifetime of the upper state is shortened by quenching the
state with a laser connecting it to a fast decaying state.

Experimentally, sideband cooling to the motional ground state was demonstrated first by
the NIST-group with a \new{single} mercury ion \citep{Diedrich1989} and then later with various other ion
species (\citet{Monroe1995a,Roos1999,Peik1999}; for a review see \citet{Leibfried2003b}). For ion strings sideband cooling works similarly as for single ions. The multiple normal modes can be cooled sequentially \citep{Monroe1995a,Roos1999}. However, for ion crystals, the Lamb-Dicke factors $\eta_i$ tend to be smaller  due to the increased mass as compared to single ions (for a definition of $\eta$ see Sec.~\ref{sec:basic-hamiltonian}) and thus the cooling process is slowed down.

\begin{figure}[tb!]

\begin{center}
\includegraphics[width=0.7\textwidth]{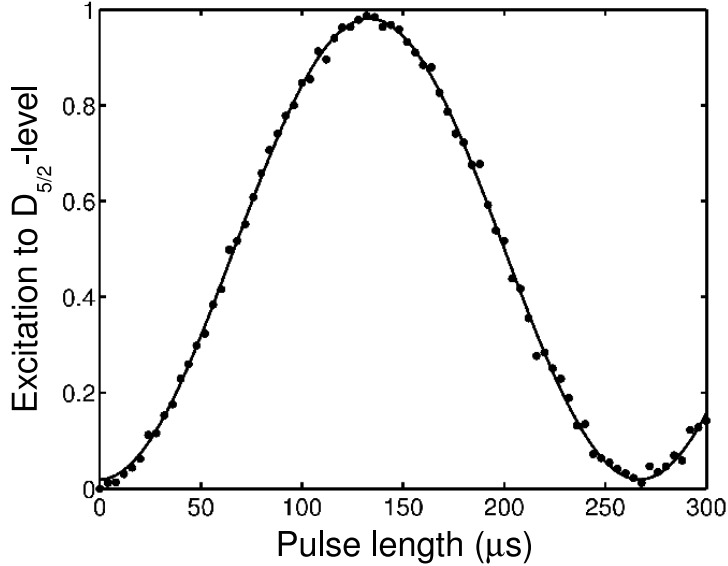}
\end{center}
\caption{Rabi oscillation on the blue sideband of the center-of-mass mode. The data were taken
on a string of two $^{40}$Ca$^+$ ions whose center-of-mass mode was cooled to the ground state. Only one of the ions was addressed. The population oscillates between the $\ket{S,0}$ and the $\ket{D,1}$ state of the addressed ion.\label{fig:Blue-Rabi-oscillations}}
\end{figure}

\label{sec:blue-sideband-flops}
Once the motion of a particular mode of the ion string is cooled to the ground state, irradiation on the respective blue sideband leads to Rabi oscillations (assuming the electronic degree of freedom of the ion is also in the ground state).
 Fig.~\ref{fig:Blue-Rabi-oscillations} shows such a Rabi oscillation on the
blue sideband of the center-of-mass mode of a string of two $^{40}$Ca$^+$ ions.

Another promising route to cool ions below the Doppler limit is based
on electromagnetically induced transparency (EIT) \citep{Roos2000,Morigi2000}. Here two interfering laser fields are used to shape the atomic absorption spectrum in such a way that sharp features appear in it which can be used to suppress heating effects on the carrier and blue sideband transitions while still maintaining absorption on the red sideband transitions. In general, EIT is
a very versatile tool as the free parameters of the two laser intensities and detunings from the atomic resonance allow one to adjust the atomic absorption spectrum to the actual needs (equilibrium temperature and cooling speed). Furthermore, \citet{Roos2000} achieved simultaneous cooling of up to three modes close to the ground state with a single setting.

% Finally, the phenomenon of EIT can  also be used to discriminate between two levels even when the natural linewidth of the relevant transititons are bigger than thefrequency splitting of the qubit. This has been demonstrated by \citet{McDonnell2004}.

\subsubsection{Cirac-Zoller gate}\label{sec:cz-gate}
Conceptually, the Cirac-Zoller phase gate is the simplest of the two-qubit gates presented here and is therefore discussed first  \citep{Cirac1995}. It requires single ion addressing and ground-state cooling of the
bus mode. On the other hand, for the M{\o}lmer-S{\o}rensen type gates \citep{Soerensen1999,Milburn2000,Leibfried2003a}  both ions are illuminated simultaneously and
the ion string have to be cooled only into the Lamb-Dicke limit.
 The geometric phase gate \citep{Leibfried2003a} is closely related to the M{\o}lmer-S{\o}rensen gate. However, it will be presented separately as its implementation looks different.

Cirac and Zoller proposed the following procedure to perform a two-qubit gate between
two ions in an ion string \citep{Cirac1995}: First the quantum information of the first
qubit is transferred to the motional degree of freedom of a mode (the bus mode) of the ion crystal. Importantly, the resulting motional state affects not
only the addressed ion itself but the ion string as a whole. Therefore on a second ion, operations conditioned on the motional state of the ion string can be carried out. Finally the quantum information of the motion is mapped back onto the first ion of the string.

% This procedure relies strongly on the ability to couple the electronic and motional
% degree of freedom.
% Such a coupling can be achieved by irradiating
% an ion with a laser detuned from the carrier transition by the bus mode frequency (see Sec.~\ref{sec:blue-sideband-flops}, c.f. Figs.~\ref{fig:1-ion-spectrum}, ~\ref{fig:ion-tensor-oscillator}).
% The scheme relies also on the knowledge of the initial vibrational state of the ion string. The bus mode is usually cooled into its ground state to achieve a well-defined motional state.

\begin{figure}[tb!]
\begin{center}
\includegraphics[width=1\textwidth]{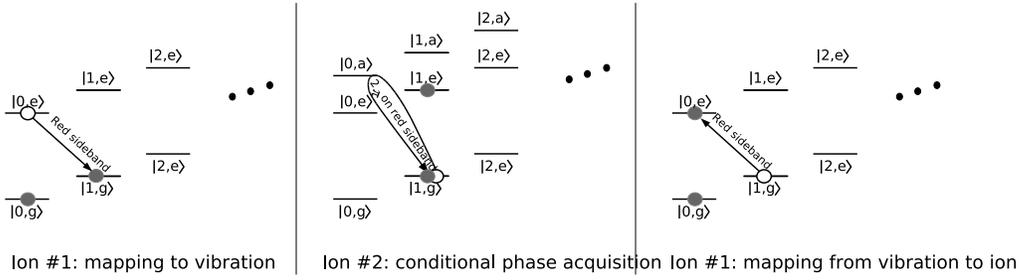}
\end{center}
\caption{\label{fig:CZ-NOT-levels} Graphical representation of the three steps to perform
a phase gate between two ions with an electronic ground state $\ket{g}$, an excited state $\ket{e}$ and an auxilary state $\ket{a}$.}
\end{figure}
The individual steps of the Cirac-Zoller phase gate are as follows:
 A laser pulse directed to the first ion with length $\theta=\pi$ and frequency $\omega_{\rm qubit}-\omega_{\rm trap}$ (red sideband) moves
all population present in the $\ket{0,e}$-state  to $\ket{1,g}$-state  (see left panel in Fig.~\ref{fig:CZ-NOT-levels}). However, if the first ion is in the $\ket{0,g}$-state, no state transfer happens.
Thus, we have effectively mapped the
quantum information from the electronic degree of freedom to the motion. Note that the coupling strength on the sideband depends strongly on the vibrational excitation. Therefore, this procedure works only if one knows the vibrational state such that one can properly adjust the laser intensity to achieve a $\pi$ pulse. With the quantum information of the first ion in the motion, we can exploit the common motion of the two ions  and perform a conditional operation on a the second ion. We follow here the original proposal \citep{Cirac1995} and use a
$2\pi$ rotation between $\ket{1,g}$ and a third auxiliary state $\ket{0,a}$ on the red sideband (see center panel in Fig.~\ref{fig:CZ-NOT-levels}). Importantly, only the $\ket{1,g}$ state is coupled to another level; for the states $\ket{0,e}$, $\ket{1,e}$ and $\ket{0,g}$, however, the coupling vanishes as there are no levels present with the appropriate energy. Effectively, we
have therefore performed the following operation:
\begin{equation}
\begin{array}{rcr@{}l}
\label{eq:cz-phase-gate}
~ & 2\pi  \\
\ket{0,e} & \longrightarrow & & \ket{0,e}  \\
\ket{1,e} & \longrightarrow & & \ket{1,e}   \\
\ket{0,g} & \longrightarrow & & \ket{0,g}  \\
\ket{1,g} & \longrightarrow & -&\ket{1,g}  \\
\end{array}
\end{equation}
This implements a phase shift of the second ion conditioned on the motional state. Finally, another $\pi$ pulse on the red sideband addressed to the first ion maps the quantum information present in the motion back onto the first ion and in total a phase
gate between the two ions is performed. To turn this phase gate into a controlled-NOT operation, one can  apply an $R^C(\pi/2,\varphi_i)$ pulse ($i={1,2}$)  to the target ion on the carrier transition (see Eq.~\ref{eq:single-qubit-operation}) before and after the phase gate. Choosing a particular phase relation $\varphi_2-\varphi_1$ either an ordinary controlled-NOT gate or a zero controlled-NOT operation is obtained.

The heart of this procedure, i.e. the dynamics presented in Eq.~\ref{eq:cz-phase-gate},
was demonstrated by the NIST group with a single beryllium ion \citep{Monroe1995}. The Innsbruck group implemented the complete protocol with individual addressing
of two calcium ions \citep{Schmidt-Kaler2003,Schmidt-Kaler2003a}. Sec.~\ref{sec:cz-experiments} describes both experiments.

\citet{Jonathan2000} present a generalization of this gate where they propose to drive the system so strongly that due to AC-Stark shifts  the eigenstates of different vibrational quantum  numbers (the dressed states) get the same energy. Thus a motional dynamics can be achieved just with carrier pulses of appropriate phase. Essentially, this proposal trades faster gate speeds against more sensitivity to laser intensity noise as the energy of the dressed states depends strongly on the laser power.

\subsubsection{M{\o}lmer-S{\o}rensen gate}
\label{sec:Moelmer-Soerensen-gate}
\begin{figure}[tb!]
\begin{center}
\includegraphics[width=0.5\textwidth]{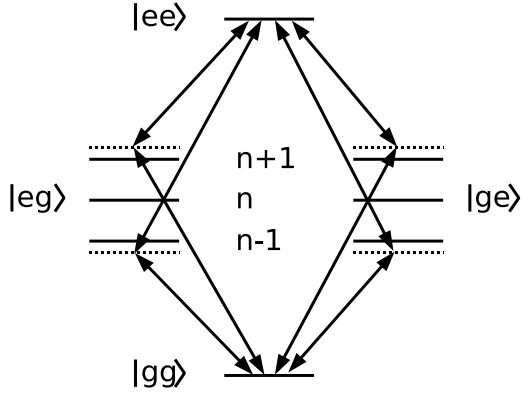}
\end{center}
\caption{\label{fig:moelmer-soerensen} Energy-level diagram of two trapped ions illustrating the principle of the S{\o}rensen and M{\o}lmer gate. The bus mode is populated with $n$ phonons. Two laser beams tuned close to the blue and red sideband, respectively, drive the system via the dashed virtual levels between the $\ket{n,gg}$ and $\ket{n,ee}$ state. A similar process takes place if the ion string is either in the $\ket{n,eg}$ or in the $\ket{n,ge}$ state.}
\end{figure}

Another possibility to implement a two-qubit gate is to use laser radiation
tuned close to the motional sidebands \citep{Soerensen1999,Moelmer1999,Soerensen2000,Solano1999}.
The basic principle is as follows: both ions are irradiated with a bichromatic laser field
with frequencies $\omega_{0} \pm (\omega_{\rm qubit} + \delta)$, tuned close to the red and the blue sideband of a collective mode,
respectively (see
Fig.~\ref{fig:moelmer-soerensen}). 
The  two frequencies sum up to twice the qubit frequency $\omega_{\rm qubit}$, each laser field itself, however, is not resonant to any level. Thus both ions can change their state only collectively and choosing an interaction time appropriately, the dynamics
\begin{eqnarray}\label{eq:ms-dynamics}
\ket{\rm ee} & \rightarrow & (\ket{\rm ee}+ i\ket{\rm gg})/\sqrt{2} \nonumber \\
\ket{\rm eg} & \rightarrow & (\ket{\rm eg}+ i\ket{\rm ge})/\sqrt{2} \nonumber \\
\ket{\rm ge} & \rightarrow & (\ket{\rm ge}+ i\ket{\rm eg})/\sqrt{2} \nonumber \\
\ket{\rm gg} & \rightarrow & (\ket{\rm gg}+ i\ket{\rm ee})/\sqrt{2}
\end{eqnarray}
is achieved.
%We note that AC-Stark shifts and the phase difference of the light field at the ion positions can lead to additional phase factors between the superpositions on the right hand side.
To see that this gate leads to a universal set of gates, we introduce the new basis $\ket{\pm}_i=(\ket{\rm e}_i\pm \ket{\rm g}_i)/\sqrt{2}$.
$\ket{\pm}\ket{\pm}$ are eigenstates of the unitary operation described by Eq.~\ref{eq:ms-dynamics} and transform as $\ket{++}\rightarrow\ket{++}$, $\ket{+-}\rightarrow i\ket{+-}$, $\ket{-+}\rightarrow i\ket{-+}$, $\ket{--}\rightarrow\ket{--}$ by the action of the gate, where we omitted a global phase factor of $e^{-i\pi/4}$ on the right hand sides. This transformation is a conditional phase gate up to single-qubit phase-shifts and is known to be universal together with single-qubit operations.

The M{\o}lmer-S{\o}rensen gate has the particular feature that it
does not require individual addressing of the ions and that it
does not fail completely if the ion string was not cooled to the
ground state. As will be described in
Sec.~\ref{sec:entangled-states}, the NIST group entangled up to
four $^9$Be$^+$ ions with this gate type \citep{Sackett2000}.
Furthermore, \citet{Haljan:2005b} created all four Bell states \new{and implemented Grover's seach algorithm with
$^{111}$Cd$^+$ ions using this gate operation (see also Refs.~\citet{Lee2005,Brickman2005,Brickman:2007phd})}.

Very recently, the application of a M{\o}lmer-S{\o}rensen gate to
optical qubits excited on dipole-forbidden transitions has been
analyzed \citep{Roos:2008a}. It was shown that fast, high-fidelity
gate operations are achievable by smoothly switching on and off
the bichromatic laser fields inducing the gate action. The
experimental implementation of this technique has resulted in the
creation of Bell states with a fidelity of 99.3(1)\%
\citep{Benhelm:2008b}.

\subsubsection{Geometric phase gate}
\label{sec:geometric-phase-gate}
\begin{figure}
\begin{center}
\includegraphics[width=0.9\textwidth]{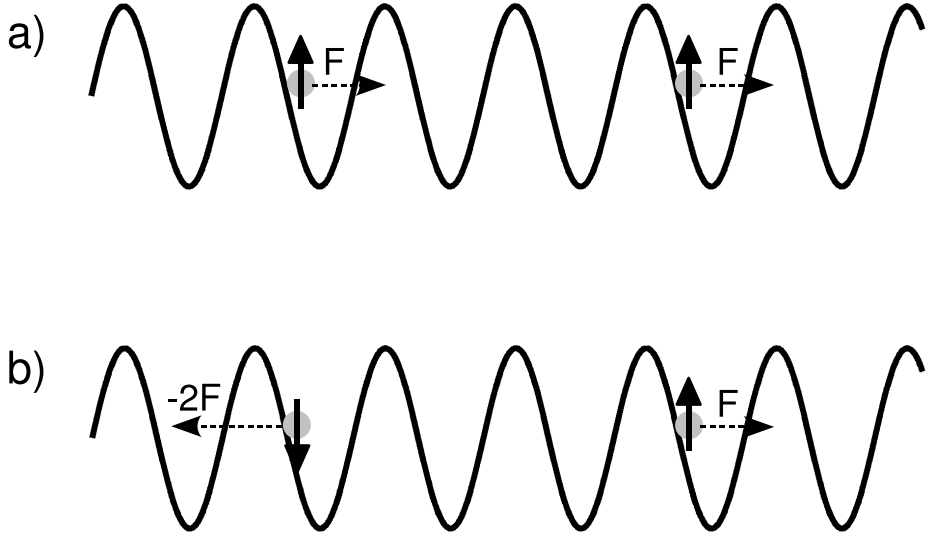}
\end{center}
\caption{\label{fig:GeometricPhaseGate} Force on two ions in a standing wave of two laser fields. a): both ions spin up. Both ions experience the same force b) one ion spin down, the other spin up. Both ions experience opposing forces. Tuning the frequency difference of the two laser fields close to the breathing mode frequency, in a) the motion of the ion string cannot be excited efficiently, while in b) the breathing mode is excited.}
\end{figure}

%Another variant of M{\o}lmer-S{\o}rensen gates is the so-called geometric phase gate \citep{Leibfried2003a}.
The so-called geometric phase gate uses also two laser fields
irradiating multiple ions at the same time. An interesting feature of this gate is that ideally during the gate  operation the electronic state of the ions is not touched. Only a force dependent on the electronic states is applied such that for the various  electronic states different phases are acquired \citep{Milburn2000}. In the scheme implemented by the NIST group \citep{Leibfried2003a} two non-co-propagating laser fields create a standing wave (see Fig.~\ref{fig:GeometricPhaseGate}). The difference of the two laser frequencies is tuned closely to one of the axial frequencies which leads to a walking wave. Thus each ion experiences a periodic AC-Stark shift and a force depending on the slope of the spatial variation of the AC-Stark shift. Most importantly the size and even the direction of the force can depend on the electronic state of the ion. Choosing the distance between the ions such that each ion experiences the same phase of the walking wave for a given time (see Fig.~\ref{fig:GeometricPhaseGate}),
ions are pushed in the same direction if they have the same electronic state. In this
way the breathing mode cannot be excited. However, if the ions are
in different electronic states the forces on the ions are not the same and the
breathing mode can be excited. The detuning of the drive from the motional mode is chosen such that the
phase between the motion and the drive changes its sign  after half the gate time. In this way, the ion string is driven back to the original motional state after the full gate time. This ensures that the motion  is disentangled  from the electronic state after the gate operation. The intermediate energy increase as compared to situations where the motion is not excited leads to the desired phase factor and the ions have picked up a phase factor that depends non-linearly on the internal states of both ions.

\citet{Leibfried2003a} implemented this gate with a fidelity of 0.97, limited mainly by the spontaneous decay from the $P$ manifold of the Be$^+$ ions during the gate operation. This exceptional fidelity was reached because the gate avoids a number of imperfections in the first place. We mention here the absence of off-resonant excitations on the carrier transition. Furthermore, the gate execution time of about 10~$\mu$s is two to three orders of magnitude faster than the corresponding coherence time of the qubits and thus decoherence effects are small. To obtain a differential force on the ions, the NIST group used two laser beams detuned blue from the  $S_{1/2} \rightarrow P_{1/2}$~transition by $2\pi \times 82$~GHz (fine structure splitting of the $P$ state is 200~GHz). For the ratio of the forces on the $\ket{\!\!\downarrow} = \ket{F=2,m_f=-2}$ and the $\ket{\!\!\uparrow} = \ket{F=1,m_f=-1}$ state this yields: $F_\downarrow$=-2$F_\uparrow$.
%, whereas the corresponding Stark shifts average out over the total gate time as the forces oscillate sinusodial.
Additional Stark shifts can be efficiently suppressed by choosing almost perpendicular and linear polarizations for the laser beams \citep{Wineland2003}.
Finally, \citet{Leibfried:2007} discuss a version of this gate where the laser intensities impinging on the ions are
controlled by transporting the ion crystal through the laser fields. Controlling the
interaction of the ions by transport offers also the possibility to use spatially modulated magnetic fields to create the
state dependent oscillating force \citep{Leibfried:2007}.

These three gate types have different strengths and weaknesses. Both, the geometric phase gate and the M{\o}lmer-S{\o}rensen gate do not need
single ion addressing. While this is often advantageous, it has the inconvenience that it is not straightforward
to carry out these gates between specific ions in a string. Using segmented traps for moving and splitting ion strings (see Sec.~\ref{sec:moving-ions}) resolves this issue. Another route to introduce ion specific gates is to  hide certain ions with single ion operations (see Sec.~\ref{sec:selective-readout}) such that these ions are not affected by the multi-qubit gate operation. The Cirac-Zoller gate on the other
hand demands single ion addressing but allows for a straightforward implementation
of quantum algorithms.

Another aspect is speed. The geometric phase gate can be executed quite fast as  the laser can be tuned such that off-resonant
transitions are quite unlikely. The Cirac-Zoller and M{\o}lmer-S{\o}rensen  gates on the other hand require a laser tuned to or tuned close to a sideband transition. This automatically implies that the laser is detuned only by (approximately) the trap frequency from the strong carrier transition. Thus it seems that the gate speed has to be much slower than a trap frequency (see Sec.~\ref{sec:AC-Stark-limit}).
However, for special temporal and spectral arrangements of the laser field, it is possible to suppress the spectral contribution on the carrier transition so that gate times close to the trap period seem feasible.

Finally, it is important whether the gate works efficiently with magnetic field insensitive transitions. Qubits encoded in superpositions of levels connected via a magnetic field insensitive  transition  provide very long coherence times of many seconds (Sec.~\ref{sec:clock-transition}). In its originally proposed form, the geometric phase gate is very inefficient on magnetic field insensitive transitions because hyperfine states with a similar magnetic moment appear to experience similar AC-shifts \citep{Langer2006phd}. However, this could be circumvented by using a pair of laser beams tuned close (as compared to the qubit difference frequency) to a spectrally narrow transition to induce the state-dependent force \citep{Aolita2007b}. In this way one induces a spectral rather then a polarization dependent fore.  Another option is to recode the qubits for the conditional phase gate from a magnetic insensitive coding to a different coding \citep{Langer2006phd}. The M{\o}lmer-S{\o}rensen gate, on the other hand, has been already applied on magnetic field insensitive transitions \citep{Haljan:2005b}. The Cirac-Zoller gate (either with composite pulses described in Sec.~\ref{sec:composite-pulses} or a third magnetic field insensitive transition as available for instance with the D manifold in $^{43}$Ca$^+$) can be also directly applied in these situations.
% to temporarily transfer the quantum information into a magnetic field sensitive manifold for the geometric gate operation.

\subsubsection{Other gate \new{types}}
\label{sec:other-gates}
Another possibility to employ AC-Stark shifts for a two-qubit gate was demonstrated by \citet{Brune1994} with Rydberg atoms passing through a microwave cavity. We summarize here the ion trap version implemented by \citet{Schmidt-Kaler2004}. In these experiments, a laser was tuned close to the axial motional sideband of a single ion. The Rabi frequency of the blue motional sideband $\Omega_+$ depends on the phonon occupation number $n_z$ such that the resulting AC-Stark shift of the electronic levels due to the sideband resonance depends on the motional state:
\begin{equation}\label{eq:AC-Stark}
\Delta E = \hbar \frac{\Omega_+^2}{2\Delta} =   \hbar \frac{\eta_i^2\Omega^2}{2\Delta}(n_z+1)\;,
\end{equation}
where we used Eq.~\ref{eq:blue-rabi-frequency}.
Rotating the frame by  $\exp(\frac{\eta_i^2\Omega^2}{2\Delta} t)$ removes the phase evolution of the two electronic levels in
motional ground state $n_z=0$. The relative
phase of the two electronic states in the first excited state $n_z=1$, however,
evolves as  $\varphi=\frac{\eta_i^2\Omega^2}{2\Delta} t$.
Thus, choosing the interaction time $t=\frac{2 \pi \Delta}{\eta_i^2\Omega^2}$, the phase gate operation ${\rm diag}(1, 1, -i, i )$ is implemented, where we used the basis $\{\ket{e,0},\ket{g,0},\ket{e,1},\ket{g,1}\}$.

\begin{figure}[tb!]
\begin{center}
\includegraphics[width=0.6\textwidth]{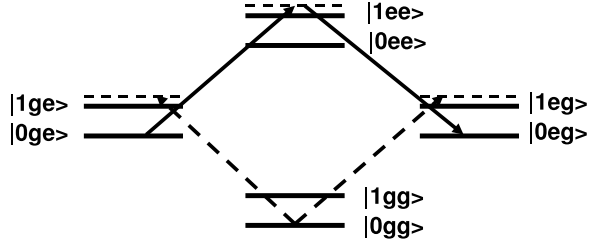}
\end{center}
\caption{\label{fig:quantstark-gatter} Simplified level scheme of two trapped $^{40}$Ca$^+$ ions including the motional state of the bus mode (from \citet{Schmidt-Kaler2004}). The virtual levels are represented by dashed lines. A monochromatic laser beam can only induce transitions between the $\ket{ge}$ and the $\ket{eg}$ state (solid arrows). In addition, the $\ket{gg}$ state acquires an additional phase factor due to an AC Stark effect (dashed arrows).}
\end{figure}

This gate can also be generalized to multiple ions \citep{Schmidt-Kaler2004}. Here all ions simultaneously interact with the laser beam. The situation is very reminiscent of the M{\o}lmer-S{\o}rensen gate: taking as an example a two-ion crystal, only transitions between the
eigenstates  with the same number of excited ions can be induced, e.g. between the $\ket{ge}$ and the $\ket{eg}$ (see Fig.~\ref{fig:quantstark-gatter}). All other basis states acquire only a phase factor which can be corrected for with single-qubit operations.
Using this method starting from the $\ket{eg}$ state, the Bell
state $(\ket{ge}+\ket{eg})/\sqrt{2}$ was generated with a fidelity
close to 0.9 by applying a pulse corresponding to a $\pi/2$
dynamics with trapped $^{40}$Ca$^+$ ions \new{(H. H\"affner {\it et al.}, Innsbruck, unpublished)}.

Another gate type uses the dependence of the carrier Rabi
frequency from the motional state (see
Eq.~\ref{eq:carrier-rabi-incl-all-modes}). Proposed by
\citet{Monroe1997}, it was implemented later by the same group on
a single $^9$Be$^+$ ion \citep{DeMarco2002}. Essentially, a laser
pulse on the carrier transition is applied, with the Lamb-Dicke
parameter chosen such that for one motional state an even number
of Rabi oscillations occurs, while for the other motional state an
odd number of Rabi cycles occurs. This way the phase of an
electronic state is only flipped in the latter case, while it
remains unaffected in the former case. This implementation of a
controlled-NOT operation requires, however, a reasonably large and
tunable Lamb-Dicke factor of the bus mode $\eta_b$ (the gate time
is quadratic in $1/\eta_b$), while either all other motional modes
are cooled to the ground state or their Lamb-Dicke factors are
much smaller than $\eta_b$.

Finally, there have been quite a number of other promising gate
proposals which  are not implemented as of yet. Some of them rely
---as some of the already presented gates--- on state dependent
AC-Stark shifts, albeit in the static regime
\citep{Cirac2000,Steane2004,Staanum2004}. The general idea here is
that an inhomogeneous laser beam changes the distance between the
ions depending on their electronic state.  Simulations show that
the fidelities can be quite high and that the experimental
requirements are not very demanding. For instance, only moderately
low temperatures are required. Alternatively, the state dependent
potential can be created with strong magnetic field gradients
\citep{Mintert2001}. Creation of sufficently strong field gradients might be eased by implementing these gates in microfabricated ion traps where smaller ion-surface distances and microstructured current carriers would facilitate the generation of the required static or dynamic field gradients \citep{Leibfried:2007,Chiaverini:2008,Ospelkaus:2008}.

Another gate class, which recently received much attention, uses
short but strong laser pulses to kick the ion string strongly
\citep{Garcia-Ripoll2003}. Most interestingly, the corresponding
gate operation times can be  shorter than even one trap period.
Choosing proper phases and amplitudes of the pulses, the ions are
kicked in such a way that they acquire a state dependent phase due
to their motion. Another gate variant uses a continuous
irradiation with fast phase modulations \citep{Garcia-Ripoll2005}.
Based on these ideas, \citet{Duan2004} propose quantum computation
in an array of trapped ions where the fast gates are used to
induce a next neighbor interaction.

\subsection{Apparative requirements}
Experimental quantum computation requires exceedingly long coherence times as well as an exquisite control over the qubits. The former is achieved by
decoupling the qubits from the environment, while the latter requires a well-defined and switchable interaction with the environment. In ion trap quantum computing, the qubits are held in free
space with electromagnetic forces which hardly couple to the ion's internal degree of freedom. The control is achieved with strong laser fields such that during their interaction they still can be treated classically and  entanglement between the laser field and ions is negligible. Thus the seemingly contradictory requirements ---decoupling from the environment and controlled interactions--- can be fulfilled.

\begin{figure}[tb!]
\begin{center}
\includegraphics[width=0.8\textwidth]{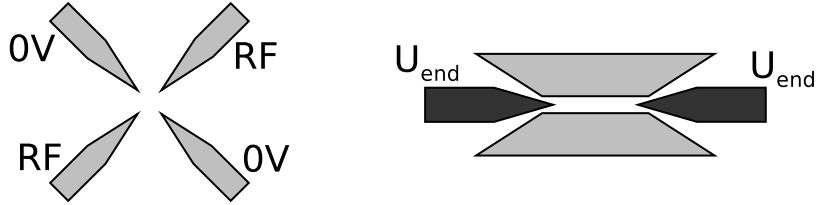}
\end{center}
\caption{\label{fig:trap} Front (left) and side (right) view of the Innsbruck trap (from \citet{Schmidt-Kaler2003a}). It consists of four blades (light gray) and two end caps (dark gray). A radio-frequency drive is applied to two opposing blades while the other blades are held at ground. This provides confinement perpendicular to the trap axis. The two end caps held at a positive DC-potential $U_{\rm end}$ prevent the ions to escape along the axis.}
\end{figure}

For ion-trap quantum computing usually so-called linear Paul traps are used where the ions can form a linear ion chain  \citep{Raizen1992b,Drees1964}.
In these devices, a radio-frequency potential is applied to two electrodes which are parallel to the axis of the trap (see Fig.~\ref{fig:trap}).
These electrodes create an oscillating two-dimensional quadrupole potential that is translation invariant along
the trap axis. If the frequency of the RF field  is sufficiently large, the ions experience an effective restoring force to the center axis. Additionally, static electric fields confine the ions along the trap axis.
The trap resides in an ultra-high vacuum vessel to reduce collisions with residual molecules and atoms as much as possible. For typical experiments, it is sufficient to describe the time dependent confining forces of the ion trap as if they resulted from a static three-dimensional harmonic potential. If the confinement perpendicular to the trap axis (radial direction) is much larger than the confinement along the trap axis, cold ions form a linear chain. Typical trap frequencies for the radial frequencies are between 4 and 10~MHz, while axial frequencies range mostly between 0.5 and 5~MHz.

For the control of electronic and motional states of the ions, lasers with high frequency and intensity stability are used. Acousto-optical modulators allow one to control both the frequency and the intensity. Typically, the lasers are either referenced to an ultra-stable cavity or to a molecular transition. Ions tend to have higher energy splittings as compared to atoms. Therefore often lasers emitting in the ultraviolet are required. In the next paragraphs, we briefly describe the experimental setups used in the NIST and the Innsbruck experiments.

\subsubsection{The NIST setup}\label{sec:NIST-setup}
\begin{figure}[tb!]
\begin{center}
\includegraphics[width=0.8\textwidth]{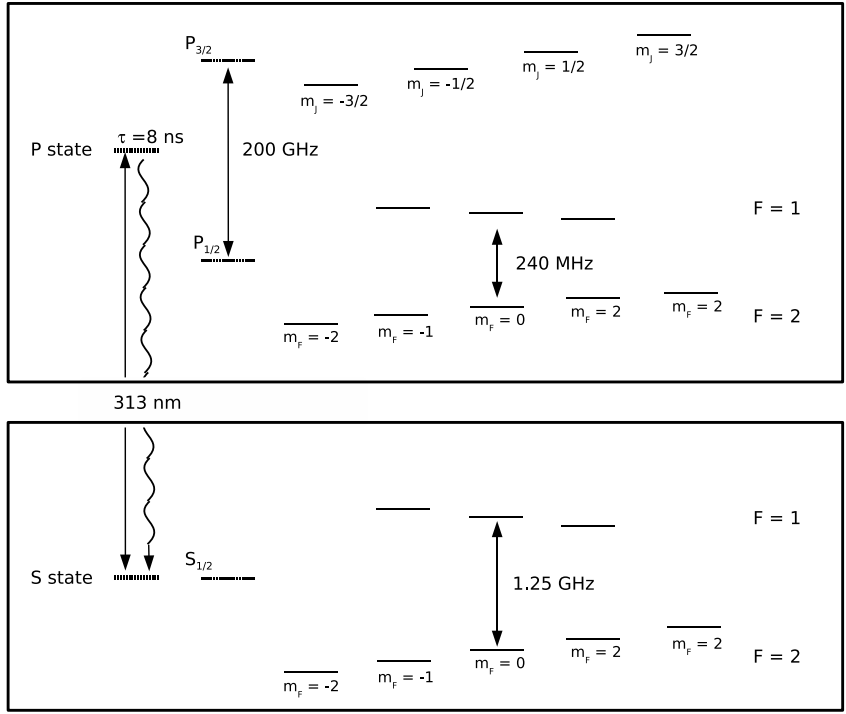}
\end{center}
\caption{\label{fig:beryllium-level-scheme} Energy level scheme of a $^9$Be$^+$ ion (nuclear spin: 3/2). The hyperfine splitting of the P$_{3/2}$ state is smaller than 1~MHz and not shown.}
\end{figure}

Details of the NIST setup can be found in the PhD theses of \citet{Langer2006phd}, \citet{Kielpinski2001phd} and \citet{King1999phd}.
The NIST group uses $^{9}$Be$^+$ ions whose level scheme is depicted in Fig.~\ref{fig:beryllium-level-scheme}. Qubits are encoded in the hyperfine manifold of the S$_{1/2}$ electronic ground state split by 1.25~GHz. For most of the experiments discussed here,
the qubit was encoded in the $\ket{F=2,m_F=-2}\longleftrightarrow\ket{F=1,m_F=-1}$-transition as the levels are easily prepared and distinguished from each other.
Doppler cooling at 313~nm (laser power $\sim \mu$W at a waist of $\sim 25 \: \mu$m) and optical pumping with $\sigma^-$ polarized light on the  $\ket{S_{1/2},F=2,m_F=-2}\longleftrightarrow\ket{P_{3/2},F=3,m_F=-3}$ transition along the quantization axis given by a weak magnetic field (typically $B\sim 1$~mT) initializes the ions in the $\ket{F=2,m_F=-2}$ state. Finally, pulsed resolved
sideband cooling on the $\ket{F=2,m_F=-2}\longleftrightarrow\ket{F=2,m_F=-1}$ transition is used to prepare the ions' motion with a probability of about 0.99 in the ground state \citep{Wineland1998}, before the quantum information is manipulated.

The qubit states are coupled with a Raman transition via the P manifold. The NIST group
generates the necessary Raman beams from the same laser source such that the relative phase of the beams
is well-defined. Therefore the dominant decoherence source is dephasing due to magnetic field
fluctuations (e.g. due to the mains supply at 60~Hz) causing a qubit lifetime on the order of a few
milliseconds. However, recently the NIST group used a magnetic field insensitive transition and measured coherence times on the order of a few seconds (see Sec.~\ref{sec:clock-transition}).

 Read-out is performed again on the cycling transition  $\ket{S_{1/2},F=2,m_F=-2}\longleftrightarrow\ket{P_{3/2},F=3,m_F=-3}$ with $\sigma^-$ polarized light. This light does not couple efficiently to the $\ket{S_{1/2},F=1,m_F=-1}$ state as it is detuned from any possible transition by a little bit more than 1~GHz. In order to avoid pumping into the bright state by the still present off-resonant excitations, the $\ket{S_{1/2},F=1,m_F=-1}$ population is
 transferred with two $\pi$ pulses to the  $\ket{S_{1/2},F=1,m_F=+1}$ level. In this state, absorption of a single off-resonant photon cannot lead to a population of the $\ket{S_{1/2},F=2,m_F=-2}$. Thus the ion remains dark. The theoretical analysis by \citet{Langer2006phd} shows that with this method detection errors can be kept smaller than $10^{-4}$. At the moment, due to stray light background and imperfections in the $\pi$ transfer pulses, typically detection efficiencies of about 99\% are attained in the NIST setups \citep{Langer2006phd}.

% As the NIST group tends to use small traps relatively large heating rates of a few phonons/10ms are present

The small atomic mass of beryllium allows for
high trap frequencies and comparatively large Lamb-Dicke factors. Both factors alleviate a strong coupling to the motional degree of freedom and thus facilitate fast multi-qubit operations. In order to initialize, manipulate and detect the quantum states, light sources in the ultraviolet at 313~nm are required.
%Currently, these are generated with frequency doubled dye lasers.

The NIST group uses various microstructured traps. Many of these traps have multiple trapping zones, which ease the scaling of ion trap quantum computers to larger ion numbers \citep{Kielpinski2002}. In addition, segmented traps allow for single-qubit addressing without tightly focused laser beams (see Sec.~\ref{sec:teleportation}) and a separate loading zone avoiding patch effects (see Sec.~\ref{sec:motional-coherence}).
%Nevertheless, various other schemes are used to achieve effective single ion addressing (c.f. Sec.~\ref{sec:ion-addressing}, \citet{Wineland1998}).

% More recently the NIST group also employed planar surface traps (see Sec.~\ref{sec:planar-traps}). Already two trap designs, one  based on gold on a quartz substrate\citep{Seidelin2006} and the other one based solely on silicon \citep{Britton2006} trapped successfully.

\subsubsection{The Innsbruck setup}
The experimental setup used by Innsbruck group is described in Refs.~\citep{Schmidt-Kaler2003a,Gulde2003phd}. As qubits, superpositions of the S$_{1/2}$ ground state and the metastable D$_{5/2}$ state of $^{40}$Ca$^+$ are used (see Fig.~\ref{fig:ca-40-level-scheme}). The  D$_{5/2}$ state has a lifetime
$\tau \simeq$ 1.16~s.

 A magnetic field of 300~$\mu$T lifts the degeneracies of the Zeeman manifolds.
For the experiments, the entire quantum register is prepared by Doppler cooling, followed by sideband
 cooling to the motional ground state. Normally,  only the center-of-mass mode
($\omega_{\rm CM}\approx 1.2$~MHz) is cooled to the ground state. The ions' electronic qubit states are
initialized in the S$_{1/2} (m_j= - 1/2$) state by
optical pumping with $\sigma^-$ light. Then each
ion-qubit is individually manipulated by a series of laser pulses on the ${\rm S}
\equiv {\rm S}_{1/2}$ (m$_j$=-1/2) to ${\rm D}
\equiv {\rm D}_{5/2}$ (m$_j$=-1/2) quadrupole transition near 729~nm. In order to guarantee phase coherent manipulation, the laser frequency is  stabilized to about 50~Hz on a time scale of 1 minute by locking it to an ultra-stable reference cavity with a similar design as presented by \citet{Notcutt2005}. This time scale corresponds to the typical duration of a full set of quantum computing experiments. On time scales of a few seconds even a 3~Hz linewidth has been observed. The slow drift of the reference cavity ---typically about 1~Hz/s--- is monitored every few minutes by interrogating the qubit transition and compensated with a feedback loop. Thus many thousands of experiments under comparable conditions are feasible.

Spectroscopy on a transition more sensitive to magnetic field fluctuations (S$_{1/2}$ (m$_j$=-1/2) $\rightarrow$ D$_{5/2}$ (m$_j$=-5/2)) is used to monitor slow magnetic field drifts continuously. In addition, a passive magnetic shield reduces the magnetic field fluctuations on timescales longer than 1~s by about a factor of 30 and by more than 2 orders of magnitude for frequencies higher than 10~Hz. The total field amplitude noise is on the order of a few nT. Thus, typically coherence times of about 5~ms on the  S$_{1/2}$ (m$_j$=-1/2) $\rightarrow$ D$_{5/2}$ (m$_j$=-5/2) transition and 15~ms on the  S$_{1/2}$ (m$_j$=-1/2) $\rightarrow$ D$_{5/2}$ (m$_j$=-1/2) transition are achieved. The laser driving these transition is tightly focused onto individual ions in the string with a waist size of 2~$\mu$m (inter-ion distance $\sim5\:\mu$m).

While the NIST group uses predominantly global addressing and state read-out, the Innsbruck-group uses tightly focused laser beams to address individual ions (see Sec.~\ref{sec:ion-addressing}). Thus in principle any
quantum algorithm can be implemented  straightforwardly, only limited by the decoherence time. The trade-off of this method, however, is that the axial trap frequency can not be increased too much, since then the ions move closer to each other, thwarting single ion addressing. A consequence of a lowered trap frequency is a reduced speed of the entangling operations on the sidebands.

\subsubsection{Composite pulses \new{and optimal control}}\label{sec:composite-pulses}
Quantum algorithms are usually implemented by a sequence of a few fundamental gates. Many of those gates can be carried out with single laser pulses. However, using a set of pulses sometimes offers an advantage over using single pulses (as demonstrated already in Sec.~\ref{sec:ion-addressing}). In NMR, the composite pulse technique is well known and allows for the compensation of many systematic effects like intensity mismatch and detuning errors \citep{Freeman1997,Levitt1979,Levitt1986}. The spin echo sequence discovered by \citet{Hahn1950} is such a sequence with which a constant detuning between the excitation and the transition can be removed to a large extent.

In addition to the many sequences discovered and used in NMR, there are a few sequences which are important in the context of ion traps. We present here two of those which were used in the implementation of the Deutsch-Josza algorithm \citep{Gulde2003} (see Sec.~\ref{sec:dj-algorithm}) and were described first by \citet{Childs2000}. The first sequence uses four sideband pulses \citep{Childs2000,Gulde2003,Schmidt-Kaler2003a} to implement a phase gate in the computational subspace $\{\ket{D,0},\ket{S,0},\ket{D,1},\ket{S,1} \}$. The advantage over the method laid out in Sec.~\ref{sec:cz-gate} is that no third  level is required to implement the gate.
% As this auxiliary level almost inevitably has a very different magnetic moment as compared to the other two levels, the gate becomes more robust against magnetic field fluctuations.
To achieve the desired gate with just two levels, cleverly chosen pulse lengths and phases avoid
leakage into higher phonon number states. Using the definitions from Eq.~\ref{eq:Rblue}, the pulse sequence (to be read from right to left) is:  $R^+(\pi/2,\pi/2)R^+(\pi/\sqrt{2},0)R^+(\pi/2,\pi/2)R^+(\pi/\sqrt{2},0)$.
Having Fig.~\ref{fig:ion-tensor-oscillator} in mind, we analyze the effect of this pulse sequence on the four physical eigenstates $\{\ket{D,0},\ket{S,0},\ket{D,1},\ket{S,1} \}$. The $\ket{D,0}$ state is not affected at all and therefore $\ket{D,0} \rightarrow \ket{D,0}$.
\begin{figure}[tb!]
\begin{center}
\includegraphics[width=0.98\textwidth]{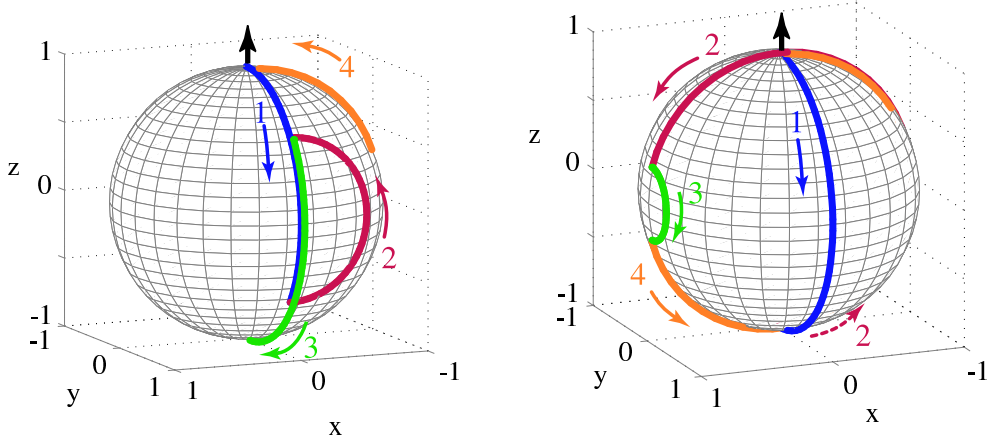}
\end{center}
\caption{\label{fig:comp-phase-gate} Evolution of the Bloch vector during the composite phase gate in the $\ket{S,0}\leftrightarrow\ket{D,1}$ (left) and $\ket{S,1}\leftrightarrow\ket{D,2}$ manifold (right) (from \citet{Schmidt-Kaler2003a}).}
\end{figure}
Fig.~\ref{fig:comp-phase-gate} illustrates the evolution of the respective Bloch vector in the $\ket{S,0}\leftrightarrow\ket{D,1}$ and $\ket{S,1}\leftrightarrow\ket{D,2}$ manifold, respectively. Because the couplings and thus the effective pulse lengths differ by a factor of $\sqrt{2}$  between the two manifolds, the Bloch vector
follows different paths. Still, it reaches always its original position when the pulse sequence is finished. Using Eq.~\ref{eq:Rblue} and Eq.~\ref{eq:Rblue}, one can show that for the three remaining cases, each time a phase factor of -1 is picked up and thus the diagonal matrix ${\rm diag}(1,-1,-1,-1)$ is implemented.

Similarly, three blue sideband pulses can be employed to implement a SWAP operation between an electronic and a motional degree of freedom of trapped ions \citep{Gulde2003}: $R^+\!({\pi}/{\sqrt{2}},0)R^+\!({2\pi}/{\sqrt{2}},\!\varphi_{\rm swap})R^+\!({\pi}/{\sqrt{2}},0)$, where $\varphi_{\rm swap}=\arccos\left(\cot^2(\pi/\sqrt{2})\right)$.
This pulse sequence was used by \citet{Gulde2003} to implement the Deutsch-Josza algorithm \citep{Deutsch1989}.
Already these two pulse sequence examples demonstrate that composite pulses are
a quite versatile tool.

\new{Especially, gradient ascent pulse engineering (GRAPE) developed in the context of NMR carries the idea of composite pulses to its extreme \citep{Khaneja2005}. Here a pulse sequence thought to implement a particular unitary operation is split into many pulses. Then a special algorithm is used to vary  amplitudes and phases of the pulses to perfect the desired unitary that is optimal with respect to certain quality criteria (e.g. execution time) with a special algorithm \citep{Khaneja2005}. In addition, various boundary conditions (e.g. experimental constraints like finite pulse rise times) can be included in form of cost functions. Thus,
significant performance improvements in terms of speed and reduced susceptibility to experimental imperfections can be achieved. Most interestingly, the sensitivity to
control parameters can be minimized with GRAPE, too. However, note that in spite of the fact that the execution time can be reduced considerably, composite pulses lead usually to a larger total pulse area. Thus decoherence effects which scale with the pulse area might become relevant. One such source of decoherence is spontaneous emission during Raman transitions (see Sec.~\ref{sec:spontaneous-emission}).}

\new{First steps in applying such optimal control techniques to trapped ions have already been taken. \citet{Timoney:2008} deviced pulse sequences to robustly perform $\pi/2$ and $\pi$ rotations between two hyperfine qubits of a single $^{171}$Yb$^+$. Creating the states $\ket{0}+e^{i\varphi}\ket{1}$ and $\ket{1}$ from $\ket{0}$, this work also experimentally demonstrates the robustness of these pulse sequences as compared to simple $\pi/2$ or $\pi$ pulses, respectivvely, with respect to intensity and detuning errors. }

\new{Furthermore, \citet{Nebendahl:2008} modify a GRAPE algorithm to construct a controlled-NOT operation from a global M{\o}lmer-S{\o}rensen interaction and single qubit operations. The algorithm allows for the optimization of whole algorithms. Taking as an example a quantum error correction scheme for bit flips based three qubits and two ancilla qubits, the original length of more than 100 pulses was reduced to 34 pulses. }

\section{Decoherence in ion trap quantum computers}
\label{sec:decoherence}
This section describes the most relevant decoherence mechanisms for ion trap quantum computers. For further discussions, we refer to Ref.~\citet{Wineland1998}.
\subsection{Sources of imperfections in ion trap quantum computers}
\subsubsection{Bit-flip errors}\label{sec:bit-flips}
Bit-flip errors occur when a process transfers populations between the physical eigenstates of the qubit. Usually, the physical eigenstates are the energy eigenstates of the system and bit-flips are
connected to radiation or absorption of photons. Thus, bit-flip errors are usually caused by spontaneous emission.
%, if all other sources of photons like the manipulation lasers are switched off properly.
 The frequency differences of hyperfine and Zeeman qubits
are quite small and therefore the time constants for spontaneous emission are usually longer than a year and thus are of no relevance if the ions are not irradiated with electromagnetic radiation. For optical qubits usually superpositions of the ground state and the metastable state $D$-level in earth-alkali elements are used. Typical life times of the $D$-levels
are about 1 second \citep{Barton2000,Kreuter2005a,Letchumanan2005}, which is long as compared to the gate time of less than 1~ms. Therefore, most current experiments are not yet limited by bit-flip errors during their free evolution.
Bit-flip errors on the motional degree of freedom are discussed in section~\ref{sec:motional-coherence}.
\subsubsection{Dephasing}
The phase evolution of a superposition often depends on a classically well-defined parameter such as the magnetic field. Ignoring the time evolution of the  classical parameter leads to dephasing of the superposition.
For instance, a superposition consisting of two levels with differing magnetic moments experiences energy shifts due to a (fluctuating) magnetic field. Thus the phase of the superposition depends on the particular magnetic field history and dephasing occurs. Similarly, dephasing also takes place for the ion motion if the trap frequency is not constant, e.g. due to
voltage fluctuations on the trap electrodes (see section~\ref{sec:motional-coherence}).

\begin{figure}[tb!]
\begin{center}
\includegraphics[width=0.98\textwidth]{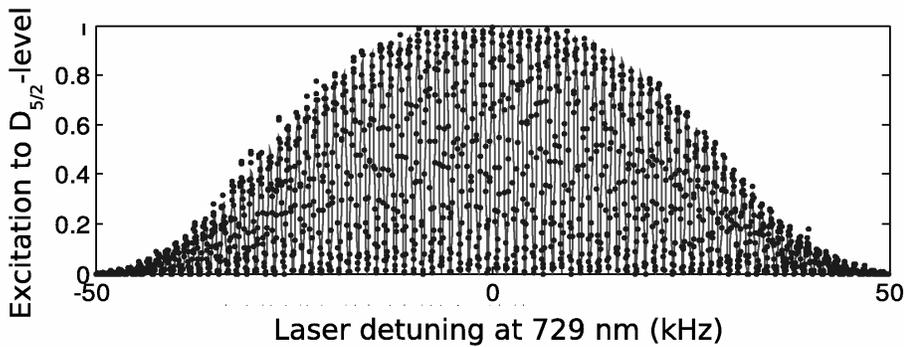}
\end{center}
\caption{\label{fig:ramsey-fringes} Ramsey fringes with a free precession time of $T=100 \:\mu$s \new{(Chwalla {\it et al.}}, Innsbruck, unpublished). The data points indicate the excitation probability to the metastable $D$ level in a single $^{40}$Ca$^+$ ion.}
\end{figure}

  A typical experimental sequence to test\label{sec:phase-coherence}
the phase coherence consists of  two $\pi/2$ pulses separated
by a waiting time $T$ (a Ramsey experiment). During the waiting time,
 any  difference between the atomic transition frequency  and the laser frequency will lead
to an evolution of their relative phase. The second $\pi/2$ pulse will then rotate the atomic state according to the phase difference either towards the excited or the ground state. This effect can be
seen in Fig.~\ref{fig:ramsey-fringes} when the population of the ion oscillates with the detuning of the laser frequency from the atomic transition (Ramsey fringes). The contrast $\left(\frac{{\rm max}-{\rm min}}{{\rm max}+{\rm min}}\right)$ in the center of this curve is in the following termed Ramsey contrast.

Experimentally, fluctuations of the atomic resonance frequency, e.g. typically due to fluctuations of the magnetic field  and of the laser frequency, lead to dephasing. Sometimes it is useful
to distinguish between fast fluctuations (the relative frequency changes during the
waiting time), slow fluctuations (the relative frequency is constant in each experiment, but
not the same in the next experiment(s)) and an intermediate regime. It can be shown that with an increasing
waiting time, fast fluctuations lead to an exponential decay of the Ramsey contrast, while slow fluctuations
lead to a Gaussian decay of the Ramsey contrast \citep{Sengstock1994}. Furthermore, the slow fluctuations
can be compensated for by spin-echo techniques \citep{Hahn1950}.

Currently, the coherence time of most qubits is limited by magnetic field fluctuations to a few milliseconds. Usually Fourier components at multiples of the mains frequency contribute most to the magnetic field changes. Synchronizing the experiment with the phase of the mains supply counters time variations of the magnetic field effectively.

A more generic way to reduce magnetic field fluctuations is
shielding of ambient magnetic fields with $\mu$-metal and/or active cancellation with a feedback loop. Currently, the Innsbruck group uses an aluminum/$\mu$ metal shield which suppresses the magnetic field fluctuations by more than two orders of magnitude. However, in most cases a finite magnetic field is needed to lift the degeneracy of the electronic levels. Therefore, a very stable magnetic field has to be created inside the magnetically shielded region without saturating the $\mu$-metal. The standard procedure is to use a pair of Helmholtz coils through which a current is passed. As it is very difficult to stabilize currents to better than $10^{-6}$, in the future it might be worth while going through the trouble to use superconducting solenoids. Here magnetic field stabilities of better than $10^{-11}$ at 6~T have been attained. These experiments were made possible by choosing a particular geometry of the superconducting coils to shield the external magnetic fields \citep{vanDyck1999}, especially of high importance for small fields. It remains to be seen to what extent these exceptional field stabilities can be obtained at the relatively small fields of $1$~mT as required for quantum computation.

A more elegant solution to reduce dephasing due to magnetic field fluctuations is to use qubit levels having the same magnetic moment. For this especially ions
with a hyperfine structure have interesting levels. Obvious choices for the qubit transitions are of the form $m_F=0 \rightarrow m_F=0$, which experience only a quadratic Zeeman-effect at small magnetic fields. However, at the magnetic fields required to lift the Zeeman degeneracies, a considerable
linear Zeeman-effect is present.
\label{sec:clock-transition} Therefore, it seems advantageous to work with stronger magnetic fields
where transitions with a purely quadratic Zeeman-effect can be found.

While magnetic field insensitive transitions were extensively
explored in microwave precision experiments
\citep{Bollinger1991,Thompson1990}, in the context of quantum
computing it has been implemented only very recently by the NIST
group on $^9$Be$^+$ \citep{Langer2005}, by the Oxford
\citep{Lucas:2007} and the Innsbruck groups on $^{43}$Ca$^+$
\citep{Benhelm2007,Benhelm:2008c} and by the Ann-Arbor group on
$^{111}$Cd$^+$ \citep{Haljan:2005b}. To achieve a reasonable
spatial selectivity, the qubits are manipulated with an optical
Raman drive (c.f. Sec.~\ref{sec:ion-qubits}). However, care has to
be taken to maintain the phase coherence between both laser fields
on time scales of several seconds. Using co-propagating laser
beams simplifies this obstacle considerably as both beams
propagate along the same path such that effects of mirror
vibrations as well as of air currents cancel. However, a
co-propagating geometry does not allow for easy coupling to the
motional degree of freedom. On the other hand, if only phase gates
are used as two-qubit gates, the phase stability only has to be
maintained during each phase gate operation.

\citet{Langer2005} demonstrated coherence times $\tau > 10$~s using a magnetic-field-independent hyperfine transition in  $^9$Be$^+$ at a magnetic field of $B_0 \simeq 0.01194$~T. In these experiments, Ramsey spectroscopy was carried out on the $\ket{F=2,m_F=0}\longleftrightarrow\ket{F=1,m_F=-1}$ qubit transition to measure the phase coherence. The
optimal magnetic field was determined by measuring this transition frequency ($\sim 1$~GHz) as a function of the magnetic field. The minimum of the resulting parabola (second order derivative $B_2\simeq0.305\:{\rm Hz}/\mu{\rm T}^2$, $B_0 \simeq 0.01194$~T) corresponds to the desired magnetic field with the least magnetic field dependency. The coherence time
of this qubit is limited by slow drifts of the ambient magnetic field within the typical measurement times of a few hours especially for scans with long Ramsey waiting times.
These experiments demonstrate a qubit memory error rate on the order of $t_{0}/\tau \approx 10^{-5}$ where the time scale $t_0$ is set by the detection time of $t_{0}=200\:\mu$s, which is the longest operational time of the NIST ion-trap quantum computer.

The Oxford group carried out coherence measurements on
$^{43}$Ca$^+$ with a microwave drive \citep{Lucas:2007}. They used
the $\ket{F=3,m_F=0} \leftrightarrow \ket{F=4,m_F=0}$ transition
in the hyperfine ground state manifold of  $^{43}$Ca$^+$ at small
magnetic fields ($B\approx 0.178$~mT) and observed a dephasing
time  of $1.2\:(2)$~s. Additionally, they investigated the
coherence properties in a spin echo configuration and could not
detect any decay of the Ramsey contrast on time scales of up to
1~s. From this they deduce a spin-echo dephasing time of larger
than 45~s. Similar results were obtained by the Innsbruck group in
microwave-induced Ramsey experiments on the $\ket{F=3,m_F=0}
\leftrightarrow \ket{F=4,m_F=0}$ clock transition of
$^{43}$Ca$^+$. For Ramsey interrogation periods $\tau=1\,$s, the
Ramsey contrast was still 88\% while for $\tau=50\,\mu$s, a
contrast of 97\% was found. The experiment was carried out at a
field of 0.05\,mT \new{\citep{Benhelm:2008c,Benhelm:2008phd}}.

\subsubsection{Imperfect control}\label{sec:imperfect-control}
Other serious sources of decoherence are imperfect realizations of the intended
gate operations, usually caused by fluctuating or insufficiently calibrated control parameters. Typical candidates for these parameters are intensity and frequency fluctuations of the
laser and beam pointing instabilities. In addition, pulses
on a particular ion can have unwanted side effects on the ion itself (AC-Stark shifts, off-resonant excitations) or on neighboring ions (addressing errors). Many of these errors can be greatly reduced with composite pulse and optimal control techniques (c.f.~Sec.~\ref{sec:composite-pulses}). We list here some of these imperfections:
\begin{itemize}
\item {\bf Pulse length errors} arise from intensity fluctuations of the laser beam, beam pointing instabilities or just miscalibration. All of these reasons are almost equally relevant in current experiments. Common to all of them is also
that the fluctuations  take place at frequencies below a kHz, such that they can be considered
constant during the execution of a pulse sequence. Relative amplitude fluctuations of $10^{-2}$ are typical.

Insufficient cooling can also lead to effective intensity fluctuations. During each run, the phonon numbers $n_m$ have different values (which corresponds to a finite temperature of
the ion crystal) and thus for a non-vanishing Lamb-Dicke factor the Rabi frequencies are different (see Eq.~\ref{eq:carrier-rabi-incl-all-modes} and Eq.~\ref{eq:blue-rabi-incl-all-modes}). Interestingly, for an increasing number of ions the Innsbruck group observes that this effect is reduced. In fact, in their current experiments it is only relevant for single ions. As detailed by \citet{Wineland1998}, there are two reasons for this:
\begin{enumerate}
 \item With an increasing ion number the Lamb-Dicke factor for each mode tends to get smaller.
 \item The contributions of the increasing number of modes averages, such that the variance of the
effective Rabi frequency narrows.
\end{enumerate}

\item {\bf Detuning errors}
take place when the qubit transition frequency is miscalibrated or the drive frequency fluctuates slowly as compared to the duration of the coherent manipulation. Furthermore, magnetic field fluctuations have the same effect and usually are also slow as compared to  the coherent manipulation time. The effect of a detuning error is a constant phase evolution during the experiment. A simple and effective method to remove such a constant phase evolution is the so-called spin-echo method \citep{Hahn1950}. The idea is that after half the evolution time the roles of the upper
and the lower qubit level are exchanged. Thus the phase rewinds during the second half and ---if the detuning is constant--- arrives at zero after the complete evolution time. Often this method can be implemented quite straightforwardly. An example in the ion trap context can be found  in \citet{Leibfried2003a}. However, usually during algorithms the qubits state is changed. In this
case either more spin-echo sequences might have to be used (e.g. for each free evolution one as in \citet{Barrett2004}) or  an effective spin echo sequence can be found. For instance, \citet{Riebe2004} optimized the time position of the population inverting pulses both in simulations and in the actual experiment.
\item {\bf Addressing error} While addressing a single ion with a focussed laser beam, residual light might affect other ions in the trap and thus perform an undesired unitary evolution. See Sec.~\ref{sec:ion-addressing} for more details.
\item {\bf Off-resonant excitations} also limit the obtainable fidelity. Off-resonant excitation is usually a problem if one drives a weak transition in the presence of a nearby strong transition (c.f. Eq.~\ref{eq:off-resonant-excitations}). Exactly this situation occurs in ion traps when driving the
sideband transition \citep{Steane2000}. The transition matrix element of the sideband transition is weaker by a factor $\eta$
than the one of the carrier transition as can be
inferred from Eq.~\ref{eq:blue-rabi-incl-all-modes} and Eq.~\ref{eq:carrier-rabi-incl-all-modes}. Thus strong laser fields are required to obtain a reasonable
gate speed which can then yield high gate fidelities in the presence of dephasing mechanisms.

 However, the strong laser field, characterized by the Rabi frequency $\Omega$, leads to off-resonant excitations on the carrier transition (see Eq.~\ref{eq:off-resonant-excitations}).
The Innsbruck experiments suffered particularly from this effect \citep{Schmidt-Kaler2003}.
Quantum mechanically, the off-resonant excitation can be understood as Rabi oscillations induced by a non-adiabatic switching of the energy eigenbasis while the laser power is changed. Thus a system initially being in an energy eigenstate finds itself not any longer in
an eigenstate of the Hamiltonian and consequently oscillations between the newly populated energy eigenstates occur.

Off resonant excitations can be greatly reduced with pulse shapes which have no spectral Fourier components at the carrier-transition. In the simplest case the laser pulse powers are switched smoothly such that during the smooth turn on and off the system follows adiabatically.

\item {\bf Unwanted AC-Stark shifts}\label{sec:AC-Stark-compensation} have a similar origin as off-resonant excitations (see Eq.~\ref{eq:carrier-Stark-shift}), however, affect the phase of the qubit rather than the population. Here the carrier transition nearby leads to an AC-Stark shift of the qubit levels and thus the qubit phase evolves \citep{Haeffner2003a}.
In principle, this phase evolution can be measured and then taken into account in the algorithm to
be implemented. In the Innsbruck setup the problem, however, with this approach is that the acquired phase shift during a controlled-NOT operation is typically on the order of  $20\pi$. In order  to obtain a reasonable phase stability of 0.1~$\pi$, one needs an intensity reproducibility of better than 0.005.
To achieve this intensity stability of a particular polarization at the ion position with a narrow beam waist is quite demanding. To relieve this stringent condition, a second light field can be used which induces an AC-Stark shift of the same magnitude but of opposite sign \citep{Haeffner2003a,Kaplan:2002}. This field can be most conveniently derived by driving the acousto-optical modulator (AOM) used to control the qubit-control field with a second RF signal. Thus the two laser fields, for qubit manipulation and for AC-Stark shift compensation, are generated simultaneously by the same AOM. Both light fields pass then along the same path to the ions such that they pick up virtually the same intensity, polarization and beam pointing fluctuations, removing the AC-Stark shifts to a large extent.

In the Innsbruck experiments, also a considerable
AC-Stark shift appears due to dipole-allowed transitions \citep{Haeffner2003a}. For the blue sideband,
the presence of the dipole-allowed transitions cancels partially the shift induced by the presence of the carrier transitions, whereas for the red sideband
they add. Thus in the Innsbruck experiments for coupling the ions to the motion, the blue sideband is preferred over the red one.
Another possibility to reduce AC-Stark shifts is to use the polarization degree of freedom. The NIST group tunes the polarization of the Raman-laser beam pair to minimize the shift \citep{Wineland2003}.
\label{sec:spontaneous-emission}
\item For Raman-driven qubits, spontaneous decay from the levels used to couple the two qubit levels has to be considered. \citet{Ozeri2005} show that using very large detunings this decoherence effect can be reduced sufficiently, however, at the expense of requiring large laser powers. For a detailed discussion of these issues, we refer here to \citet{Ozeri2007}.
\end{itemize}
Most of these decoherence sources can be minimized by changing external parameters. For example the laser intensity can be reduced such that AC-Stark shifts become negligible at the expense of slow gates. Slow gates in turn increase the susceptibility to dephasing due to fluctuating magnetic fields and laser frequencies. Similarly high trap frequencies allow for faster gates \citep{Steane2000}, but make good addressing of the individual qubits more difficult. In the experiments therefore often a compromise has to be made to keep all decoherence mechanisms reasonably small. Finally,
%\subsubsection{Other }

The previous paragraphs listed the most common important imperfections. However this list is of course incomplete:
for instance, a finite residual temperature of the ion string leads to single and two-qubit gate errors. Especially, the Cirac-Zoller gate is very susceptible to imperfect cooling. This error source is very special, since the combined fidelity of two concatenated gates is not the product of the individual gate fidelities as discussed in Sec.~\ref{sec:process-tomography}. Therefore, a full simulation of the whole algorithm is required to
make reliable predictions on its performance.

% : depending on whether  ground state cooling was successful, all gates will either succeed or fail at the same time (provided motional heating is neglegible). Therefore, a memory effect is present which leads to correlated gate failures. This gives a hint that it is non-trivial to estimate the total fidelity of an
% algorithm from the individual gate fidelities \citep{Riebe2006}. Instead, for each gate the exact nature of all relevant
% error sources must be known to find a reliable estimate.Needless to say that
% a full simulation of the whole algorithm is quite useful here. But it is also
% clear that this is only possible for a small number of qubits.

\subsection{Motional coherence}\label{sec:motional-coherence} In the Cirac-Zoller proposal, the quantum information is temporarily stored in the motional degree of freedom. Some other gate types, like the
M{\o}lmer-S{\o}rensen gate require the ion string to be only well within the Lamb-Dicke regime  (Sec.~\ref{sec:Moelmer-Soerensen-gate}). Both gate types are affected by population changes  (motional heating) and dephasing (e.g. trap frequency fluctuations) during the gate operation.
Dephasing is often caused by slow drifts of the trap voltage on the order of a few Hertz, whereas  motional heating can be caused by electromagnetic background radiation at the trap frequencies. This background stems predominantly from voltage fluctuations in material close to the trap. The most fundamental source of these voltage fluctuations should be the thermal motion of the electrons inside the conductors. This mechanism has been thoroughly investigated, both theoretically and experimentally, by \citet{Wineland1975} with electrons in Penning traps.
The Johnson noise heating power $P$ is given by:
\begin{equation}\label{eq:Johnson-noise}
 P_{\rm noise} = kT \Delta \nu\:,
\end{equation}
where $kT$ is the thermal energy and $\Delta \nu $ is the frequency bandwidth in which the ion accepts the power. The time $\tau$  in which one motional quantum of energy $E_{\rm q}=h\nu$ is generated is  given by
\begin{equation}\label{eq:Johnson-heating}
{\tau^{-1}}=\frac{P_{\rm noise}}{E_{\rm q}}=\frac{kT\Delta \nu}{h\nu}=\frac{kT}{hQ} \:,
\end{equation}
\begin{figure}
\begin{center}
\includegraphics[width=0.7\textwidth]{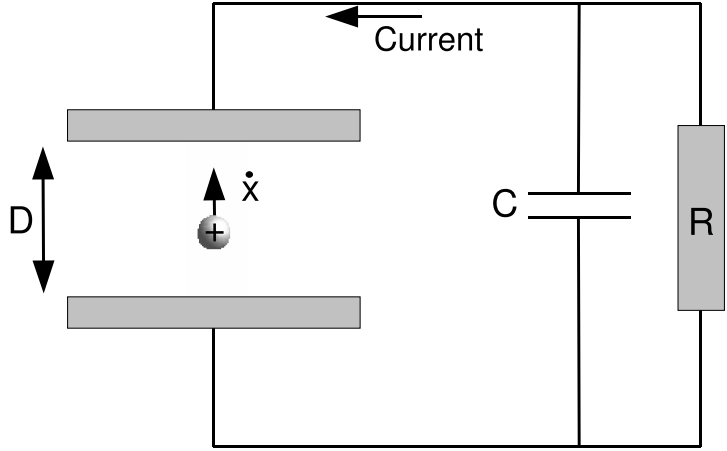}
\caption{\label{fig:lumped-circuit} Lumped circuit model of a single ion at velocity $\dot{x}$ interacting with the trap electrodes. The connection between the trap electrodes is mainly characterized by the resistivity $R$ and capacitance $C$ between the electrodes.}
\end{center}
\end{figure}

where we introduced the quality factor $Q$ of the ion motion. This quality factor can be derived from the dissipated power $P_{\rm dis}=I^2 {\rm Re}Z$ of the current induced by a single ion with an energy $E_{\rm ion}$ at the real part of the impedance Re$Z$. For this we assume a lumped circuit model (see Fig.~\ref{fig:lumped-circuit}) where the ion induces a current $I=q\dot{x}/D$  with $q$ being the charge and $\dot{x}$ the velocity of the ion \citep{Shockley1938}. The characteristic dimension $D$ is on the order of the ion-electrode distance. Using this, we obtain for the quality factor:
\begin{equation}\label{eq:Q-value}
 Q=\frac{E_{\rm ion}}{P_{\rm dis}/\nu}=\frac{m \dot{x}^2 \nu}{I^2 {\rm Re}Z} = \frac{m \nu D^2}{q^2  {\rm Re}Z}\:.
\end{equation}
Inserting Eq.~\ref{eq:Q-value} into Eq.~\ref{eq:Johnson-heating}, we arrive for the time in which one motional quantum is acquired at
\begin{equation}\label{eq:heating}
 {\tau^{-1}}=\frac{kT}{h\nu} \frac{q^2  {\rm Re}Z}{m  D^2}\:.
\end{equation}
For typical values of $D=100\:\mu$m,
 ${\rm Re}Z=1\:\Omega$ at
 room temperature, the expected heating time from Johnson noise is $\tau \sim 200$~s/quantum and thus very small.

% We finally arrive at
% \begin{equation}\label{eq:heating-rate}
%
% \end{equation}

In order to measure a heating rate, one can cool the ion (string) to the ground state, wait for some time to allow for some heating and then probe the strength of the motional sideband. Assuming a thermal distribution, the mean phonon number is deduced by employing  Eq.~\ref{eq:blue-rabi-incl-all-modes}.
Depending on the thermal excitation, the Rabi oscillations on the blue sideband speed up and eventually degrade (c.f.~Sec.~\ref{sec:blue-sideband-flops}).
Repeating this procedure for various waiting times yields the heating rate.
\citet{Seidelin2006}, \citet{Epstein2007} and \citet{Wesenberg2007} developed and applied another method which is based on the strength of the ion fluorescence. The basic idea is that the ion motion leads to Doppler broadening of the absorption spectrum. The change in fluorescence is recorded as a function of a waiting time when cooling is switched off. The latter method is less sensitive than the first one, however, does not require manipulation of the sidebands.

The NIST group observed heating rates of a
few phonons per ms  \citep{Leibfried2003b,Turchette2000} in various traps.
 This is orders of magnitude more than what is expected from
fundamental electrical noise in the trap electrodes (Eq.~\ref{eq:heating}).
Patch charges on the trap electodes have been suspected to cause these excessive heating rates \citep{Wineland1998,Turchette2000}. The
former publication also discusses various other sources of heating in detail. Patch charges can be generated when an electron beam is
used to ionize the atoms during trap loading. Indeed many experiments suggest that using photo ionization techniques to produce the
ions \citep{Kjaergaard2000} can help to reduce the heating rate. A reason for the seemingly reduced heating rate could be the
reduced vapor pressure of the atoms during photo ionization as compared to the less efficient electron beam ionization. Thus a much reduced atom flux can be used which reduces deposition of atoms on the trap electrodes. Furthermore,
ionization with a laser produces only a minimum of charged particles whereas the electron beam can charge any insulating layer on the
trap electrodes. While photoionization seems to lead to reduced heating rates, no experiment with laser cooled ions was reported as of
yet where the fundamental thermal noise given in Eq.~\ref{eq:heating} dominated the heating rate. Only in some Penning trap
experiments using a resonance circuit to enhance the real part of the resistivity in Eq.~\ref{eq:heating}, thermal noise plays the
dominant role \citep{Wineland1975,Haeffner2003b}.

\begin{figure}
\begin{center}
\includegraphics[width=0.8\textwidth]{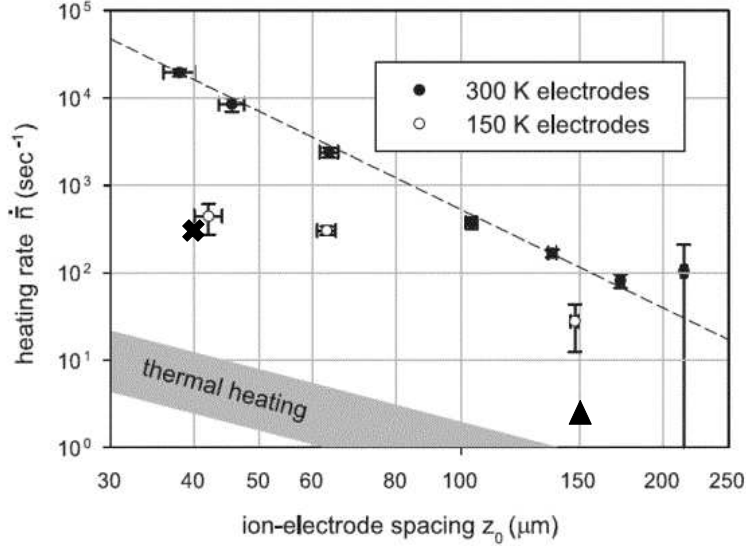}
\caption{\label{fig:motional-heating}  Single ion heating rate as function of ion-electrode distance (a measure of the trap size) (from \citet{Deslauriers:2006a}). The trap consists of
two needles to which the radio frequency is applied to. \new{Furthermore, two heating rates from planar surface traps are added: the cross shows the heating rate of a $^{25}$Mg$^+$ ion in a room temperature trap (electrode material: gold) \citep{Epstein2007}, while the triangle shows the heating rate of a single Sr$^+$ ion where the silver trap electrodes were held at 6~K \citep{Labaziewicz:2008a}}. }
\end{center}
\end{figure}

The Ann-Arbor group observed much reduced heating rates by cooling the trap electrodes (see Ref.~\citet{Deslauriers:2006a} and Fig.~\ref{fig:motional-heating}).
In the course of cooling the trap electrodes from 300~K to 150~K, the
heating rate dropped by more than one order of magnitude. This strong
dependence on the temperature hints at a thermally activated process causing
 the unexplained heating in ion traps. In addition, the authors
measured heating rates as a function of the trap size $d$. From Eq.~\ref{eq:heating}, a  $1/d^2$ scaling is expected while one expects a ${1}/{d^4}$ dependence in the case of heating due to moving patch charges  \citep{Turchette2000,Epstein2007}. The Ann-Arbor group extracted from the data in Fig.~\ref{fig:motional-heating} an exponent of 3.5 \citep{Deslauriers:2006a} only slightly different from the postulated exponent of 4  \citep{Turchette2000}.

The MIT-group has investigated various planar traps made of silver electrodes on a quartz substrate close to 4~K as well as one trap at room temperature \citep{Labaziewicz:2008a}.
They measured a heating rate as low as 2~quanta/s for trap sizes on the order of 100~$\mu$m while a similar trap at room temperature had a devastating heating rate seven orders of magnitude larger. Even the extremely small heating rates for the 4~K experiments cannot be explained by Johnson noise, only. They also measured heating rates for three different trap sizes and found them to be consistent both with a $1/d^2$ and a $1/d^4$ scaling. In addition, a strong dependence of the heating rate on the annealing temperature used in the fabrication process was found. Furthermore, the MIT group investigate the heating rate of the ions as a function of the electrode temperature $T$ \citep{Labaziewicz:2008b}. Above $T=40$~K, they found that the heating rate is proportional $T^\beta$ with $2<\beta<4$ depending on the trap.
Overall, these findings suggest that with improved fabrication methods, smaller heating rates can be achieved. Furthermore, the NIST group observed heating rates of 300 quanta/s for a $^{25}$Mg$^+$ ion 40~$\mu$m above the gold surface of a planar trap \citep{Epstein2007}. This heating rate is significantly smaller than what one would expect from the MIT measurements and supports the conclusion that choice of materials and fabrication methods are very important to achieve small heating rates.

To speed up quantum gates and to ease cooling, there is a strong tendency towards constructing small ion traps. On one hand, small traps with characteristic dimensions of a few tens of microns allow for large trap frequencies on the order of 10~MHz, on the other hand they seem to lead to inacceptable heating rates.  Therefore, heating in ion traps is not only an interesting topic on its own but needs a lot of attention from a technological point of view. Cooling the traps to 4~K seems to offer a solution to the heating problem.

Many two-qubit gate implementations, however, store quantum information in the motional degree of freedom. Therefore, dephasing of the motional states must be also taken into account. To measure the motional coherence, a superposition of two motional states can be created whose phase coherence is tested after some waiting time. In the Innsbruck experiments, the pulse sequence (read from right to left) $R^+(\pi,0)R^C(\pi/2,0)$ creates the state $(\ket{D,0}+\ket{D,1})/\sqrt{2}$. The inverse pulse sequence  $R^C(\pi/2,\varphi)R^+(\pi,0)$ closes the interferometric procedure after some waiting time $T$. From the contrast of the interference fringes obtained by varying $\varphi$ with waiting time $T$, the coherence
time can be directly deduced.
One might be tempted to use just a pair of two $R^+(\pi/2)$ pulses to implement the Ramsey experiment. In this case, however, motional as well as electronic dephasing mechanisms lead to decoherence of the intermediate state $(\ket{S,0}+\ket{D,1})/\sqrt{2}$. The  former pulse sequence, however, is insensitive to phase decoherence of the electronic qubit, and waiting times $T$ of many tens of milliseconds are possible. Thus, with the former pulse sequence the trap frequency can be easily determined with an accuracy of a few Hertz.

\begin{figure}
\begin{center}
\includegraphics[width=0.4\textwidth]{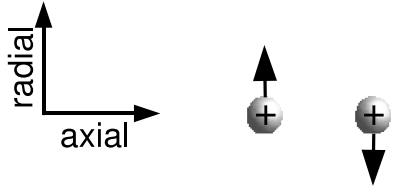}
\end{center}
\caption{\label{fig:breathing-dephasing} An anti-correlated radial motion (excited rocking mode) of a two-ion crystal changes the mean distance between the two ions.}
\end{figure}

Using this method, the Innsbruck group observed a motional
coherence time on the order of 100~ms \citep{Schmidt-Kaler2003b}
on the center-mass-mode, being consistent with expected voltage
fluctuations on the order of $10^{-5}$. Furthermore, the Oxford
group observed on a single $^{40}$Ca$^+$ion a motional coherence
time of $182\:(36)$~ms, limited most likely by motional heating of
about 3~quanta/s \citep{Lucas:2007}. However, the Innsbruck group
found that for the axial breathing mode (and other higher axial
modes) coherence times of about 5~ms are more typical
\citep{Roos:2008a}. Thus it must be concluded, that on the
breathing mode, a dephasing mechanism is present which can be
neglected for the center-of-mass mode. \citet{Roos:2008a} show,
both theoretically and experimentally, that for a two-ion crystal
the breathing mode frequency depends on the motional state of some
of the radial modes. The basic mechanism responsible is that an
anti-correlated motion (the rocking mode) along a radial direction
changes the mean distance between the ions (see
Fig.~\ref{fig:breathing-dephasing}). Thus, the repelling force
between the ions is reduced leading to a reduction of the axial
breathing mode frequency with increasing excitation of the rocking
mode. Therefore, if the rocking modes are not in a well-defined
state, the breathing mode frequency is different for each
experimental realization which in turn is interpreted as dephasing
of the breathing mode.

% In order to analyze the origin of this decoherence, the Innsbruck group
% measured the coherence of the breathing mode for various trap frequencies.
% It appears that the decoherence mechanism becomes much stronger
% ($\tau_{\rm dec}<1$~ms) when the ratio of the breathing modes
% frequency and the radial mode is 0. Still up to now the mechanism of the reduced coherence of the higher order modes is not identified. \remark{Sollen wir diesen Abschnitt drin lassen, abschwächen oder ganz herausnehmen}

% \subsection{Decoherence free subspaces}
% DFS spaeter oder hier

\subsection{Modelling ion trap quantum computers}
The influence of each imperfection on the performance of the quantum computer can be reliably estimated with numerical simulations. In this way the necessary steps to improve the performance can be analyzed and planned in detail. For ion trap quantum computers this procedure is relatively straightforward as the Hamiltonians are well-known (see Eq.~\ref{eq:ion-trap-hamiltonian}) \citep{Wineland1998,Leibfried2003a,Jonathan2000}. Here we describe the simulations as appropriate for the Innsbruck set-up. With small modifications these simulations should also be applicable to other ion trap set-ups.

Eq.~\ref{eq:ion-trap-hamiltonian} is a good starting point to model ion trap quantum computers. The Hamiltonian is first generalized to multiple ions:
\begin{eqnarray}\label{eq:ion-trap-simulation}
 H&=&\hbar \sum_{n,m} \Omega_n\left\{\sigma_{+}^{(n)}e^{-i\Delta t} + \sigma_{-}^{(n)}e^{i\Delta t}\right. \\ \nonumber
 & & \left.+ i \eta_{nm} (\sigma_{+}^{(n)}e^{-i\Delta t} - \sigma_{-}^{(n)}e^{i\Delta t})\left(a_m e^{-i\omega t} + a_m^\dagger e^{i\omega t} \right)\right\}\:.
\end{eqnarray}
Here the indices $n$ and $m$ denote the various ions and motional modes taken into account, respectively, and $\eta_{nm}$ accounts for the different coupling strengths of the ions to the motional modes \citep{James1998}.

For the simulations the initial state vector is first transformed into the rotating frame of the laser field such that Eq.~\ref{eq:ion-trap-simulation} becomes time independent and the Hamiltonian can be directly integrated.  A quantum algorithm usually consists of laser pulses with different frequencies. Therefore, this procedure has to be carried out for each pulse, separately. For the bi-chromatic laser fields employed for the AC-Stark shift compensation (c.f.~Sec.~\ref{sec:AC-Stark-compensation}) and M{\o}lmer-S{\o}rensen-gates (Sec.~\ref{sec:two-qubit-gates}) this method fails. In these cases the differential equation can be numerically integrated. In the following we describe how almost all experimental imperfections were incorporated into the simulations:
\begin{description}
\item[AC-Stark effects] due to the carrier transition are described by Eq.~\ref{eq:ion-trap-simulation} and appear with increasing $\Omega_n$ (see Sec.~\ref{sec:AC-Stark-limit}). Therefore, these shifts are automatically included in the simulation. AC-Stark shifts due to other (dipole) transitions are not taken into account by Eq.~\ref{eq:ion-trap-simulation}. However, in the experiments a second light field is used anyways to minimize the total effect of all AC-Stark shifts. Therefore the light shifts must be artificially removed from the Hamiltonian to match the experiments.

\item[Off-resonant excitations] (c.f. Eq.~\ref{eq:off-resonant-excitations}) are described by Eq.~\ref{eq:ion-trap-simulation} and are therefore automatically included.

\item[Laser freqency noise and magnetic field noise] are divided into fluctuations slower and faster as compared with typical coherent manipulation times (on the order of 1~ms in current experiments). In the experiments usually slow fluctuations dominate and the laser detuning remains to a large extent constant during coherent state manipulation. The observed increase in the coherence time ---when spin echo sequences are used--- supports this conjecture. These slow fluctuations can be modeled by running the simulations for several detunings from the qubit transition and by averaging the measured populations. On the other hand, fast fluctuations can be taken into account by transforming Eq.~\ref{eq:ion-trap-simulation} into a master equation. In this case, the  dimension of the state space describing the system is squared as compared to the Schr\"odinger approach and thus this method starts to get tedious already for simulations of a five-ion algorithm. In fact, even today's supercomputers cannot hold the complete density matrix of an arbitrary twenty qubit system in their memory.

\item[Laser intensity fluctuations] are assumed to be constant during the coherent state manipulation and therefore can be modeled by simulating the algorithms a few times for various laser intensities.

\item[Addressing errors] are described by setting the Rabi frequencies $\Omega_n$  in Eq.~\ref{eq:ion-trap-simulation} to the corresponding values. However, it should be noted that
there is an additional degree of freedom connected to addressing errors: the phase of the laser at an ion position when directed on this particular ion and when mainly directed on another ion is in general not the same. This is due to the different paths the light field takes in these two cases.  Usually, in the simulations all phases are taken to be the same as they seem to give an upper bound for most algorithms.

\item[Imperfect ground state cooling] can be taken into account by running the simulation with the different initial states and averaging the results appropriately.
\end{description}
These simulations allow one to reduce the sensitivity of quantum algorithms to the respective imperfections. In the Innsbruck experiments, especially the influence of laser freqency and addressing errors issues was reduced considerably by optimizing the implementations taking the simulations as a guide.

Finally, we document some of the imperfections not taken into account in the simulations for the Innsbruck experiments. In Eq.~\ref{eq:ion-trap-simulation} each ion was approximated as a two level system. For $^{40}$Ca$^+$ the $S_{1/2}$-ground state is split into two Zeeman levels and the upper qubit level  $D_{5/2}$ into six levels. The level separation is about 5~MHz and therefore not much larger than the involved trap frequencies. Therefore, off-resonant excitations of the additional transitions are possible. Furthermore, AC-Stark effects arise due to the presence of the other Zeeman levels as well as due to dipole transitions \citep{Haeffner2003a}. Finally, the AC-Stark effect was canceled with a second off-resonant laser field.  Treating this second light field  leads to a time dependent Hamiltonian and was therefore usually not taken into account in the simulations.

\section{\new{Key} experiments}
\subsection{Cirac-Zoller-type gates}
\label{sec:cz-experiments}
The NIST group demonstrated the central part of the Cirac-Zoller gate on a single $^9$Be$^+$ ion by implementing the operations displayed in Eq.~\ref{eq:cz-phase-gate} \citep{Monroe1995}. The phase gate was turned into a controlled-NOT operation by inserting two Ramsey $\pi/2$ pulses, one before and one after the phase gate, to verify the quantum nature of the phase gate. The first two motional excitations $n=\{0,1\}$ (see Fig.~\ref{fig:ion-tensor-oscillator}, $\omega_{\rm trap}=2\pi\times 11$~MHz) served as the control bit, while the target bit was represented by superpositions of the $\ket{F=2,m_f=-2}$ and $\ket{F=1,m_f=-1}$ states. For the auxiliary state, the $\ket{F=2,m_f=0}$ state was used.

The complete Cirac-Zoller gate was finally implemented by \citet{Schmidt-Kaler2003}: two $^{40}$Ca$^+$ ions were addressed individually using a laser beam with a waist of about 2$\mu$m. First, the quantum state of ion~\#1 was mapped onto the breathing mode with a blue sideband pulse $R^+(\pi)$. Then, a controlled-NOT gate with the motion as the control bit and
the electronic state of ion~\#2 as the target bit was carried out, before the motional state was mapped back onto ion~\#1. However, to implement the conditional phase shift, the Innsbruck group did not use an auxiliary level but the composite pulse sequence $R^+(\pi/2,\pi/2)R^+(\pi/\sqrt{2},0)R^+(\pi/2,\pi/2)R^+(\pi/\sqrt{2},\pi)$ discussed in Sec~\ref{sec:composite-pulses}. In the first experiments  \citep{Schmidt-Kaler2003}, the controlled-NOT mapped the four logical eigenstates and one superposition state  to their desired respective output states with an average fidelity of 0.73(2) (coherence time $\tau \sim$800~$\mu$s, addressing error $\epsilon \sim$0.05). After several improvements of the experimental set-up ($\tau \sim$2~ms, $\epsilon \sim 0.03$, more flexible computer control, etc.), fidelities as large as 0.91.0(6) were observed \citep{Riebe2006} where here the fidelity is defined as an average of the overlap of the produced output states with the ideal output. Additionally, this work implemented the conditional phase shift with the pulse sequence $R^+(\pi/2,0)R^+(\sqrt{2}\pi,\pi/2)R^+(\pi/2,\pi)$ yielding a fidelity of up to 0.926(6).

\subsection{Entangled states with trapped ions}
 One important application of quantum computers in basic research is the generation of interesting quantum  states  as for instance the first deterministic generation of entangled particles with $^9$Be$^+$ ions \citep{Turchette1998}. \new{In addition, new applications of entangled states especially for metrology appear constantly \citep{Blatt:2008}.}
%  % That means, that
% the experimenter could push a button and know that after a well specified time
% an entangled state is available for further experiments. This is to contrasted to
% e.g. photon experiments where the state is not only generated randomly but where
% also within the generation process the state is destroyed by the detection.
Entangled states play an important role in discussions on the
foundations of quantum mechanics. Especially since Bell formulated
inequalities which could distinguish between so-called local realistic theories and quantum theories \citep{Bell1965,Bell1971,Clauser1969}, physicists were keen to produce these states and to check the predictions of quantum mechanics. Since then, there have been numerous experiments  demonstrating a violation of a Bell inequality. Almost all of these experiments were carried out with photons (for a summary see e.g.  \citet{Clauser1978}, \citet{Weihs1998} and \citet{Tittel1998}). Photons naturally explore the non-locality of entanglement and thus violations over distances of many kilometers were established. However, current detection efficiencies of photons are not high enough to close the so-called 'detection loophole', i.e. one must assume that the detected particles represent a fair sample of all particles emitted by the source \citep{Clauser1969}. Therefore there is a large interest in testing Bell-inequalities with trapped ions where the
detection fidelities approach unity. Such an experiment was conducted
with $^9$Be$^+$ ions by \citet{Rowe2001} and closed
the detection loophole. However, the ions were not detected outside their respective light cone, i.e. the detection time was longer than the time it takes light to travel between the ions. Thus there could still exist a combined detection-locality loophole. This loophole could be excluded for instance by creating entanglement between ions or atoms separated by several kilometers.

Another interesting aspect of the ion trap experiments is that the entangled states are produced deterministically. That means that in contrast to the photon experiments, the entangled states are created on demand. More importantly, the entangled states are not destroyed during their creation such that they can be used for further experiments.

%  For the most basic setup, this is not the case for photons. Here only a coincidence of detected photons signalizes an entangled state. Thus usually the photon state is destroyed before the presence of an entangled state is known.

Current research is directed towards entangled states with more than two particles. Already for three qubits, two classes of entanglement appear: GHZ states \citep{Greenberger1989} and W states \citep{Duer2000,Zeilinger1992}. These two classes of entanglement cannot be transformed into each other by local operations and classical communication, i.e. with single-qubit operations and measurements of the individual qubits \citep{Duer2000}. Both classes are not only maximallay entangled but also violate Bell-type inequalities.

GHZ states are states of the form
 \begin{equation}\label{GHZ-state}
\ket{\rm GHZ_N}=(\ket{00 \cdots 0}+\ket{11 \cdots 1})/\sqrt{2}\:.
\end{equation}
GHZ states with many qubits can be interpreted as Schr\"odinger-cat states. For this, e.g. the first qubit is treated as a separate degree freedom and all other qubits as a single quantum system, i.e. the cat system. Then the state of the first qubit indicates the state of the second ''macroscopic'' system. Another use of GHZ states is by encoding quantum information in a superposition of the form $\alpha \ket{000}+ \beta \ket{111}$. If a single qubit flips, the original state can still be recovered with so-called quantum error correction protocols \citep{Shor1995,Steane1996}.

W states are states of the form \begin{equation}
\label{W-state} \ket{\rm W_N}=\left(\ket{0\cdots 001} +
\ket{0\cdots 010}+\ket{0\cdots 0100} +\cdots +
\ket{10\cdots 0}\right)/\sqrt{N}\:.
\end{equation} They are remarkably stable against various decoherence sources: they are intrinsically stable against collective dephasing mechanisms \citep{Roos2004a} and even loss of qubits does not completely destroy the entanglement present in them.

\label{sec:entangled-states}
A four ion GHZ state was first produced by \citet{Sackett2000} using a M{\o}lmer-S{\o}rensen type gate (see Sec.~\ref{sec:Moelmer-Soerensen-gate}). The produced  GHZ-state
was analyzed by applying $R^{C}(\pi/2,\varphi)$-pulses to all ions
simultaneously and measuring the number of fluorescing ions
with a photo multiplier. Here, we will illustrate this procedure used by \citet{Sackett2000} for the two particle  Bell state $(\ket{00}+\ket{11})/\sqrt{2}$:
\begin{equation}\label{eq:parity-oscillations}
\begin{array}{rll}
\ket{00}+\ket{11}
 & & \xrightarrow{R^C_2(\pi/2,\varphi),R^C_1(\pi/2,\varphi)} \\
 & & (\ket{0}+ie^{i\varphi}\ket{1})\;(\ket{0}+ie^{i\varphi}\ket{1}) \;+\;
 (\ket{1}+ie^{-i\varphi}\ket{0})\;(\ket{1}+ie^{-i\varphi}\ket{0})\\
&=& (1-e^{-2i\varphi})\ket{00} + ie^{i\varphi}(1+e^{-2i\varphi})\ket{01} \\ & & +\; ie^{i\varphi}(1+e^{-2i\varphi})\ket{10} + (1-e^{-2i\varphi})\ket{11}\:,
\end{array}
\end{equation}
where we used Eq.~\ref{eq:single-qubit-operation} in the first step and omitted the normalization factors. To find an estimate for the fidelity of the original Bell state, it is useful to introduce the parity operator $P$ which is defined as $P=P_{00}-P_{01}-P_{10}+P_{11}$. Here $P_{xy}$ are the probabilities to find the ions in state $\ket{xy}$.  Plotting the expectation value of the parity, we see that it oscillates twice as fast as the phase $\varphi$ of the analyzing pulses. This behavior can also be interpreted as a consequence of the doubled energy difference between the $\ket{00}$ and the $\ket{11}$ state as compared to the single ion case.

To find the fidelity of the Bell state, two sets of experiments can be carried out: first a Bell state is created and the populations $P_{00}$ and $P_{11}$ are recorded. In a second set of experiments, the maximum ($\max P$) and minimum ($\min P$) of parity oscillations as described in Eq.~\ref{eq:parity-oscillations} are determined. The overlap $F$ of the experimentally produced state with a state of the form $(\ket{00}+ e^{i \phi}\ket{11})/\sqrt{2}$ is given then by $F=(P_{00}+P_{11})/2+(\max P - \min P)/4$ \citep{Sackett2000}.
No  individual addressing of the ions is required in the analysis procedure. In addition, the parity can be inferred from the global fluorescence of the ion string. Thus, this analysis method is relatively simple and efficient.
Furthermore, it can be generalized to GHZ states with an arbitrary number of ions and is thus very useful to gain information on the generated GHZ states, without individual qubit addressing and read-out.

This analysis technique was also used to verify the creation of a three particle GHZ-states with fidelities of up to 0.89 \citep{Leibfried2004}. The quality of the GHZ state was high enough, that from the resulting generalized Ramsey fringes (c.f.~Eq.~\ref{eq:parity-oscillations}) the phase could be estimated $1.45$ times more accurately
than using three uncorrelated particles. Estimating the phase of superpositions is quite important in frequency measurements. The high fidelity of the GHZ states was made possible by using the geometric phase gate (Sec.~\ref{sec:geometric-phase-gate}). Encouraged by this, the NIST group applied this gate also to create four-, five- and six-particle
entangled GHZ-states, with lower bounds for the fidelities of  $0.76\;(1), 0.60\;(2)$ and $0.509\;(4)$, respectively \citep{Leibfried2005}. As for GHZ states, a fidelity above 0.5 implies automatically the presence of genuine $N$-partite entanglement, the latter experiments demonstrated up to six partite GHZ-like entanglement.
% %
%  the Innsbruck-group uses a slightly different approach to entangle ions. Here laser pulses specifically addressing each time a single ion create various entangled states. The benefit of straightforward single ion addressing is a very high flexibility. Thus any
% % quantum algorithm can be implemented  most straightforwardly, only limited by the decoherence time. The trade-off of this method, however, is that the axial trap frequency can not be increased too much, since then the ions move closer to each other, thwarting single ion addressing. A consequence of a lowered trap frequency is a reduced speed of the entangling operations on the sidebands.

While the NIST group uses predominantly global addressing and state read-out, the Innsbruck group entangled ions
mainly  with laser pulses addressed to individual ions.
For instance, a Bell can be created in
the following way (please note that the right-most ion is the first one)
%(see Sec.~\ref{sec:ibk-definitions}
% ); this method has been applied to an atom-cavity system first by \citet{Hagley1997} (see also \citep{Raimond2001}).)
 \citep{Roos2004b}:
\begin{eqnarray}
\ket{SS,0} & \stackrel{R^+_1(\pi/2,\varphi+\pi/2)}{\longrightarrow} & (\ket{SS,0}+e^{i \varphi}\ket{SD,1})/\sqrt{2} \nonumber \\
& \stackrel{R^C_2(\pi,0)}{\longrightarrow} & (\ket{DS,0}+e^{i \varphi}\ket{DD,1})/\sqrt{2}\nonumber \\
& \stackrel{R^+_2(\pi,0)}{\longrightarrow} & (\ket{DS,0}+e^{i \varphi}\ket{SD,0})/\sqrt{2}\label{eq:Bell-generation}
\end{eqnarray}
The success of the Bell-state generation is usually verified using a procedure called state tomography (see Sec.~\ref{sec:tomography}).
The laser phase offset $\varphi$ of the first pulse determines phase of  the Bell state. In addition, an additional $R^C_2(\pi,0)$-pulse on the second ion transfers the $(\ket{DS,0} + e^{i \varphi}\ket{SD,0})/\sqrt{2}$ state to
$(\ket{SS,0} +  e^{i \varphi}\ket{DD,0})/\sqrt{2}$.
Thus, using this toolbox all four Bell-states can be created in the same set-up.

\begin{figure}
\begin{center}
\includegraphics[width=0.9\textwidth]{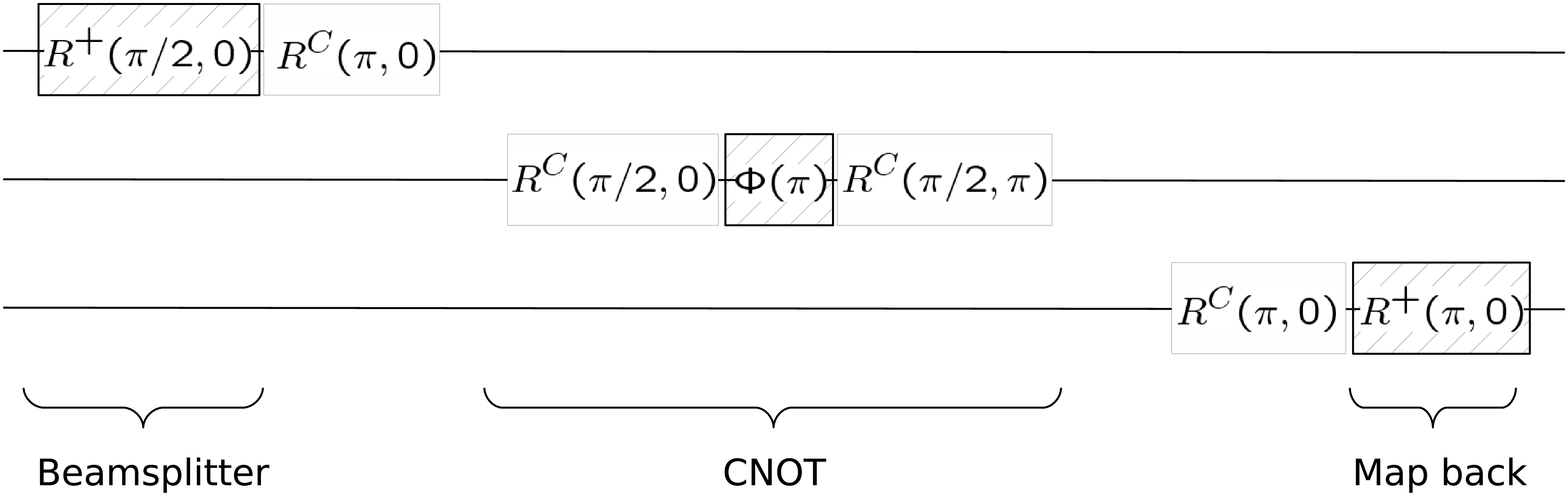}
\end{center}
\caption{\label{fig:GHZ-generation}Pulse sequence to generate a GHZ state. The controlled-NOT operation is implemented with a composite phase gate $\Phi(\pi)$ (see Sec.~\protect\ref{sec:composite-pulses}) sandwiched in between two $R^C$-pulses on the center ion. Hatched areas indicate sideband pulses.}
\end{figure}

Furthermore, the Innsbruck group used the flexibility of the entangling method to create three particle GHZ and W states \citep{Roos2004a}. The idea to create a GHZ state \citep{Cirac1995}, is to apply a controlled-NOT gate while the motional degree is still in a superposition of $\ket{0}$ and $\ket{1}$ (see Eq.~\ref{eq:Bell-generation}). For this CNOT-operation, the motion is the control bit and the new ion is the target bit (the second line in Fig.~\ref{fig:GHZ-generation}) \citep{Rauschenbeutel2000,Raimond2001}).
% \begin{table}
% \begin{tabular}{|l||l|l|l|l|}
% %\hline
% \hline
%  ion \#1  & $R_1^+(\pi/2,0)$ & $R_1^C(\pi,\pi/2)$ & & \\\hline
%  ion \#2  & $R_2^C(\pi/2,0)$ & & & \\\hline
%  ion \#2  & $R_2^+(\pi,\pi/2)$ & $R_2^+(\pi/\sqrt{2},0)$ &
%  $R_2^+(\pi,\pi/2)$ &  $R_2^+(\pi/\sqrt{2},0)$  \\\hline
%  ion \#2  & $R_2^C(\pi/2,\pi)$ & & & \\\hline
%  ion \#3  & $R_3^C(\pi,0)$ & $R_3^+(\pi,0)$ & &  \\\hline
% \end{tabular}
% \caption{\label{GHZ-generation} Pulse sequence to create GHZ-state. A controlled NOT gate on ion \#2 in combination with the precsription given in Eq.~\ref{eq:Bell-generation} is used.}
% \end{table}
Inserting more and
more CNOT's, this GHZ-state generation method is straightforwardly generalized to more particles.
%
% The total
% pulse area on the blue sideband used for the CNOTs scales linear with the
% number of ions. However, as the Lamb-Dicke factor $\eta$ is proportional
% to $1/\sqrt{N}$, the time required to realize the pulse area arccording to
% Eq.~\ref{eq:blue-rabi-frequency} goes as $\sim \sqrt{N}$. Therefore the total
% creation time for this
% method scales as $N^{3/2}$. Using the  M{\o}lmer-S{\o}rensen gate (Sec.~\ref{sec:Moelmer-Soerensen-gate}) or the geometric phase gate (\ref{sec:geometric-phase-gate}), only the Lamb-Dicke factor plays a role. Thus the creation time scales as $\sim \sqrt{N}$, which makes the latter gates quite favorable for creating large scale entanglement. A further advantage of the latter approaches is, that they do not require individual ion addressing. For the geometric phase gate, however, the ions have to have particular relative positions in the walking standing wave. In the standard approach where axial modes are used this might be difficult to achieve for large ion numbers. Nevertheless \citet{Leibfried2005} realized a six-ion GHZ state with a geometric phase gate.

\begin{figure}
\begin{center}
\includegraphics[width=0.9\textwidth]{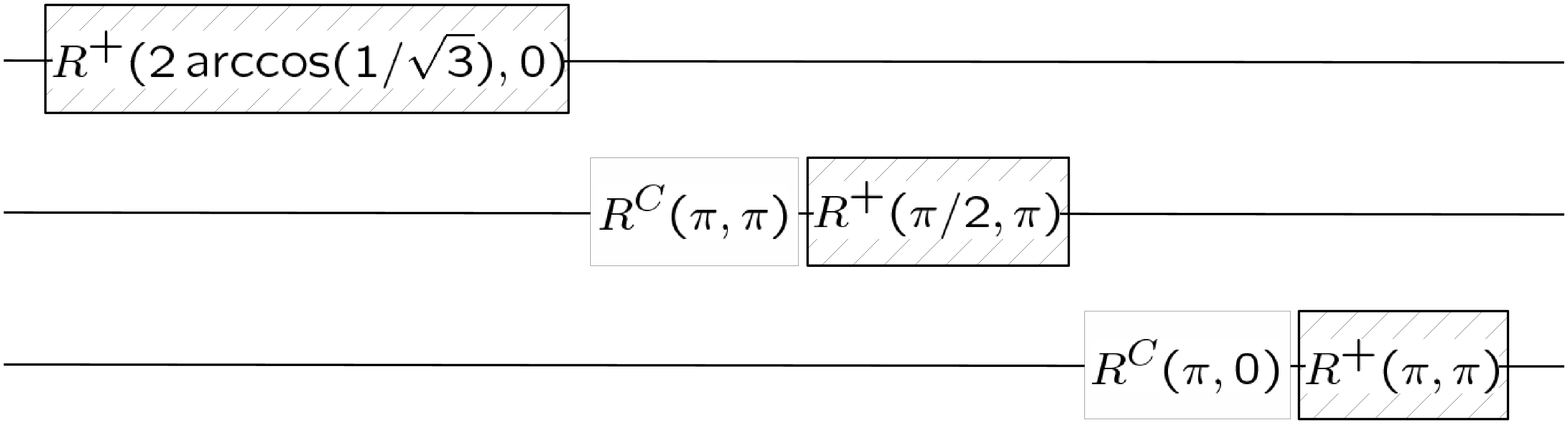}
\end{center}
\caption{\label{fig:W-generation}Pulse sequence to create the W state $(|DDS\rangle + |DSD\rangle + |SDD\rangle)/\sqrt{3}$ from the $\ket{SSS}$ state. Hatched areas indicate sideband pulses.}
\end{figure}

In order to create a three-ion W~state $\ket{\rm W_3}=(\ket{DDS+\ket{DSD}+\ket{SDD}})/\sqrt{3}$, Eq.~\ref{eq:Bell-generation} can be generalized differently: The length of the first  blue sideband pulse is adjusted such that the state $(\ket{SSS,0}+\sqrt{2}\ket{SSD,1})/\sqrt{3}$ is created and ions \#2 and \#3 are flipped to obtain $(\ket{DDS,0}+\sqrt{2}\ket{DDD,1})/\sqrt{3}$. Then the phonon is shared between the remaining ions \#2 and \#3 to create $(\ket{DDS,0}+\ket{DSD,0}+\ket{SDD,0})/\sqrt{3}$. Fig~\ref{fig:W-generation} shows the corresponding pulse sequence.
% \begin{table}
% \label{tab:W-generation}
% \begin{tabular}{|l||l|l|}
% %\hline
% %Describtion & Pulses\\
% \hline
% ion \#1 (beam splitter) & $R_1^+(2 \arccos{(1/\sqrt{3})},0)$ & \\\hline
% ion \#2 (beam splitter) & $R_2^C(\pi,\pi)$ & $R_2^+(\pi/2,\pi)$ \\\hline
% ion \#3 (map) & $R_3^C(\pi,0)$ & $R_3^+(\pi,\pi)$
% \\\hline
% \end{tabular}
%\end{table}
It can also be directly generalized to $N$ ions by adjusting the pulse area of the  first beam splitter to $\arccos{(1/\sqrt{N})}$ and then sharing the phonon excitation among the other ions equally \citep{Haeffner2005a}. Interestingly, the required blue sideband pulse area grows only logarithmically with the ion number and thus the generation time grows sublinear with the number of ions. Therefore, this scheme opens the possibility to generate large entangled states. In experiments, W states ranging from four up to eight ions have been created \citep{Haeffner2005a}. Using a technique called state tomography (see Sec.~\ref{sec:tomography}), the experimentally obtained states have been fully charactized (Fig.~\ref{fig:w8}). Analysing the measured density matrix showed that the generated states indeed carried genuine $N$-particle entanglement.

We add that  an even faster scheme to produce W~states was
proposed which does not require individual addressing in the
entangling procedure \citep{Retzker:2007,Solano2005}. The idea
here is that first a $\ket{DD\cdots D,1}$ state is created. Then a
$R^+(\pi,0)$ pulse addressed to all ions is supposed to generate
the desired W~state by mapping the phonon to the electronic state
of one of the ions, i.e. creating a symmetric superposition with
exactly one electronic state flipped. \citet{Retzker:2007} further
generalized this procedure to W~states with more than one
excitation (Dicke states).

Entangled states have also been produced with trapped $^{111}$Cd$^+$ and  $^{40}$Ca$^{+}$ ions by the Ann-Arbor \citep{Haljan:2005b} and the Oxford groups \citep{Home:2006a}, respectively. For the two qubit levels, the Ann-Arbor group used  the $\ket{F=0, m_F=0}$ and $\ket{F=1, m_F=0}$ state of the ground state of $^{111}$Cd$^+$, taking advantage of its insensitivity to the Zeeman effect in first order. The geometric phase gate (Sec.~\ref{sec:geometric-phase-gate}) does not work efficiently on magnetic field insensitive transitions \citep{Langer2006phd}.
%\footnote{However, one can imagine to leave the magnetic field insensitive manifold temporarily for the gate.}
Instead, the Ann-Arbor group used a M{\o}lmer-S{\o}rensen gate (Sec.~\ref{sec:Moelmer-Soerensen-gate}) to entangle the two ions \citep{Haljan:2005a,Haljan:2005b}. Furthermore, a tomographic state characterization was applied to evaluate the produced states and the degree of entanglement thoroughly (see Sec.~\ref{sec:tomography}). To achieve the required
individual addressing capability, a combination of ion selective AC-Stark shifts and microwave fields was used (c.f. Sec.~\ref{sec:ion-addressing}).

%Using an inhomogeneous laser beam detuned by about 200~GHz, a differential
% phase shift of $\pi$ can be imprinted onto the two ions within 10~$\mu$s. Thus one ion can acquire effectively a phase of $\pi$, while the other one has acquired a multiple of $2\pi$. Inserting this
% between two microwave pulses of length $\pi/4$ realizes thus an effective $\pi/2$-pulse on the latter
% ion, while the two microwave pulses cancel each other on the first ion. In this way arbitrary single qubit operations can be carried out on a small number of ions \citep{Lee2006phd}. A similar method \citep{Schaetz2004} to address single ions has been earlier applied in \citet{Rowe2001}.

The Oxford group created Bell states with two trapped
$^{40}$Ca$^{+}$ ions \citep{Home:2006a}. They encoded the quantum
information in the Zeeman manifold of the $S_{1/2}$-ground state,
thus effectively using the direction of the valence electron's
spin. Adopting the geometric phase gate
(Sec.~\ref{sec:geometric-phase-gate}) for  $^{40}$Ca$^{+}$, they
created the Bell state
$(\ket{\!\!\uparrow\uparrow}-\ket{\!\!\downarrow\downarrow})/\sqrt{2}$.

In Innsbruck, a M{\o}lmer-S{\o}rensen gate was used to prepare a
pair of $^{40}$Ca$^{+}$ ions in the entangled state
$\psi=(|SS\rangle+i|DD\rangle)/\sqrt{2}$, where
$|S\rangle\equiv|S_{1/2},m=1/2\rangle$ and
$|D\rangle\equiv|D_{5/2},m=3/2\rangle$. Using a gate time
$\tau=50\,\mu$s with the laser light being smoothly switched on
and off within $2.5\,\mu$s, a Bell state fidelity of 0.993(1) was
achieved \citep{Benhelm:2008b} when the ions were cooled to the
ground state of the motional mode mediating the coupling.
Moreover, uneven multiples $k=1,3,5,\ldots,21$ of the gate were
used to create entangled states. For $k=21$, the state fidelity
was still 0.8. Finally, the gate yielded Bell states with a
fidelity of 0.96 even with the ion string cooled only to the
Doppler limit ($<n_{\rm bus}> = 17$).

\subsection{Decoherence free subspaces}
\label{sec:decoherence-free-subspaces}
Laser frequency and magnetic field fluctuations are usually the dominant decoherence mechanisms in ion traps. Both mechanisms lead to fluctuations of the phase between the laser and the atomic polarization and thus to dephasing of each qubit, however, to a good approximation by the same amount for all qubits.
If one encodes a single qubit in two ions in such a way that the two phase evolutions
cancel each other, the original qubit is protected from this global dephasing and the quantum information is encoded into a decoherence free subspace (DFS).
In particular, superpositions of the form
$\alpha \ket{01} +
\beta \ket{10}$ are transformed by the global single-qubit phase-change $\ket{1}\rightarrow e^{i\phi}\ket{1}$ in the following way:
\begin{equation}
 \alpha \ket{01} +
\beta \ket{10} \rightarrow  \alpha e^{i\phi}\ket{01}+\beta e^{i\phi}\ket{10} \;.
\end{equation}
The global phase factor $e^{i\phi}$ cannot be
observed, such that the state remains immune against collective dephasing.

This property was demonstrated  by \citet{Kielpinski2001} using an
engineered dephasing mechanism. The qubits were encoded in the
hyperfine-states of $^9$Be$^+$. As a controlled dephasing
mechanism, \citet{Kielpinski2001}  chose an unfocused off-resonant
laser beam with \lanf random\ranf intensity. The laser beam leads
for each experimental realization to a different AC-Stark effect,
however, common to both ions.
\begin{figure}
\begin{center}
\includegraphics[width=0.4\textwidth]{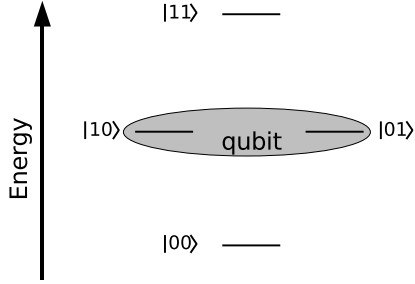}
\end{center}
\caption{\label{fig:DFS} Energy level diagram for two ions with ground state $\ket{0}$ and excited state $\ket{1}$. A collective-dephasing free qubit is formed by the degenerate logical basis $\{\ket{10},\ket{01}\}$.}
\end{figure}

In an environment with natural dephasing, \citet{Roos2004b}
observed also much increased lifetimes of qubits encoded in a DFS
as compared to single ion qubits. In these experiments, the
coherence of a state formed by two $^{40}$Ca$^+$ ions
$\ket{\Psi_{\rm qubit}}=(\ket{DS}+e^{i\varphi}\ket{SD})/\sqrt{2}$
was retained for 1~s, the dominant decoherence mechanism being
spontaneous decay of the $D_{5/2}$ level. For comparison, laser
frequency noise and magnetic field fluctuations led to a single
ion coherence time of 1~ms.

Furthermore, both the Innsbruck \citep{Haeffner2005b} and the NIST
\citep{Langer2005} groups encoded quantum information in the
ground states of two-ion strings with $^{40}$Ca$^+$ and
$^{9}$Be$^+$, respectively. Coherence times of 34~s and 7~s,
respectively, were measured. In both cases, fluctuations of the
magnetic field gradient were believed to be the reason for the
decoherence in the DFS. For these experiments, extreme care must be
taken to switch off the laser light properly. Not only residual
light scattering rates on the order of 0.1~photon/s  destroy the
coherence, but also fluctuating differential AC-Stark shifts on
the order of 1~Hz destroy the phase coherence of the entangled
states. Finally, we note that these experiments demonstrated also
extremely long lived entanglement  of up to 20 seconds between two
parties separated by 5~$\mu$m.

Apart from robustly encoding quantum information, decoherence-free
subspaces have also found an application in quantum metrology. In
\citet{Roos2006}, a Bell state was encoded in a combination of
Zeeman sublevels of the $D_{5/2}$ level of two $^{40}$Ca$^+$ ions.
The state was decoherence-free with respect to fluctuations of the
magnetic field but sensitive to energy level shifts caused by
static electric field gradients. In this way, the quadrupole
moment of the metastable state could be determined with high
precision by monitoring the Bell state's phase evolution over a
duration orders of magnitude longer than the single-qubit
coherence time.

% % \begin{figure}
% % \includegraphics[width=0.4\textwidth]{long-lived-entanglement.eps}
% % \caption{\label{fig:long-lived-entanglement} \citep{Haeffner2005b} ???}
% % \end{figure}

% Finally, we note that the geometric phase gate is an excellent candidate to implement a two qubit gate for quantum information encoded in a decoherence free subspace as discussed in Sec.~\ref{sec:decoherence-free-subspaces} \citep{Aolita2007}. This is because here the states acquire only phases during the gate operation and the electronic state remains unchanged at all times. Thus it is automatically guaranteed, that the decoherence free subspace is not left.

\subsection{State tomography}\label{sec:tomography}
Quantum state tomography \citep{QuantumStateTomography2004} is a
measurement technique that provides access to all the information
stored in density matrices describing pure and mixed quantum states. It
requires the quantum state of interest to be available in many
copies. While the basic measurement principle dates back fifty
years \citep{Fano1957}, experimental implementations of quantum
state tomography started only in the 1990's
\citep{Smithey1993,Dunn1995,Leibfried1996}. Tomographic measurements
of systems composed of qubits have been implemented in experiments
with nuclear magnetic resonance, photons, trapped ions and
superconductors \citep{Chuang1998b,White1999,Roos2004b,Steffen2006a}.

Noting that the density matrix of a single qubit can be represented by
\begin{equation}
\rho=\frac{1}{2}(I+\sum_\alpha \langle\sigma_\alpha\rangle
\sigma_\alpha), \label{eq:densitymatrix}
\end{equation}
we see that the density matrix of a
single qubit can be inferred by measuring the expectation values
$\langle\sigma_\alpha\rangle$, ($\alpha=x,y,z$), of the Pauli spin
matrices.
The measurement of $\sigma_z$ is accomplished by projecting the
qubit onto its energy eigenstate basis. For the measurement of
$\sigma_{x,y}$, an additional $\pi/2$-pulse of suitable phase
precedes the projective measurement. The tomographic procedure can
be easily extended to systems of several qubits by measuring
 the joint spin expectation values
$\sigma_{\alpha_1}^{(n_1)}\otimes\sigma_{\alpha_2}^{(n_2)}\otimes\ldots\sigma_{\alpha_k}^{(n_k)}$
where $\sigma_{\alpha_j}$ denotes a spin component of qubit $n_j (\sigma_{\alpha_j} \in \{ I, \sigma_x, \sigma_y, \sigma_z  \})$.
This way, the determination of the density matrix of an N-qubit
system requires the measurement of $4^N$ expectation values. As
some of the operators commute, a total of $3^N$ measurement bases
is necessary. While in principle the number of measurements could
be reduced by projecting onto mutually unbiased bases
\citep{Wootters1989}, this procedure is of no practical importance in current
ion trap experiments as it would demand high-fidelity entangling
gate operations for mapping the required bases to product state
bases.

 A slight complication arises since in every experimental
implementation of quantum state tomography, expectation values are
never exactly determined but only estimated based on a finite
number of measurements. The na\"{\i}ve replacement of the expectation
values $\langle\sigma_\alpha\rangle$ in Eq.~\ref{eq:densitymatrix} can
give rise to unphysical density matrices with negative
eigenvalues. This problem is avoided by employing a maximum
likelihood estimation of the density matrix \citep{Hradil1997,
James2001} that makes use of the estimated expectation values for
searching in the set of meaningful density matrices the 'most
likely one'. The maximum likelihood algorithm identifies the
sought-for density matrix with the one that maximizes the
probability of observing the experimentally recorded set of
measurement results. Even though maximum likelihood estimation has
been criticized \citep{BlumeKohout2006} for being less accurate
than Bayesian estimation techniques
\citep{QuantumStateTomography2004}, it has the practical merit of
being easily implemented in experiments.

\begin{figure}
\begin{center}
\includegraphics[width=0.99\textwidth]{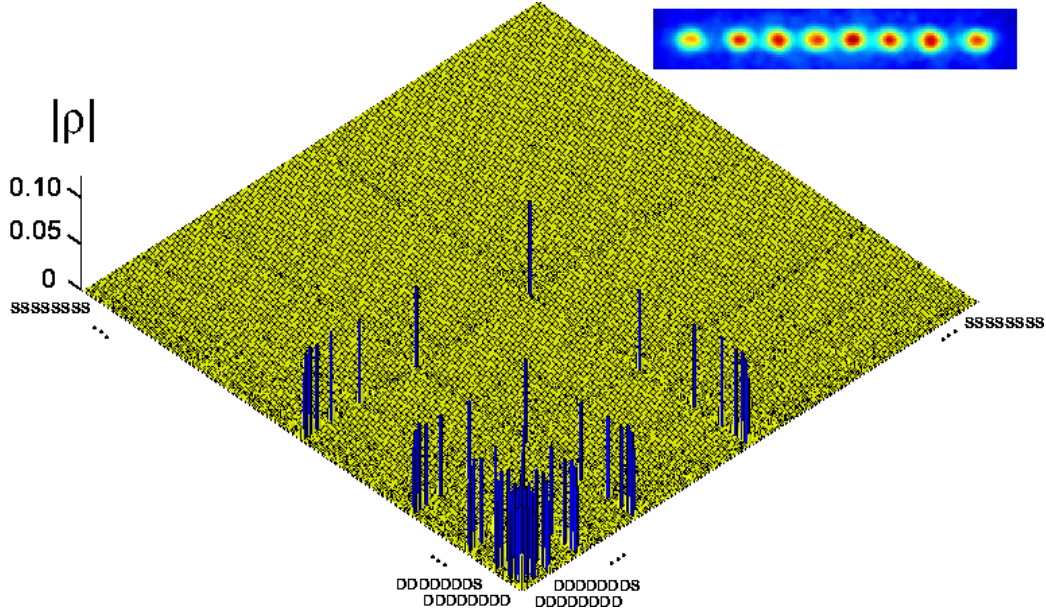}
\end{center}
\caption{\label{fig:w8} Absolute values of the density matrix of an eight-ion W state (from \citet{Haeffner2005a}).  \new{In addition, in the upper right corner the CCD-image of an eight-ion string is displayed.} For the state tomography, the experiment was repeated 100 times for each of the 6561 measurement settings. The total measurement time amounted to more than 10~h. The fidelity of the W state was determined to 0.722(1). }
\end{figure}

Starting with \citet{Roos2004b}, almost all Innsbruck experiments made heavy use of quantum state tomography. Even an eight particle W state has been fully characterized \citep{Haeffner2005a}. Figure~\ref{fig:w8} shows the experimentally obtained density matrix of an eight-ion W state. A particular merit of quantum state tomography is that all physically available information on the quantum register is extracted. Thus all aspects of the generated states can be thoroughly analyzed without taking new data.

An alternative to the above described method, was demonstrated by the Oxford group  \citep{Home:2006a}. They used a refined, albeit partial
tomographic procedure: instead of choosing three measurement settings for each qubit (either one of the pulses $R^C(0,0),R^C(\pi/2,0)$, or $R^C(\pi/2,\pi/2)$ preceding the qubit detection),
they choose to apply before the detection the pulses $R^C(\theta_i,\varphi)$ with $\theta_i$ either $0$, $0.66 \pi$, or $0.54 \pi$ and evenly distributed phases $\varphi \in [0, 2\pi[$. In this way the reconstruction is less biased and thus more robust against systematic errors or equivalently against decoherence.

\subsection{Selective read-out of a quantum register} \label{sec:selective-readout}
For some quantum algorithms like teleportation and most error-correction protocols a part of the quantum register has to be read out while leaving the rest of the register intact.
Both the Innsbruck and the NIST group succeeded in this task. \citep{Barrett2004} employed segmented traps to separate the ions to be read out from the
ions which should remain coherent. Now one set of ions can be
illuminated safely with detection light while the other ions are left dark.

\begin{figure}
\begin{center}
\includegraphics[width=0.7\textwidth]{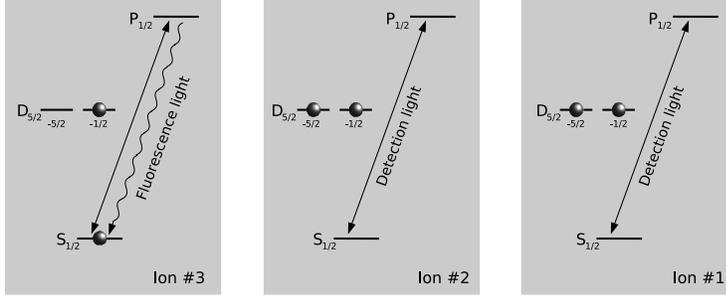}
\end{center}
\caption{\label{fig:hiding} Partial level scheme of the three Ca-ions (from \citet{Roos2004a}). Only ion \#3
is read out. Ion \#1 and \#2's quantum information is protected in the Zeeman
manifold of the D$_{5/2}$-level, namely the $m_J=-1/2$ and $m_J=-5/2$ levels. }
\end{figure}

The Innsbruck group chose a different route to selectively read out the quantum register \citep{Roos2004a,Riebe2004}.
Qubits were protected from being
measured by transferring their quantum information to superpositions of levels
which are not affected by the detection, that is, a light scattering process on the ${\rm S}_{1/2}\rightarrow {\rm P}_{1/2}$-transition in Ca$^+$. In the experiments, a $\pi$ pulse on the ${\rm S}_{1/2}\rightarrow {\rm D}_{5/2}\:(m_J=-5/2)$-transition transfers the quantum information into the $\{{\rm D'} \equiv {\rm D}_{5/2}$ ($m_j=-5/2$), ${\rm D} \equiv {\rm D}_{5/2}$ ($m_j=-1/2) \}$ manifold.
%As the auxiliary level, the D$_{5/2}$  ($m_j=-5/2$) level rather than the $m_j=-3/2$ or $m_j=+1/2$ levels was chosen, because for the
%particular orientation of the magnetic field  and the light polarization of the qubit laser only $\Delta m = 2$ transitions were allowed.
Fig.~\ref{fig:hiding} shows two ions which are protected from the detection light at 397~nm and the third ion with the original encoding which is measured.
After the selective readout, a second set of $\pi$-pulses
on the D{'} to S transition transfers the quantum information back to the
original computational subspace \{D,S\}.

\begin{figure}
\begin{center}
\includegraphics[width=0.9\textwidth]{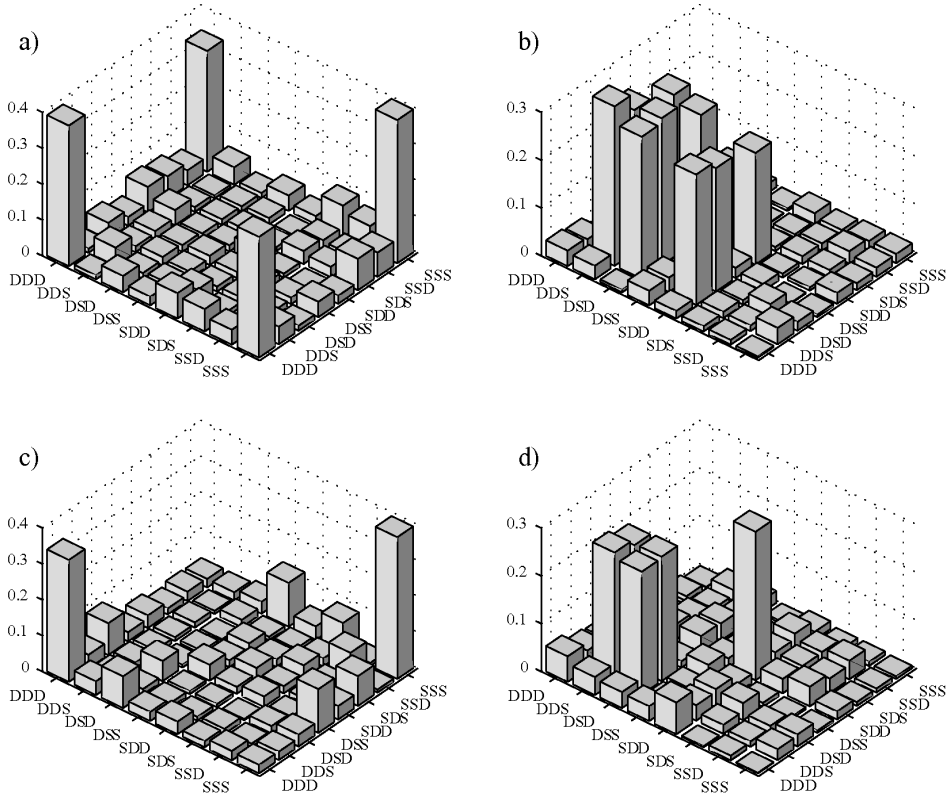}
\end{center}
\caption{\label{fig:GHZ-W-states} %\getpermission
Absolute values of density matrices after measuring ion \#3 (from \citet{Roos2004a}). (a) shows the density matrix of a GHZ-state before measuring and (c) after ion \#3 is measured. The same for a W-state ((b) before and (d) after the measurement of ion \#3).}
\end{figure}

It is interesting to apply the selective read-out to an entangled
qubit register and to demonstrate the collapse and even partial collaps of a wave function. For this \citet{Roos2004a} first prepared a three-ion GHZ- and a W-state
and then detected one of the ions  while the quantum information of the other ions was still protected
in the D-level. Fig.~\ref{fig:GHZ-W-states} shows the results of these measurements.
The quantum nature of the GHZ-state was completely destroyed by measuring a single constituent, i.e. it was projected into a mixture of $|SSS\rangle$
and $|DDD\rangle$ (Fig.~\ref{fig:GHZ-W-states}a and Fig.~\ref{fig:GHZ-W-states}c). By contrast, for the W-state, the quantum register remained partially
entangled as coherences between ion \#1 and \#2 persisted after measuring ion \#3 (Fig.~\ref{fig:GHZ-W-states}b and Fig.~\ref{fig:GHZ-W-states}d).
% Note that
%related experiments have been carried out with mixed states in NMR \citep{Teklemariam2002} and with photons \citep{Eibl2004}.

\subsection{Conditional single-qubit operations}
\label{sec:conditional-read-out}
One can take the partial read-out of a quantum register one step further and perform operations conditioned on the read-out result. As will be discussed in Sec.~\ref{sec:teleportation}, both the NIST and the Innsbruck group demonstrated this procedure within their respective teleportation experiments \citep{Barrett2004,Riebe2004}.
Furthermore, the Innsbruck group employed conditional operations  to deterministically transfer a three-particle GHZ-state with local operations  into a two-particle Bell state \citep{Roos2004a}.
This procedure can also be regarded as an implementation of a
three-spin quantum eraser as proposed by \citet{Garisto1999}.
%For this, first a GHZ-state is produced and then one of the qubits is rotated with a $\pi/2$-pulse prior to the measurement. In this way part of the quantum nature of the GHZ-state is  preserved in the measurement procedure.

\begin{figure}
\begin{center}
\includegraphics[width=0.7\textwidth]{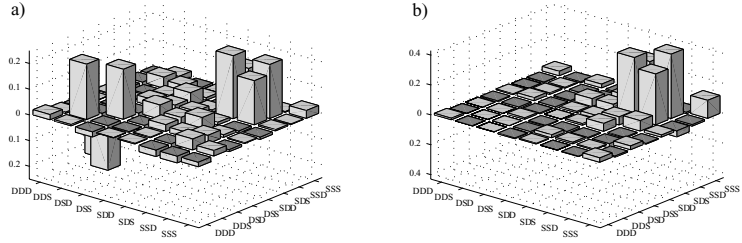}
\end{center}
\caption{\label{fig:GHZ-trafo}
Density matricies during the individual steps of the deterministic generation of a Bell state from a GHZ-state (from \citet{Roos2004a}).
(a) Real part of the density matrix elements of the system after ion \#1 of the GHZ-state $(|DSD\rangle + |SDS\rangle)/\sqrt{2}$ has been measured in a rotated basis. (b) Transformation of the GHZ-state   $(|DSD\rangle + |SDS\rangle)/\sqrt{2}$ into the bipartite entangled state   $|S\rangle(|DS\rangle+|SD\rangle)/\sqrt{2}$ by conditional local operations. Note the different vertical scaling of (a) and (b).}
\end{figure}

In the experiment by \citet{Roos2004a}, first the GHZ-state   $(|DSD\rangle +
|SDS\rangle)/\sqrt{2}$ was created. Application of $R_3(\pi/2,3\pi/2)$ yielded the GHZ state ${|D\rangle (|SD\rangle - |DS\rangle) + |S\rangle (|SD\rangle + |DS\rangle)}/2$. Measuring ion~\#3, projected ions \#1 and \#2
either onto $(|SD\rangle - |DS\rangle)/\sqrt{2}$ or onto $(|SD\rangle + |DS\rangle)/\sqrt{2}$ with ion~\#3 indicating in which of the two states the first two ions were (see Fig.~\ref{fig:GHZ-trafo}a). This mixture of the two Bell states can then be transferred to a pure Bell state by inducing a phase shift of $\pi$ on ion \#2 (pulse sequence $R^C_2(\pi,\pi/2)R^C_2(\pi,0)$) if, and only if, ion~\#3 was measured to be in the $D$-state. In addition, the state of ion \#3 was reset to $|S\rangle$.
Figure~\ref{fig:GHZ-trafo} shows the intermediate result before applying the conditional rotation as well as the resulting Bell state. The bipartite entangled state $|S\rangle (|SD\rangle + |DS\rangle)/\sqrt{2}$
was produced with fidelity of 0.75.

\subsection{Process tomography}\label{sec:process-tomography}
Process tomography is a method to characterize a quantum mechanical evolution \citep{Chuang1997,Poyatos1997}. Measurements are made to determine how an arbitrary input state, characterized by the density matrix $\rho_{\rm in}$, is transformed by the quantum process. The output density matrix $\rho_{\rm out}$ of the process can be expressed as
\begin{equation}\label{eq:xi}
\rho_{\rm out} = \sum_{i,j=0}^{2^N-1} \chi_{ij} \hat{A}_i \rho_{\rm in} \hat{A}_j\:.
\end{equation}
Here $ \chi_{ij}$ is the so-called process matrix, $N$ is the number of qubits and the operators $\hat{A}_i$ form a basis of the space of the $2^N \times 2^N$ matrices. All relevant information on the quantum process is
contained in the process matrix $\chi_{ij}$. In the standard procedure, the $4^N$ separable states ---$
%\bigotimes\limits_i^N
\{\ket{0}_i,\:\ket{1}_i,\:(\ket{0}_i+\ket{1}_i)/\sqrt{2},\:(\ket{0}_i+i\ket{1}_i)/\sqrt{2}\}$ for a single qubit--- are prepared and then the output of the process is characterized each time with a full state tomography ($3^N$ measurement settings). Inverting Eq.~\ref{eq:xi} yields the process matrix $\chi_{ij}$.

 Such a process tomography has been  carried out
for characterizing quantum gates in NMR \citep{Childs2001} and in linear-optics quantum
computing  \citep{Obrien2004,Kiesel2005a}. For ion traps a one-qubit process (the teleportation of a qubit \citep{Riebe2007}) and two-qubit processes (a CNOT and its square \citep{Riebe2006}) have been characterized.

Knowing the process matrix $\chi_{ij}$ for all basic operations of a quantum computer is a very good basis for estimating the computer's performance. However, there are a few caveats: the number of necessary measurements to determine the process matrix $\chi_{ij}$ scales quite dramatically and thus it becomes quickly impractical to characterize processes with numerous qubits. Already for a four-qubit process, 20736 measuring settings would be required summing up to about 24~hours measurement time with the current parameters of the Innsbruck experiment (100~experiments/setting, 25 repetitions/s). In addition, it is not clear that the subsequent application of two processes corresponds to the product of the process matrices. This assumption usually holds only if the relevant environment is time invariant, i.e. the process interacts only with a bath without memory. For example, for the Cirac-Zoller gate, the phonon mode might keep the memory about the failure of an earlier gate operation and thus induce a failure of the next gate. Furthermore, the realization of the process can depend on whether it is executed at the beginning or the end of an algorithm. In particular, experiments are often triggered to the phase of the power line to reduce dephasing due to magnetic field fluctuations caused by 50~Hz or 60~Hz noise and their multiples. Executing a certain gate operation a few milliseconds earlier or later within the experimental sequence leads easily to a change of the qubit resonance frequency of 100~Hz and the realization of the process becomes time dependent.

% Thus in total, the usefullness of quantum process tomography for quantum computing might be limited. The future will show to what extent quantum process tomography will be used to improve and characterize quantum computers.

Process tomography, as presented above, requires at least $4^N3^N$ measurement settings and is thus quite costly.
Above, we have restricted ourselves to the separable operators $\hat{A}_j$. Using entangled auxiliary qubits and/or measuring in non-separable bases, the number of settings could be reduced, however, scales still exponentially in the qubit number $N$ \citep{Mohseni2006}.  Furthermore, most likely the total number of measurement runs has to be on the same order of magnitude as in the standard method to obtain a similar accuracy. Therefore, we conclude that a full quantum process tomography of a large quantum systems will be not practical.

Finally, we note that there  exist other approaches to estimate
the fidelity of quantum processes. For instance,
\citet{Knill:2007} employ long sequences of randomly chosen gates.
The main idea is that while the result of each gate sequence is
known in an ideal implementation, noise leads to deviations from
the expected results. Measuring the deviations, the average
fidelity of the gate operations can be inferred. Choosing random
sequences guarantees that the gate operation is investigated with
various input states  and in various combinations.

\section{Algorithms with trapped ions}
\subsection{Deutsch-Josza algorithm}\label{sec:dj-algorithm}
The Deutsch-Josza (DJ) algorithm detects the parity of an unknown function \citep{Deutsch1989,Nielsen2000}. Concentrating on a single bit, there exist four different functions which map one (qu)bit with value $a=\{0,1\}$ onto another one. These functions can be divided into
%(see Tab.~\ref{tab:dj-logic})
 constant (even) ($f_1(a)=0$ and $f_2(a)=1$) and balanced (odd) functions ($f_3(a)=a$ and $f_4(a)={\rm NOT} \: a$). With a classical machine, it is necessary to call $f_n$  at least twice to decide whether $f_n$ is odd or even, i.e. one needs to calculate $f_n(0)$ and $f_n(1)$. However, formulating the procedure  quantum mechanically, the question whether $f_n$ is constant or balanced can be decided by calling it only once.
% \begin{table}
% \begin{tabular}{c||c|c|c|c}
% ~ & \multicolumn{2}{c}{Constant} & \multicolumn{2}{c}{Balanced}\\
% \hline
% $n=\{1,2,3,4\}$ & case 1 & case 2  & case 3 & case 4\\
% \hline
% \hline
% $f_n(0)$ & 0 & 1 & 0 & 1\\
% $f_n(1)$ & 0 & 1 & 1 & 0\\
% Logic & Id & NOT$_a$ & $CNOT$ & $0-CNOT$
% \end{tabular}
% \caption{\label{tab:dj-logic} Logic corresponding to the four cases in the 1 qubit Deutsch-Josza algorithm.}
% \end{table}

\begin{figure}[tb]
\begin{center}
\includegraphics[width=0.80\textwidth,angle=0]{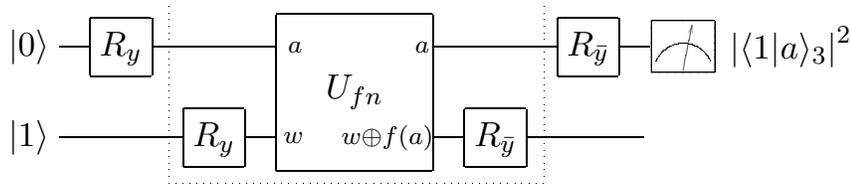}
\end{center}
\caption{\label{fig:dj-circuit} General scheme of the Deutsch-Josza algorithm (from \citet{Gulde2003}). The upper line represents the qubit holding the input variable $a$ which ---if prepared in a logical eigenstate--- does not change its value when  $U_{f_n}$ is called. The lower line holds the work qubit $w$. To this number the value of the function $f_n$ is added modulo 2. The Hadamard rotations $R_{{y}_w}$ (and $R_{\bar{y}_w}$) transfer the quantum bits into superpositions so that the inherent parallelism of quantum mechanics can be used.}
 \end{figure}

In order to formulate the problem quantum mechanically, the functions $f_n$ have to be generalized to take qubits as inputs. Within the framework of quantum mechanics all operations are
unitary and therefore another qubit (the work or auxiliary qubit) is added to allow for non-invertible functions $f_1$ and $f_2$. Rephrasing the task, qubit $\ket{a}$ holds the input variable $x$ while qubit
$\ket{w}$ (the work qubit) will receive the result of the evaluation $f_n(a)$ plus the initial value $w$ of qubit $\ket{w}$ to guarantee invertibility (see Fig.~\ref{fig:dj-circuit}). Thus, we define the unitary $U_{f_n}$ representing the implementation of the function acting on $\ket{w}\ket{a}$ with values $w$ and $a$, respectively:
\begin{equation}
U_{f_n}\ket{w}\ket{a}=\ket{f_n(a)\oplus w}\ket{a}\;.
\end{equation}
Here, $\oplus$ denotes an addition modulo 2.

%In Tab.~\ref{tab:dj-logic} the unitaries corresponding to each function $f_n$ is displayed.

The DJ-algorithm consists of the following
steps (c.f.~Fig.~\ref{fig:dj-circuit}): \begin{enumerate}
\item Initialize the system in the state $\ket{0_a}\ket{1_{\rm w}}$.
\item Transfer the input qubit $\ket{a}$ into $\left(\ket{0}+\ket{1}\right)/{\sqrt{2}}$ and the work-qubit $\ket{w}$ into $\left(\ket{0}-\ket{1}\right)/{\sqrt{2}}$ with  Hadamard operations $R_y$.
\item Call the (unknown) function with these superimposed values by implementing $U_{f_n}$.
\item Close the interferometer by applying an inverse Hadamard operation ($R_{\bar{y}}$) on $\ket{a}$.
\item Read out the result in $\ket{a}$.
\end{enumerate}

%In addition, it was implementated with fiber optics (e.g.~ \citep{Brainis2003}) and on Li$_2$ %molecules \citep{Vala2002}. were Vala et al., PRA {\bf 66}, 062316 (2002).
%\item Fiber optics $\Rightarrow$ E. Brainis et al., PRL {\bf 90}, 157902 (2003).
%\end{enumerate}
 The ion trap experiment used only a single $^{40}$Ca$^+$-ion \citep{Gulde2003}. The  internal state acted as the
 qubit $\ket{a}$  to hold the input variable for the function with the logical assignment $\ket{0}\equiv\ket{S}$, while the axial vibrational degree of freedom was used as the work qubit $\ket{w}$ (logical assignment $\ket{0}\equiv\ket{1}_{\rm ax}$ and $\ket{1}\equiv\ket{0}_{\rm ax}$). Thus, ground state cooling to $\ket{S,0}$ initialized the system in $\ket{0_a}\ket{1_{\rm w}}$, as required. A peculiarity of
encoding a qubit within the ion's motional state is that one has to take care that the system does not leave the
computational subspace $\{\ket{S,0}, \ket{D,0}, \ket{S,1}, \{\ket{D,1}\}$. In the experiment, this was achieved with the composite pulse techniques described in Sec.~\ref{sec:composite-pulses} \citep{Childs2000}. Furthermore, single-qubit operations on the motional degree of freedom had to be carried out. For this, the quantum information in the motional degree of freedom was swapped to the electronic degree of freedom such that ordinary carrier pulses could be used. Finally, the quantum information was swapped back to the motional degree of freedom. The Hadamard rotations $R_{{y}_w}$ and $R_{\bar{y}_w}$ were absorbed into the definitions of the functions such that only for $f_2$ this swapping of the quantum information had to be employed.
% and note only that the DJ-algorithm was implemented first on an NMR system \citep{Chuang1998a}.
\begin{table}
\begin{center}
\begin{tabular}{c||c|c}
$f_n(x)$ & Logic ($R_{\bar{y}_w} U_{f_n} R_{y_w}$) & Laser pulses\\
 \hline
 \hline
$f_1=0$ & $R_{\bar{y}_w} R_{y_w}$ & --- \\[0.1cm] \hline
 & $R_{\bar{y}_w}$ SWAP$^{-1}$ & $R^+\!(\frac{\pi}{\sqrt{2}},0)R^+\!(\frac{2\pi}{\sqrt{2}},\varphi_{\rm swap})R^+\!(\frac{\pi}{\sqrt{2}},0)$\\
$f_2=1$ &  NOT$_a$  &
$R^{C}(\frac{\pi}{2},0)R^{C}(\pi,\frac{\pi}{2})R^{C}(\frac{\pi}{2},\pi)$\\
 & SWAP $R_{y_w}$ & $R^+\!(\frac{\pi}{\sqrt{2}},\pi)R^+\!(\frac{2\pi}{\sqrt{2}},\pi\!+\!\varphi_{\rm swap})R^+\!(\frac{\pi}{\sqrt{2}},\pi)$\\[0.1cm] \hline
  $f_3=x$ & $R_{\bar{y}_w} $ CNOT $ R_{y_w}$ &
$R^+\!(\frac{\pi}{\sqrt{2}},0)R^+\!(\pi,\frac{\pi}{2})R^+\!(\frac{\pi}{\sqrt{2}},0)R^+\!(\pi,\frac{\pi}{2})$
\\[0.1cm] \hline
 & & $R^{C}(\pi,0)$\\
$f_4={\rm NOT}\:x$ & $R_{\bar{y}_w} $ 0-CNOT $R_{y_w}$ &
$R^+\!(\frac{\pi}{\sqrt{2}},0)R^+\!(\pi,\frac{\pi}{2})R^+\!(\frac{\pi}{\sqrt{2}},0)R^+\!(\pi,\frac{\pi}{2})$ \\ & & $R^{C}(\pi,0)$
\\[0.1cm] \hline
\end{tabular}
\\[2mm]
\caption{\label{tab:dj-laser-table} Laser pulses for the implementation of the algorithm inside the dashed box in Fig.~\ref{fig:dj-circuit} ($R_{\bar{y}_w} U_{f_n} R_{y_w}$) on a single ion. $\varphi_{\rm swap}$ is given by $\arccos\left(\cot^2(\pi/\sqrt{2})\right)$. For the whole DJ-algorithm an $R^C(\pi/2,0)$-pulse just before and an  $R^C(\pi/2,\pi)$-pulse after implementing $R_{\bar{y}_w} U_{f_n} R_{y_w}$ is applied.}
\end{center}
\end{table}

% \begin{table}
% \begin{tabular}{c||c|c|c|c}
% ~ & \multicolumn{2}{c}{Constant} & \multicolumn{2}{c}{Balanced}\\
% ~ & case 1 & case 2  & case 3 & case 4\\
% \hline
% \hline
% expected $\left|\braket{1}{x_a}\right|^2$ & 0   &0    &1  &1\\
% measured $\left|\braket{1}{x_a}\right|^2$ & 0.019(6)&0.087(6) &0.975(4)   &0.975(2)\\
% expected $\left|\braket{1}{z_w}\right|^2$ & 1 &1 &1   &1\\
% measured $\left|\braket{1}{z_w}\right|^2$ & - &0.90(1)    &0.931(9)   &0.986(4)\\
% \end{tabular}
% \caption{\label{tab:dj-ion-results} Results of the DJ-algorithm \citep{Gulde2003}.}
% \end{table}

\begin{figure}[ht!]
\begin{center}
\includegraphics[width=0.89\textwidth,angle=0]{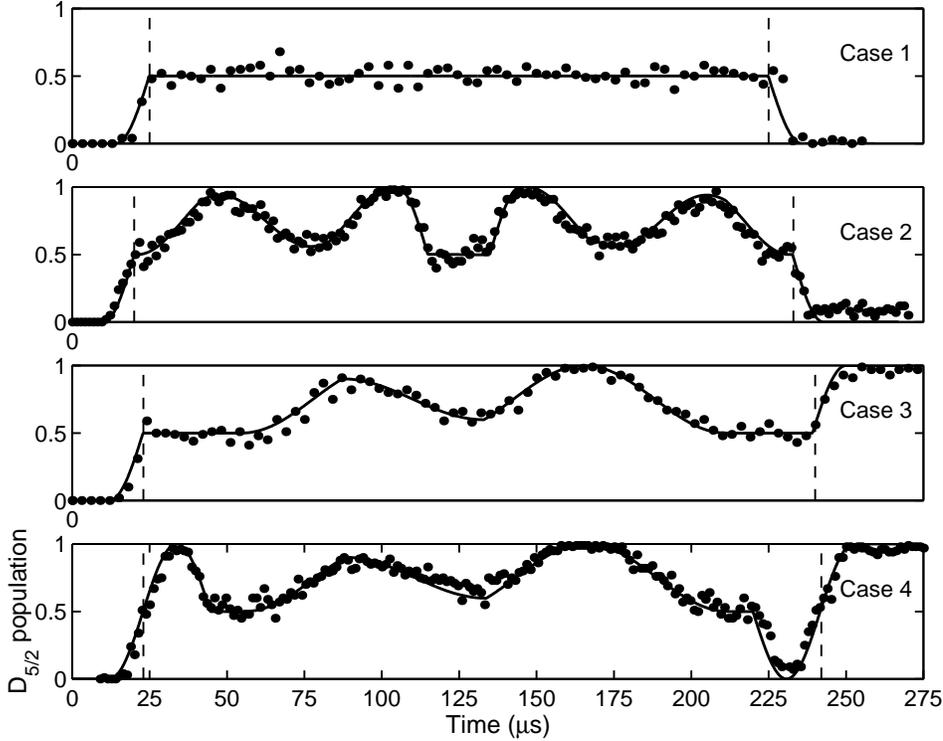}
\end{center}
\caption{\label{fig:dj-ion-result} Traces for the implementation of the DJ-algorithm (from \citet{Gulde2003}). The solid line is not a fit but a calculation based solely on
the independently measured Rabi frequencies. The vertical dashed lines enfold the pulse sequences in Tab.~\ref{tab:dj-laser-table}.
}
 \end{figure}

%Table~\ref{tab:dj-ion-results} displays the results.
%A particular feature of the ion trap experiments is, that
%the it s possible
To decide on the class of the implemented function, only  qubit $\ket{a}$ had to be measured. Finding $\ket{a}$ in $\ket{0}$ indicated that the function was even, finding $\ket{1}$ showed that the function was odd (see points at the end of the traces in Fig.~\ref{fig:dj-ion-result}). For the functions $f_1$, $f_3$ and $f_4$ the
fidelity to identify the function’s class with a single measurement exceeded 0.97,
for $f_2$ it was still above 0.9.

In order to follow the evolution of $\left|\braket{1}{a}\right|^2$, the pulse sequence was truncated at a certain time $t$ and the qubit $a$ was measured (see Fig.~\ref{fig:dj-ion-result}).
Repeating this sequence for various times, it is possible to follow the algorithm through its evolution. This procedure helps to debug algorithms and makes sure that the desired  algorithm is implemented.

\subsection{Teleportation}
\label{sec:teleportation}
In quantum teleportation, the state of a qubit is transferred from one physical system to another one. This can be achieved with the following protocol \citep{Bennett1993}: first two parties share an entangled qubit pair. The quantum information contained in an additional qubit (the source qubit) can be transferred from one party to the other party by performing a Bell-state measurement on the source qubit and one of the entangled qubits.
To conclude the transfer, the result of this Bell-state measurement is communicated via a classical channel to the receiver party and the receiver performs one of four rotations depending one the result of the Bell-state measurement.
  Thus it is possible to
transfer the information content of a qubit by
communicating two classical bits (the result of the Bell measurement) and using entanglement. Therefore, teleportation demonstrates a way to break down quantum
information into a purely classical part and a quantum part.
% and thus sheds new light on the essence of quantum information.

Another feature of teleportation is that it is not merely  a simple transmission of a quantum state:
it does not need a quantum channel to be open at the time the
transfer is carried out.  Instead it uses the non-local properties of quantum
mechanics, established by a quantum channel  prior
to the generation of the state to be teleported. Once that link has been
established, an unknown state can be transferred deterministically at any later time using
classical communication only. Especially this feature was highlighted by the two ion trap teleportation experiments \citep{Barrett2004,Riebe2004} by entangling the auxiliary and the target qubits before writing the quantum information into the source qubit. Thus these experiments demonstrate that unknown quantum information can be transferred on demand without using an active quantum channel.

% Teleportation was first demonstrated with photonic qubits \citep{Bouwmeester1997,Boschi1998,Pan2001,Marcikic2003,Fattal2004}, later  with continuous variables \citep{Furusawa1998} and in NMR \citep{Nielsen1998}. Recently also
% \citet{Sherson2006} teleported a coherent state between light and a collection of atoms.
% The two ion trap experiments \citep{Barrett2004,Riebe2004} both demonstrate deterministic teleportation on demand.

% Thus the experimenters, in contrast to previous implementations, were free to start
% the teleportation within the coherence time of the entangled state.

\begin{figure}
\centerline{\epsfxsize=5.1in\epsfbox{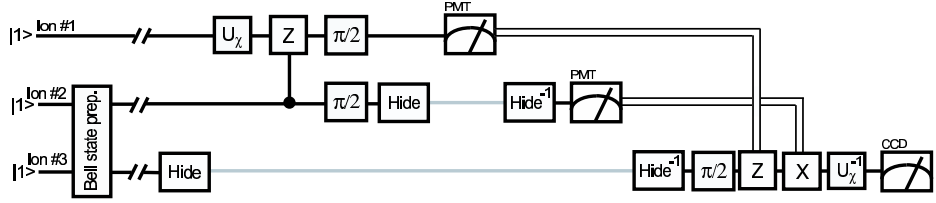}}
\caption{\label{fig:quantumcircuit} The teleportation algorithm's quantum circuit as implemented in the Innsbruck experiment (from \citet{Riebe2004}).
Double lines represent flow of classical information, whereas single lines represent flow
of quantum information. The gray lines indicate when a qubit is protected from
detection light via so-called hiding-pulses. First, ions \#2 and \#3 are entangled,
creating the quantum link between the source region (ions \#1 and \#2) and the
target ion (ion \#3). Then, after some waiting time, the state to be teleported (on ion \#1) is
prepared via the unitary operation $U_\chi$. A controlled phase gate together with detection via a
photomultiplier tube (PMT) implements the Bell state  measurement.
}
\end{figure}

The teleportation circuit  displayed in Fig.~\ref{fig:quantumcircuit} is formally equivalent to the one proposed by Bennett et {\it
al.}~ \citep{Bennett1993}, but adapted to the Innsbruck ion-trap quantum processor.
%
% To obtain directly the fidelity of the teleportation, we perform on ion \#3  the operation $U^{-1}_\chi$, which is the
% inverse of the unitary operation used to create the input state
% $|\chi \rangle$
% from state $|S\rangle$ (see pulses \#9 and \#34 in Table~\ref{tab:teleportation-pulsesequence}).
% The teleportation is successful if and only if ion \#3 is always
% found in $|S\rangle$. The teleportation fidelity, given by the overlap
% {$F=\langle S|U^{-1}_\chi\rho_{\rm exp}U_\chi|S\rangle$}, is
% plotted in Fig.~\ref{fig:teleportation-fidelities} for all four test states \{$|S\rangle$,
% $|D\rangle$, $|S\rangle+|D\rangle$, $|S\rangle+i|D\rangle$\}.
The Innsbruck group reached fidelities of about 0.75 \citep{Riebe2004}  and 0.83 \citep{Riebe2007}, while the NIST group measured a fidelity of 0.78 \citep{Barrett2004}.
Teleportation based on a
completely classical resource instead of a quantum entangled resource yields a
maximal possible fidelity of 0.667  \citep{Massar1995}.
We note that to rule out out hidden variable theories, a fidelity in excess of
0.87 is required \citep{Gisin1996}.

%(dashed line in Fig.~\ref{fig:teleportation-fidelities}).

% One shortcoming of the Innsbruck experiments was that only four test states
% were used to verify the teleportation procedure. To characterize a single qubit quantum process this is enough, however, based on the knowledge of the particular test states, classical strategies can be devised which explain the Innsbruck results. Therefore the Innsbruck group repeated in the same year (\remark{wollen wir das verraten}) their experiments with six test states  and achieved an average fidelity in the order of 83\%.

To emphasize the role of the shared entangled pair as a resource, in the Innsbruck experiments, a delay between the creation of the Bell state and the state preparation of the input qubit was introduced. For waiting times of up to 20~ms (exceeding the time required
for the teleportation by more than a factor of 10) no significant decrease in the fidelity was observed.
For longer waiting times, the measured heating of the ion crystal of
less than 1~phonon/100~ms is expected to reduce the fidelity significantly, because
the Cirac-Zoller gate requires the center-of-mass mode of the ion string to be in the ground state. %Implementations of sympathetic cooling (see Sec.~\ref{sec:sympathetic-cooling}) could help there in the future.

\begin{figure}
\begin{center}
\includegraphics[width=0.99\textwidth]{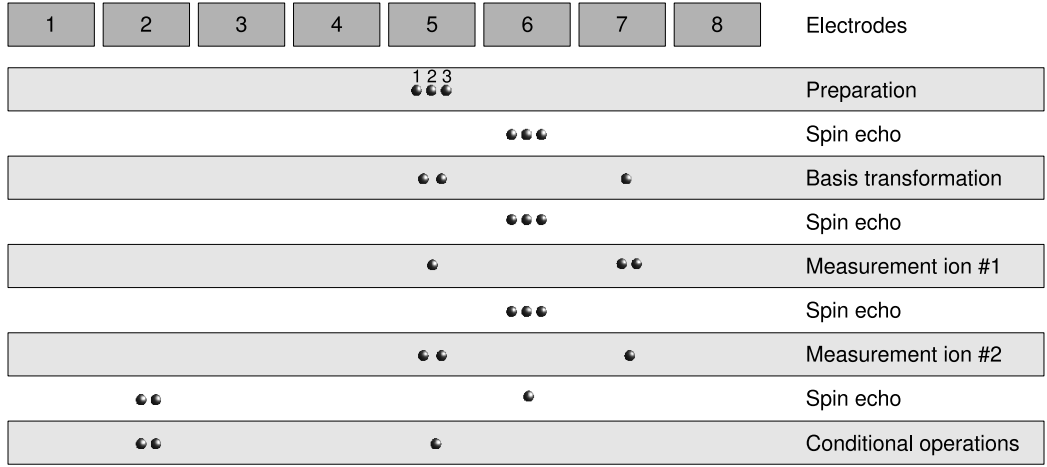}
\caption{\label{fig:teleportation-segfalle}
%\getpermission
 Position of the ions within the segmented trap during the execution of the NIST groups teleportation implementation \new{(after \citet{Barrett2004})}.}
\end{center}
\end{figure}

The implementation of the NIST group \citep{Barrett2004} demonstrates how segmented traps facilitate ion trap quantum computation. The authors use a segmented linear trap \citep{Rowe2002}, where the two DC-electrodes rails are each split into eight segments. Fig.~\ref{fig:teleportation-segfalle} shows the ion positions in each step during the teleportation procedure. By changing the potentials on the electrodes (top), the ion strings can be moved, split and merged. The teleportation algorithm was implemented in the
following way (see Fig.~\ref{fig:teleportation-segfalle}): first the leftmost and rightmost  ions (auxiliary and target, respectively) were prepared in the Bell state $\ket{\!\!\!\downarrow\downarrow}_{13}-i\ket{\!\!\!\uparrow\uparrow}_{13}$ using the geometric phase gate discussed in Sec.~\ref{sec:geometric-phase-gate} \citep{Leibfried2003a}. As the bus mode, the stretch mode was used. The center ion does not couple to the stretch mode and thus an effective two-qubit gate between the outer ions is implemented. For the experiments, it proved useful to transfer this Bell state
into a singlet state $\left(\ket{\Psi}=\ket{\!\!\uparrow\downarrow}_{13}-
\ket{\!\!\downarrow \uparrow}
\right)_{13}$ as the singlet state remains invariant under global rotations allowing for an effective single-qubit addressing of the source ion \#2. This was achieved by changing the relative position of the ions by varying the trap strength such that the subsequent laser pulses had the desired phases at the new ion positions \citep{Rowe2001}.
%$\left(\ket{Psi}=\ket{\uparrow\downarrow}_{13}-
%\ket{\downarrow \uparrow}
%\right)_{13}\otimes
%\left( \alpha\ket{\downarrow}_2+\beta\ket{\uparrow} \right)_2$.
%As no direct single ion addressing was available, the authors used the methods %presented by
%\citet{Rowe2001} to effectively address one of the ions and create the singlett %state. mehr???, wie geht das?
To implement the teleportation, the Bell state measurement has to be carried out on the source ion \#1 and the auxiliary ion \#2. For this, the target ion \#3 was first separated from the string in trap \#6 and ions \#1 and \#2 were transferred back into trap \#5. Most importantly, the stretch mode of the two ions was still close to the ground state and the required rotation into the Bell basis could be directly implemented with the geometric phase gate (Sec.~\ref{sec:geometric-phase-gate}) without the need of a sympathetic cooling step (c.f.~\ref{sec:sympathetic-cooling}). Then ion \#1 and \#2 were measured by transporting first the auxiliary ion into trap \#5, measuring there its fluorescence (see Sec.~\ref{sec:state-detection}) and pumping it into state $\ket{\!\!\!\downarrow}$. Next, ion \#2 was also transported to trap \#5 and the total fluorescence was detected, revealing the state of the source ion.  In a last step, ion \#3 was transferred into trap \#5 and conditioned on the result of the Bell measurement, the corresponding single-qubit operation was applied.

%Finally we note that in the NIST experiment made heavy use of spin-echo sequences to reduce the influence of dephasing mechanisms.

\subsection{\new{Quantum} error correction}
Classical computers use of error correction schemes intensively. It is to be expected that
quantum computers will employ error correction schemes as well. However, due to the continuous nature of quantum information, it might seem difficult to apply the ideas of classical error correction to quantum computers. However, \citet{Shor1995} and \citet{Steane1996} both found algorithms which correct errors by moving the errors from the quantum register to ancilla systems. In these procedures, a logical qubit is encoded in a number of qubits such that the two-dimensional Hilbert space of a single qubit is
embedded in a higher-dimensional space. Errors will then rotate the state vector out of the allowed subspace. Generalized measurements can project the system back to the allowed subspace and in case of small errors the original state is recovered. Quantum error codes were implemented in NMR for the first time \citep{Cory1998,Knill2001}.

Using trapped ions, \citet{Chiaverini:2004a} implemented a rudimentary quantum error correction protocol. In these experiments, the authors encoded the (arbitrary) state of a source qubit in a superposition of two distinct three-qubit states (the primary qubit + two ancilla qubits), introduced controlled errors (spin flips only) on all three of them, before they decoded the state with the inverse operation used to encode the primary qubit. Small errors occurring on the encoded state rotated the state such that after decoding, the error could be corrected for. Read-out of the ancilla qubits  provided the error syndrome, based on which the primary ion was returned in its original state. Using the language of quantum error correction, the stabilizer code $\{ZZX, ZXZ\}$  (for an introduction to stabilzer codes see for instance Refs.~\citep{Gottesman1997,Nielsen2000}) was employed in these experiments. This
encoding procedure was conveniently implemented by an entangling operation similarly to the
ones discussed in Sec.~\ref{sec:geometric-phase-gate} and used in Sec.~\ref{sec:entangled-states}. However, in this particular instance, the three ions were placed in the standing wave such that they experienced the phases
$\{\varphi,\varphi+2/3\pi,\varphi+4/3\pi\}$ in the lattice. In this symmetric, configuration the ions felt no force if they were all in the same state. In all other logical basis states the total averaged force had the same absolute value and thus the same phase was acquired.
%This interaction is capable of generating two- and three-body GHZ-like entanglement.
Most notably, the heart of the algorithm (encoding and decoding) was executed only with global qubit operations, i.e. without individual addressing. Only for the preparation of the primary qubit, individual addressing  was necessary and for detecting the error syndrome, selective state read-out was used.

The NIST-group studied the performance of the error correction for the three different input states $\{\ket{\!\!\downarrow}$, $\sqrt{0.10}\ket{\!\!\uparrow}-i\sqrt{0.90}\ket{\!\!\downarrow}$, $\sqrt{0.22}\ket{\!\!\uparrow}-i\sqrt{0.78}\ket{\!\!\downarrow}\}$ for artificial error angles $\theta_{\rm e}$ applied to all qubits simultaneously while the qubit was protected by the encoding. Technical imperfections led to a fidelity of about 0.8 even if no error was applied. Therefore the NIST group compared results where the error syndrome was used to results where the correction pulses were not used to correct the primary qubit. For the measurement eigenstate $\ket{\!\!\downarrow}$ and error angles $\theta_{\rm e}<\pi/2$, the fidelity stayed close to 0.8, whereas in the uncorrected case, the fidelity dropped to 0.5. For the two other superposition states, also a clear improvement over the uncorrected implementation was found.

One of the biggest challenges in quantum information processing will be to improve
the fidelity of an error correction algorithm such that it is below a fault-tolerant threshold. Furthermore, the qubit should never be left unprotected. This implies that the error correction has to be applied directly on the encoded qubit. Finally, it will be necessary to apply the error correction  repeatedly and to extend the algorithm to correct for spin flips as well as for phase flips.

\subsection{Semiclassical quantum Fourier-transform}
The quantum Fourier transform is the final step in Shor's algorithm to factor large integers. It is designed to find the periodicity of a quantum state \citep{Shor1994,Coppersmith1994,Ekert1996,Nielsen2000}. \citet{Nielsen2000} show
that the quantum Fourier transformation transforms an $N$-qubit register in binary notation according to ($k_i \in \{0,1\} $)
\begin{eqnarray}
& & \ket{k_{N} k_{N-1} \cdots k_2 k_1} \longrightarrow \nonumber \\
& &  \nonumber \\
& &\:\:\:\:[(\ket{0}+e^{2\pi i[0.k_1]}\ket{1}) \otimes (\ket{0}+e^{2\pi i[0.k_2k_1]}\ket{1}) \otimes \cdots\nonumber \\
& &\:\:\:\: \otimes (\ket{0}+e^{2\pi i[0.k_{N-1}\cdots k_2k_1]}\ket{1}) \otimes (\ket{0}+e^{2\pi i[0.k_Nk_{N-1}\cdots k_2k_1]}\ket{1})] / \sqrt{2^N}\:.
\end{eqnarray}
 Here
$[0.k_N k_{N-1}\cdots k_2k_1]$ stands for
$k_N/2+k_{N-1}/4+\cdots + k_2/2^{N-1}+k_1/2^{N}$. Each qubit is rotated by $R^C(\pi/2,-\pi/2)$ and has acquired a phase shift conditional on other qubits. If the quantum Fourier-transform is the last step in a quantum algorithm, one can take advantage of this structure and perform the semiclassical quantum Fourier-transform  \citep{Griffiths1996}. Starting with the first qubit $k_{i=1}$, first an $R_i^C(\pi/2,-\pi/2)$ pulse is applied to qubit $k_{i}$. Then qubit $k_{i}$ is measured and conditioned on the result $\exp(i \pi \sigma_z/2^{(j-i+1)})$ rotations are carried out on qubits $k_{j}$ for all $j>i$. Note that a $z$~rotation directly preceding a measurement does not change the measurement result, such that the last $z$~rotation before a measurement can be omitted. This procedure is then repeated for the next qubit $i$ with $i$ increased by one.

\citet{Chiaverini:2005a} implemented this procedure and tested it on a variety of separable and entangled three-qubit states. The implementation used a segmented trap, such that the state of one ion could be measured without destroying the quantum state of the others. In this way one after the other ion was measured and  appropriate single-qubit operations were carried out conditioned on the measurement result. Four states
($\ket{001}+\ket{010}+\cdots +\ket{111}$, $\ket{001}+\ket{011}+\ket{101}+\ket{111}$, $\ket{011}+\ket{111}$, $\ket{111}$)
representing all the possible periods one, two, four and eight, respectively, of three bits were successfully tested.
In addition, the entangled state $a_{001}\ket{001}+a_{011}\ket{011}+a_{100}\ket{100}+a_{110}\ket{110}$ ($|a_i|=|a_j|$) with  approximate period three was investigated. This state has the particular property that the result depends on the relative phases of the coefficients $a_{i}$ which was confirmed in the experiments, too.

\subsection{Entanglement purification}
\label{sec:entanglement-purification}
\begin{figure}
\begin{center}
\includegraphics[width=0.8\textwidth,angle=0]{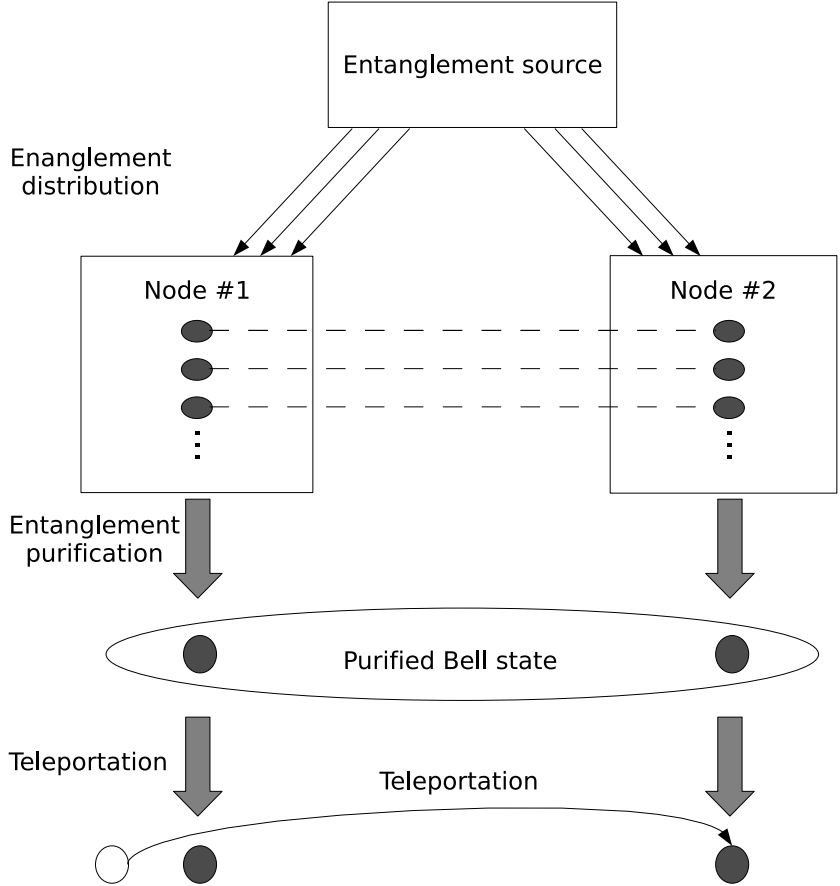}
\end{center}
\caption{\label{fig:entanglement-purification} Schematics of a high-fidelity transfer of quantum information over a noisy quantum channel. First, a number of entangled states of moderate fidelity ($>0.5$) is shared by both parties (quantum nodes). Next, local operations and measurements are used to create a single Bell state with high fidelity. Finally, this qubit pair is used to teleport quantum information from one node to the other. }
\end{figure}

High fidelity entanglement can be obtained from multiple entangled states of lower fidelity by a procedure called entanglement purification.  In the context of fault-tolerant quantum computing, entanglement purification can alleviate thus the stringent requirements for quantum communication \citep{Bennett1996,Gottesman1999}: first many entangled states of a relatively moderate fidelity are created and shared between the two communicating parties. Next, each party carries out local high fidelity quantum operations and measurements before it communicates the measurement results to the other party (see Fig.~\ref{fig:entanglement-purification}). Based on this, the parties can decide when the entanglement purification was successful and which states to keep for further use. This shared entanglement can be employed to perform a high fidelity state transfer between the two nodes via quantum teleportation (see Sec.~\ref{sec:teleportation}). In this way, two quantum nodes can be linked with high fidelity even if the direct transfer of quantum information between them limits the fidelity of Bell pairs to just above 0.5.
Together with quantum teleportation, this Bell state can be used for a high fidelity state transfer (see Sec.~\ref{sec:teleportation}).

\begin{figure}
\begin{center}
\includegraphics[width=0.8\textwidth,angle=0]{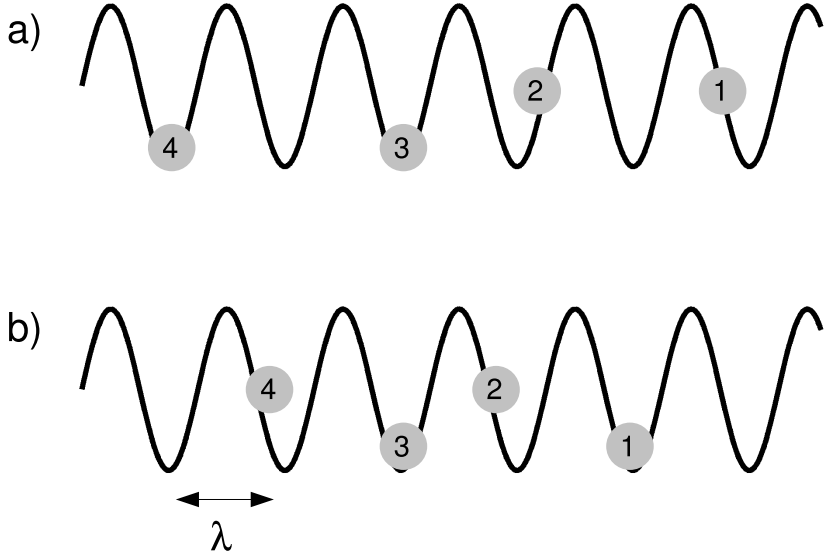}
\end{center}
\caption{\label{fig:standingwave-entpurif} Location of the ions inside the standing wave pattern for a given point in time during the phase gate operations \new{used by the NIST group for entanglement purification}. Panel a) shows the situation for entangling ion \#1 and \#2 on one side and for entangling ion \#3 and \#4  on the other side. The situation to carry out the two phase gates between ion \#1 and \#3 and between ion \#2 and \#4  is illustrated in  panel b). }
\end{figure}

\citet{Reichle2006a} demonstrated the entanglement purification procedure by purifying a two-atom Bell state out of two pairs of Bell states.
First, the NIST-group used the geometric phase gate (see Sec.~\ref{sec:geometric-phase-gate}) to entangle four ions pairwise in a single step. To achieve this, the ions were placed such that ions~\#1 and \#2 as well as ions~\#3 and \#4 were spaced each by multiples of the standing wave ($\Delta x = n \lambda$), while
between ions~\#2 and \#3 was a distance of $(n \pm 1/4) \lambda $ (see Fig.~\ref{fig:standingwave-entpurif}). 
In this way the interaction between the two pairs vanished while within each pair the ions were entangled (ion \#1 and \#2 on one side and \#3 and \#4 on the other side). To implement the purifying procedure, a controlled-NOT operation has to be implemented on each quantum node, ion \#1 and \#3 on one side and ion \#2 and \#4 on the other side. For this, the axial confinement was changed such that there was a distance of $n \lambda$ between ions~\#1 and \#3 as well as between ions~\#2 and \#4, while the distance was $(n \pm 1/4) \lambda$ between ions belonging to different nodes. Thus another geometric phase gate
executed two (local) phase-gate operations simultaneously. Measuring
ion \#1 and \#2 (originally entangled with each other) in opposite states signaled success of the purification and the entanglement
of the other entangled pair increased, while measuring them in the same state signaled failure of the protocol. In order to verify the
entanglement between ions \#3 and \#4, the NIST group split these ions off from the other two ions prior to the measurement, such that
the measurement did not destroy their coherence. In this way, a Bell state with a fidelity of $0.629 \pm 0.0015$ was obtained starting from two states of the form $\alpha\ket{\!\!\uparrow \uparrow}+\beta\ket{\!\!\downarrow\downarrow}$ with a Bell-state fidelity of $0.614 \pm 0.0015$.

\subsection{Quantum simulations}
Quantum simulations appear to be  very appealing and especially feasible
applications of quantum computers. The idea was first put forward by  \citet{Feynman1982} and can be actually thought of as the birth of the field of quantum information processing. Later Lloyd verified this suggestion \citep{Lloyd1996} and together with Braunstein showed how a quantum system with continuous variables can be simulated efficiently \citep{Lloyd1999}. Quantum simulations appear so attractive because they
require only the interactions which are present in the system to be simulated. In addition, often only moderate manipulation fidelities are necessary to obtain meaningful results.

\citet{Leibfried2002} show that the quantum dynamics of a spin-1/2 particle in an arbitrary
potential can be efficiently simulated with a single trapped ion.
In the same publication the NIST-group simulates the action
of an optical Mach-Zehnder interferometer and thus a first quantum simulation with a single ion. In this work, a beam splitter is implemented in the following way: the electronic state of the ion represents the number of photons in one of the incoming modes\footnote{Since only one photon is assumed at the input this description is complete.}, while the motional state of the ion describes the state of one of the output modes. An $R^+(\pi/2,0)$ pulse on the blue sideband can transfer photons from one mode to the other and thus implements the beam splitter. Thus, a sequence of two $R^+(\pi/2)$-pulses on the blue sideband, resembles the action of two beam splitters of
a Mach-Zender interferometer. Furthermore, using second and third order
sideband pulses (detuning $\Delta=n \omega_{\rm t}$, $n=\{2,3\}$) a non-linear interferometer can be implemented where one photon in one mode can generate two and more
photons in the other mode. This process is equivalent to parametric down conversion ---a quite important process in quantum information science with photons \citep{Tittel2001}.

In the experiments, first a single $^9$Be$^+$-ion
was cooled to the motional ground state. Next, a $\pi/2$-sideband pulse of order $n={1,2,3}$ was applied on the axial motional sideband to create
the state $(\ket{\!\!\!\downarrow,0}+\ket{\!\!\!\uparrow,n+1})/\sqrt{2}$. Then a change
in trap frequency induced a phase shift between the two eigenstates. Finally, the phase shift was analyzed with another $\pi/2$-sideband pulse of order $n$.
The phase shift acquired between the two $\pi/2$ pulses is proportional to the energy separation. Thus the fringes are more sensitive to the trap frequency
change for larger $n$. This demonstrates the enhanced sensitivity of a non-linear interferometer as compared to a linear one.

Quantum simulations with trapped ions have also been investigated in the context of quantum phase transitions. \citet{Porras2004b} propose that each ion in the crystal represents a spin. Irradiating the string for example with off-resonant laser radiation produces in each ion a differential level shift which can
be interpreted as a magnetic field acting on the spins. Interaction between the spins, i.e. an effective spin-spin interaction, can be simulated by coupling the ions to a collective vibrational mode. Thus a wide class of Hamiltonians
can be investigated efficiently. In particular, it should be possible to observe quantum phase transitions in ion traps.

 \new{A first step in this direction has already been taken by the Max Planck group in Garching by simulating the phase transition of a tiny quantum magnet consisting of two $^{25}$Mg$^+$ ions from a paramagnetic to a ferromagnetic order
\citep{Schaetz2004a,Friedenauer:2008}. The spin-spin interaction was realized by the same laser-ion coupling used in geometric phase gates (see
Sec.~\ref{sec:geometric-phase-gate}) while the action of a magnetic field in the $x$ direction was simulated by driving local
spin-flips between the hyperfine states of the ions with an RF field of frequency $\omega_{\rm qubit}$. The ground state of an ion with a magnetic field in $x$ direction is given by $\ket{\!\!\rightarrow}=(\ket{\!\!\uparrow}+e^{i\omega_{\rm qubit} t}\ket{\!\!\downarrow})/\sqrt{2}$. Tuning the ratio of the two simultaneously applied Hamiltonians adiabatically, the ground state
$\ket{\!\!\rightarrow\rightarrow}$ of the parametric phase  was transferred to the ground state $\ket{(\!\uparrow \uparrow \!\! + \!\!\downarrow \downarrow)/ \sqrt{2} }$ of the ferromagnetic phase with a fidelity of 0.88.
}

\section{Shuttling and sympathetic cooling of ions}
\label{sec:moving-ions}
Shuttling ions between various traps might relieve the requirements for scalable ion trap quantum computing considerably \citep{Wineland1998,Kielpinski2002}. In accelerator experiments, shuttling  of ions between different traps and re-cooling has been long established to slow down fast ions efficiently \citep{Herfurth2001}. Also single ions have been transported reliably between Penning traps \citep{haeffner2000}. However, for ion trap quantum computing, the requirements are more stringent: in particular, ion strings have to be separated, moved through junctions, recombined and the quantum information must be preserved during all these operations.

In this context, \citet{Rowe2002} demonstrated the reliable transport of single $^9$Be$^+$-ions over 1.2~mm within tens of microseconds in a segmented ion trap. Moreover, they showed that the coherence of the hyperfine qubit was not affected by the transport. For this they first transfer the ion in a superposition with a
 $\pi/2$-pulse, transported the ion and tested for the coherence. A contrast of $95.8\%\pm 0.8\%$ limited by magnetic field fluctuations was measured. In a second set of experiments, the NIST-group used a spin echo sequence to reduce the influence of magnetic field fluctuations. Here the ion (put again in a superposition of physical eigenstates with $\pi/2$ pulses) was transported back and forth. A spin-echo $\pi$ pulse was applied in the remote trap. The contrast of this measurement as well as of a control experiment without transport was almost 97\%, limited by imperfections of the Raman pulses and off-resonant scattering from the P-level. These results demonstrated that the transport did not affect the coherence of the quantum state.

Furthermore, \citet{Rowe2002} investigated motional heating for various transport speeds. For  this, the voltages on the trap electrodes were changed such that the trap minimum moved with time along a $\sin^2$ function. This ensures that not only the velocity but also the acceleration varies
 smoothly with time. They found that the ions can be transported between the two traps within 54~$\mu$s with hardly any observable motional heating (axial trap frequency 2.9~MHz, heating $\lesssim$~0.01~quanta/transport). Only for shorter transporting times heating of the ion motion was observed in agreement with a classical simulation.

In addition, \citet{Rowe2002} demonstrate splitting of ion strings. However, their electrode structure was not optimized for this task as the size of the separation electrodes was too large. Therefore only success rates of 95\% were achieved as well as relatively long separation times of about 10~ms were required. Furthermore, the ions heated excessively by $140\pm 70$~quanta due to the fact that during the separation the ion oscillation frequency was quite small. Later the NIST group demonstrated that these problems can be overcome \citep{Barrett2004} (c.f. Sec. \ref{sec:teleportation}). Separation times of a few 100~$\mu$s as well as small motional heating have been achieved in these experiments. In particular, after splitting one ion from a three ion crystal, the stretch mode of the remaining two-ion crystal was still in the ground state, while the center-of-mass mode had acquired one motional quantum. A theoretical study of the splitting process and the required electrode structures can be found in Ref.~\citet{Home:2006b}

Another requirement for the proposal by \citet{Kielpinski2002} is
the transport through junctions. First experiments were carried
out by \citet{Pearson2006} and \citet{Hensinger2006}. In the
former experiment, charged nano-particles were transported through
a four-way crossing of a planar segmented trap. In the latter,
Cd$^+$ ions were moved through a T-junction. Recently, the linear
transport of ions was studied within the framework of quantum
mechanics \citep{Reichle2006b}, while the non-adiabatic transport
was investigated theoretically by \citet{Schulz2006} and
experimentally by \citet{Huber:2008}. Finally, \citet{Hucul:2008}
analyzed the transport through various junction geometries
(including T junctions) quantum mechanically.

\label{sec:sympathetic-cooling}
After the transport, the ion strings might have to be re-cooled such that the subsequent operations can be carried out with high fidelity. In order to achieve this while maintaining the coherence, a viable way seems to use the strong Coulomb coupling and to cool only a part of or even only one ion of an ion string. Such sympathetic cooling was demonstrated in various experiments  \citep{Drullinger1980,Larson1986,Rohde2001,Hornekaer2001}. Furthermore it has been established in various contexts ranging from precision measurements \citep{Roth2005a} to electron cooling in accelerators \citep{Meshkov1997}.

Within the context of quantum information, \citet{Rohde2001} cooled a two ion string to the motional ground state as required for many two-qubit proposals. However, in these experiments two ions of the same species were used. As typically one of the qubit levels takes part in the cooling process, it is important to shield other nearby ions from the radiation to maintain the coherence of the qubits during the cooling process. This is very difficult if only one ion species is employed. \citet{Blinov2002} cool a $^{112}$Cd$^+$ ion via Doppler cooling of $^{114}$Cd$^+$. The isotope shift would help to preserve the coherence, however, it might be better to use a completely different ion species as done in the early experiments by the NIST group. \citet{Barrett2003} cooled Be$^+$ with Mg$^+$ to the motional ground state (vice versa only to the Doppler limit) and in separate measurements showed that even on time scales of 30~ms a qubit encoded in the hyperfine manifold of $^9$Be$^+$ is not affected by the strong cooling light for Mg$^+$. Finally, \citet{Schmidt2005} took the idea one step further and cooled and detected the internal state of Al$^+$ via Be$^+$.

While all prerequisites for quantum computing by shuttling ion strings have now been demonstrated in separate experiments, the combination of shuttling ions, splitting and re-cooling the ion strings in the same experiment and at the same time preserving the quantum information has yet to be accomplished.

\section{New trap developments}
\label{sec:trap-development}
\begin{figure}
\begin{center}
\includegraphics[width=0.6\textwidth]{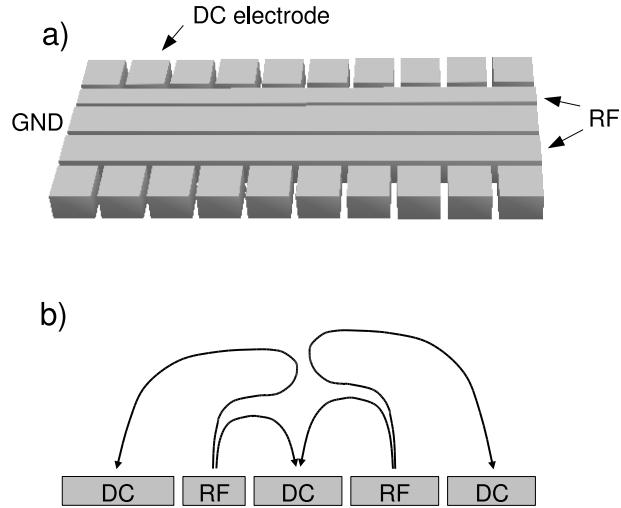}
\caption{\label{fig:planar-trap} Electrode configuration of a planar trap. Fig.~\ref{fig:planar-trap}a shows the a three-dimensional view of the asymmetric electrode arrangement, while Fig.~\ref{fig:planar-trap}b shows a side view with some of the \new{electric} field lines. }
\end{center}
\end{figure}

Parallel to the efforts to shuttle and split ion strings, in particular the NIST group has put quite some effort into developing new traps manufactured by microfabrication techniques to build a medium sized quantum computer. Microfabrication techniques allow for complicated and precise electrode structures. Trap sizes (measured as the distance between two RF electrodes) range from about 200~$\mu$m down to a few tens of $\mu$m.
 While in the beginning mainly three-dimensional designs were built using two substrates, recently a new, planar trap design was invented by \citet{Chiaverini:2005b}. \label{sec:planar-traps}
% Fig.~\ref{fig:planar-trap} shows that
In such traps the electrodes are aranged all in a plane such that a single substrate suffices to mount the electrodes. Typically there are five (four) electrode groups in such a linear surface trap (see Fig.~\ref{fig:planar-trap}): on the center electrode, ground or a small DC potential is applied. The two neighboring electrodes receive an RF potential which provides the radial confinement. The outer electrodes are often structured and various DC potentials provide the axial confinement.
The ions are trapped in the center of the field lines in Fig.~\ref{fig:planar-trap}b where the RF field vanishes. The asymmetry of the RF electrodes tilts the principle axes of the quadrupole field  such that a laser beam traveling parallel to the trap surface has a projection along both radial motional degrees of freedom and all modes can be cooled satisfactory. Two trap designs were successfully used  by the NIST group, one  based on gold on a quartz substrate \citep{Seidelin2006} and the other one based solely on silicon \citep{Britton2006} . Ions have been trapped as close as 40~$\mu$m to the surface with still reasonable heating rates of a few phonons/ms \citep{Seidelin2006,Epstein2007}. The MIT group developed a trap based on printed circuit board fabrication techniques \citep{Pearson2006}. They also built traps using silver on a quartz substrate \citep{Labaziewicz:2008a}.  For trap sizes between 75~$\mu$m and 150~$\mu$m (ion-substrate distance), heating rates between  2 and 20  phonons/s were measured at 4~Kelvin (see Sec.~\ref{sec:motional-coherence}).
\old{Finally, the American government initiated a collaboration between various ion trap groups and Lucent Technologies Bell Laboratories as well as Sandia National Laboratories. Here, very advanced fabrication techniques are used to build traps which also allow for  on-chip voltage filtering and processing.}
%Furthermore, the integration of switchable micro mirrors and focussing optics is investigated.

\section{Future challenges and prospects for ion trap quantum computing}
\begin{figure}
\begin{center}
\includegraphics[width=0.96\textwidth]{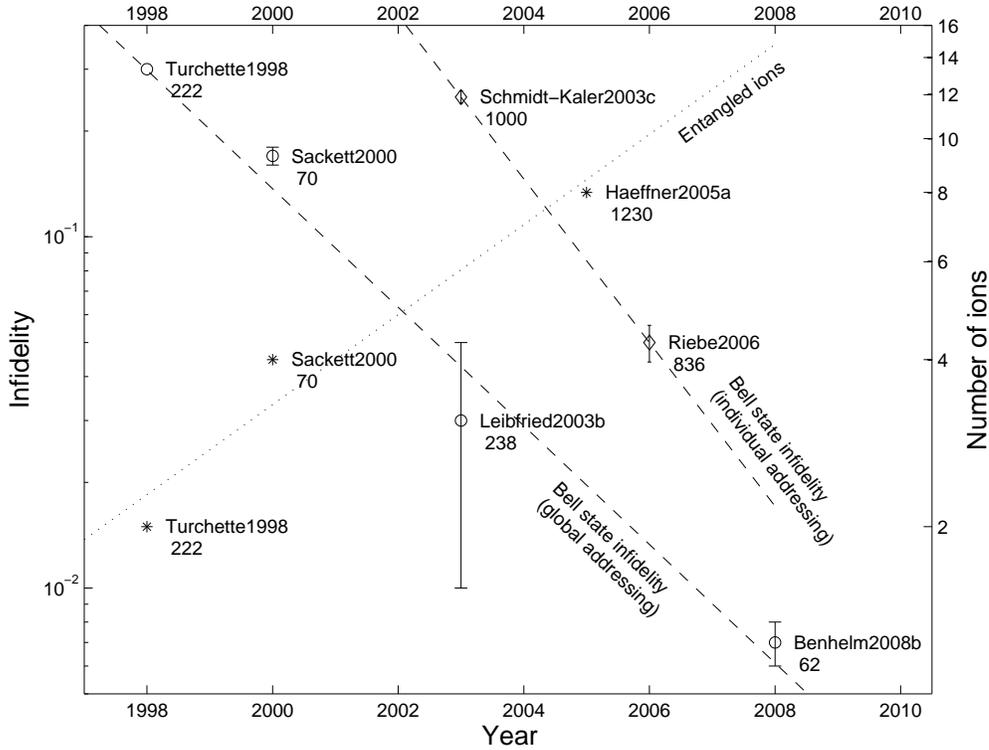}
\caption{\label{fig:two-qubit-fidelities} \new{Progress in reducing the error rate of two-qubit gates (taken from  Ref.~\citet{Benhelm:2008phd}) and in increasing the number of entangled ions.} \new{Open circles represent experiments using two-qubit gate operations with global addressing while diamonds show results based on individual addressing. The performance is measured in terms of the infidelity of produced Bell states. The stars mark the largest number of entangled ions obtained at that time. Numbers below the reference indicate the number of trap cycles required for the operation. Dashed and dotted lines indicate the trends. }}
\end{center}
\end{figure}

In order to achieve universal quantum computing, the algorithms have to be implemented in a fault-tolerant way. It is commonly accepted that this requires quantum error correction. Therefore, currently one of the most important goals  is to implement quantum error correction repeatedly with high fidelity to prolong coherence times and to correct for errors induced by the gate operations. The largest obstacle to perform a successful quantum error correction protocol seems to be the limited fidelity of the operations. The current state of the art for the control in ion trap quantum computing can be summarized as follows:
\begin{itemize}
 \item The qubit coherence times are one or two orders of magnitude longer than the  basic (gate) operations. In specific
cases coherence times longer by more than five orders of magnitude
the gate time are available (see Sec.~\ref{sec:clock-transition}).
In most current experiments, motional decoherence is not a
problem. In addition, it can be further suppressed with cooling
of the trap electrodes (see Sec.~\ref{sec:motional-coherence}).
\item Initialization accuracies are on the order of 0.999 as
discussed in Sec.~\ref{sec:optical-pumping}. Most likely they can
be improved further if necessary. \item Single qubit operation can
be carried out with fidelities exceeding 0.995 \citep{Knill:2007}.
If required, further improvements are possible with more stable
laser fields at the ion positions. \item Implementations of
two-qubit gate operations achieve fidelities of about 0.9--0.99.
Depending on the gate type, various sources limit the fidelity.
Errors are caused by off-resonant scattering, imperfect addressing
of individual qubits, insufficient cooling, laser frequency and
intensity noise. \item The read-out of a single qubit can be
performed with a fidelity of 0.999. Further improvements seem
possible (see Sec.~\ref{sec:state-detection}). \item Ion strings
can be shuttled, split and merged (see Sec.~\ref{sec:moving-ions})
with high fidelity and small decoherence.
\end{itemize}
Currently, two-qubit gate operations seem to be the main limiting factor and receive therefore most
attention both from experimenters and theoreticians. Figure~\ref{fig:two-qubit-fidelities} shows the progress of the fidelity
achieved in the last decade. Most notably \citet{Benhelm:2008b} demonstrate two-qubit gate fidelities high enough to allow in principle fault tolerant quantum computation according to the scheme proposed by \citet{Knill2005}.
In addition, one must always keep in mind that all these requirements have to be met at the same
time. Furthermore, some emphasis should be given to parallel processing  quantum information \citep{Steane2004}.
However, also initialization of the necessary ancillas, read-out, coherence times and the particular layout and the attainable degree of
parallelization are important.

Both, analytical and numerical results, indicate that operational fidelities on the order of $0.9999$/operation seem to be sufficient to achieve fault tolerance (the so-called fault-tolerant threshold), provided certain other criteria can be met, too \citep{Steane2004}: specific errors, error propagation, the allowed overhead, specific requirements, the amount of possible parallelization, amongst others have to be considered to get a full grasp on the situation at hand. Thus the concept of thresholds is oversimplifying the situation.
For instance, \citet{Knill2005} published numerical results which indicate that even error rates on the order of $10^{-2}$ are permitted, however with a huge overhead of $10^{6}$ physical qubits for one logical qubit. It seems reasonable that every operation in a quantum computer should be treated with the same meticulous attention and be implemented as perfect as possible to achieve fault tolerance while keeping the overhead as small as possible.\label{sec:threshold}

Interestingly, state transfer between interconnected ion trap quantum computers at the fault-tolerant level is not necessary as discussed in Sec.~\ref{sec:entanglement-purification} \citep{Gottesman1999,Reichle2006a}: quantum information can be teleported deterministically between two locations using a purified entangled Bell-state. Entanglement between distant ions can, for example, be generated by splitting an entangled two-ion string and transporting one of the ions or by interfering  fluorescence light from two ions on a beam splitter as demonstrated already by \citet{Maunz2007}, \citet{Moehring2007} and \citet{Matsukevich2008}.

Another important issue is the speed of the operations. Having in mind universal quantum computing to outperform classical computers, e.g. in factoring large integers, billions of operations have to be carried out. Thus, the current typical time scales for the basic operations of a few hundreds of $\mu$s seem simply too slow for factoring large numbers even if the operations are carried out to a large extent in parallel. There exist proposals for gate operations which are faster than the trapping frequencies \citep{Garcia-Ripoll2003,Garcia-Ripoll2005,Zhu2006b}, however, there are other bottlenecks such as read-out and ion string separation which might slow down the processor speed.

From this perspective, it seems attractive to work on hybrid devices  where e.g. quantum information is stored in trapped ions (incl. error correction) and most of the processing is implemented e.g. with Josephson junctions \citep{Makhlin2001,Steffen2006a,Plantenberg2007} which operate at speeds roughly three orders of magnitude faster than current ion trapping approaches. Before this can happen, however, tremendous difficulties have to be solved in transferring quantum information  between the two systems within the short coherence times of the Josephson junction qubits. Currently, there exist a few proposals to couple ion-trap and Josephson junction qubits \citep{Tian2004a,Tian2005,Soerensen2004}, but so far almost no experimental results were achievable. Furthermore, within the field of Josephson junction quantum information processing, there are some open challenges before advantage can be taken of the speed of this system. The biggest of these seems to be that coherence times of only a few $\mu$s have been achieved so far.
%In addition, Josephson-junction qubits can currently only be initialzed on time scales which are longer than the coherence time.

The original proposal of Cirac and Zoller \citep{Cirac1995} is scalable in the sense that it does not require exponentially many resources with an increasing number of qubits. Indeed, Fig.~\ref{fig:two-qubit-fidelities} shows that in the last decade larger and larger ion strings have been entangled.
\new{However, it seems impractical to construct a device for a large number of qubits by trapping all ions in the same trap because it gets more and more difficult to obtain the required strong radial confinements to work with linear ion strings at reasonably high axial trapping frequencies. Furthermore, the mode structure of the ion crystal gets more complicated with more ions as well as the speed of the sideband operations is reduced with the larger mass of the crystal. Up to date there are a couple of routes
known which potentially ease these technological challenges. Almost all of them are based}
on distributing the ions across different traps and to interconnect these traps via photons \citep{Cirac1997}, superconducting strip lines \citep{Tian2004a,Heinzen1990} or even via auxiliary ions \citep{Cirac2000}.

\label{sec:scaling}
The currently most advanced procedure, however,  is to merge and shuttle small ion strings in segmented traps \citep{Wineland1998,Kielpinski2002}. Those ideas have been studied in detail and seem to offer a practicable and viable way to scale ion trap quantum computers.  Currently, big efforts are under way in realizing this architecture (c.f. Sec.~\ref{sec:moving-ions}). Major challenges are the fabrication of such complex small ion traps combining high flexibility of ion movement (junctions), low motional heating rates and high trap frequencies. It is also strongly desirable to integrate the control electronics and optics on such ion trap devices.

In summary, \new{the basic requirements for a general purpose quantum computing device with trapped ions have been demonstrated and no fundamental road block is in sight. However, building such a device is extremely challenging.} Especially, the stringent requirements for fault tolerance and for scalability to many thousands of qubits pose huge difficulties.
However, reaching a good control over a reasonable number of qubits seems feasible in the next decade and might be of quite some interest: already with about forty qubits, physical systems can
be simulated which are intractable with current computing technology.
It remains to be seen whether and how the dream of universal quantum computing can be implemented with trapped ions.

\section{Acknowledgments}
We thank Ferdinand Schmidt-Kaler, Wolfgang H\"ansel, Stephan Gulde, Mark Riebe, Gavin Lancaster, J\"urgen Eschner, Christoph Becher, Michael Chwalla, Jan Benhelm, Umakant Rapol, Timo K\"orber, Thomas Monz, Philipp Schindler, Kihwan Kim and Piet Schmidt for their ideas, work on the ion trap apparatus and moral support. Furthermore, we thank Dietrich Leibfried for carefully reading the manuscript. H.H. was partially funded by the Marie-Curie program of the European Union. We gratefully acknowledge also support by the Austrian
Science Fund (FWF), the Army Research Office, by the Institut f\"ur Quanteninformation
Ges.mbH and by the European Commission within the QUEST, CONQUEST, QGATES and SCALA networks.

%\bibliographystyle{abbrvnat.bst}
%\bibliographystyle{elsart-harv.bst}
%\bibliography{../../../anna/all/Documentation/Bibtex/biblio-most-recent}
%\bibliography{biblio-most-recent}

\end{document}